\documentclass[11pt]{article}      
\usepackage{amsmath}
\usepackage{amssymb}
\usepackage{ntheorem}
\usepackage{graphicx}
\usepackage{float}
\usepackage{color}
\usepackage{titletoc}
\usepackage[titletoc]{appendix}
\usepackage{mathrsfs}
\usepackage{subfigure}
\numberwithin{equation}{section}
\newtheorem{thm}{Theorem}
\newtheorem{prop}{Proposition}
\newtheorem{defn}{Definition}
\newtheorem{lem}{Lemma}
\newtheorem{cor}{Corollary}
\newtheorem{exa}{Example}
\newtheorem{rem}{Remark}
\newtheorem*{asm1}{H1}
\newtheorem*{asm2}{H2}

\def\N{{\mathbb N}}
\def\Z{{\mathbb Z}}

\def\R{{\mathbb R}}
\def\C{{\mathbb C}}

\def\bb{\begin}
\def\bc{\begin{center}}
\def\ec{\end{center}}
\def\be{\begin{equation}}
\def\ee{\end{equation}}
\def\ba{\begin{array}}
\def\ea{\end{array}}
\def\bea{\begin{eqnarray}}
\def\eea{\end{eqnarray}}
\def\beaa{\begin{eqnarray*}}
\def\eeaa{\end{eqnarray*}}
\def\bpm{\begin{pmatrix}}
\def\epm{\end{pmatrix}}
\def\bsp{\begin{split}}
\def\esp{\end{split}}

\def\bx{{\bf x}}
\def\bx{{\bf x}}
\def\by{{\bf y}}
\def\bv{{\bf v}}

\def\bm{{\bf m}}
\def\bk{{\bf k}}
\def\bK{{\bf W}}

\def\bq{{\bf q}}

\def\al{\alpha}
\def\bt{\beta}

\def\La{\Lambda}

\def\da{\delta}

\def\e{\varepsilon}
\def\th{\theta}

\def\si{\sigma}

\def\om{\omega}
\def\Om{\Omega}
\def\ups{\upsilon}
\def\ol{\overline}

\def\ka{\kappa}

\def\tl{\tilde}
\def\d{\cdot}

\def\oo{\infty}
\def\f{\frac}

\def\z{\left}
\def\y{\right}
\def\q{\quad}
\def\qq{\qquad}

\def\tm{\times}
\def\wi{\widetilde}

\def\mcm{{\mathcal M}}

\def\mcp{{\mathcal P}}
\def\mcr{{\mathcal R}}
\def\ssk{{\mathcal S}^*_{\bf W}}

\def\mcx{{\mathcal X}}

\def\bu{$\bullet$\ }

\def\lto{\rightarrow}

\def\to{\lto}

\def\andq{\quad \mbox{ and } \quad}
\def\qqf{\qquad \forall}

\def\ifl{\iffalse}
\def\Proof{\noindent{\bf Proof} \quad}
\def\qed{\hfill $\Box$ \smallskip}

\def\nn{\nonumber}
\def\lb{\label}
\def\x#1{{\rm (\ref{#1})}}

\def\inn#1#2{\z\langle #1, \, #2 \y\rangle}

\def\Mfr{R_{\textbf{k}_{h}}}
\def\per{{{\bf 0},\Lambda}}
\def\kla{{{\bf k},\Lambda}}
\def\Kla{{{\bf W},\Lambda}}
\def\bz{{\bf 0}}
\def\Las{\Lambda^*}

\begin{document}
\title{Three-fold Weyl points in the Schr\"odinger operator with periodic potentials}

\author
{Haimo Guo\footnote{E-mail: guohm17@mails.tsinghua.edu.cn.}, \qquad
Meirong Zhang\footnote{E-mail: zhangmr@tsinghua.edu.cn.}, \qquad
Yi Zhu\footnote{E-mail: yizhu@tsinghua.edu.cn.}\\
Department of Mathematical Sciences, Tsinghua University, Beijing 100084, China
}

\maketitle
\tableofcontents\newpage

\begin{abstract}
Weyl points are degenerate points on the spectral bands at which energy bands intersect conically. They are the origins of many novel physical phenomena and have attracted much attention recently. In this paper, we investigate the existence of such points in the spectrum of the 3-dimensional Schr\"{o}dinger operator $H = - \Delta +V(\bx)$ with $V(\bx)$ being in a large class of periodic potentials. Specifically, we give very general conditions on the potentials which ensure the existence of 3-fold Weyl points on the associated energy bands.  Different from 2-dimensional honeycomb structures which possess Dirac points where two adjacent band surfaces touch each other conically, the 3-fold Weyl points are conically intersection points of two energy bands with an extra band sandwiched in between. To ensure the 3-fold and 3-dimensional conical structures, more delicate, new symmetries are required. As a consequence, new techniques combining more symmetries are used to justify the existence of such conical points under the conditions proposed. This paper provides comprehensive proof of such 3-fold Weyl points. In particular, the role of each symmetry endowed to the potential is carefully analyzed. Our proof extends the analysis on the conical spectral points to a higher dimension and higher multiplicities. We also provide some numerical simulations on typical potentials to demonstrate our analysis.
\end{abstract}


\section{Introduction and Notations}
\subsection{Introduction}\lb{intr}

Weyl points are singular points on the 3-dimensional spectral bands of an operator with periodic coefficients, at which two distinct bands intersect conically.  Much attention has been paid to looking for such fundamental singularities in various physical systems in the past few decades \cite{Ablowitz09,Allaire08,Castro07,Novoselov11}. They are the hallmark of many novel phenomena.  Many materials such as graphene exhibit such unusual singular points on their energy bands \cite{Castro07,Novoselov11}. These singular points carry topological charges and play essential roles on the formation of topological states, for instance chiral edge states or surface states \cite{Avila2012,Goldman2012,Graf2012,Perez2014}. In the past decade, constructing and engineering the conically degenerate spectral points become one of the major research subjects in many fields. Accordingly, understanding the existence of these points on the energy bands and their connections to interesting physical phenomena are extremely important in both theoretical and applied fields.

How to obtain and justify the existence of such degenerate points become urgent in various physical systems. For instance, it is well known that honeycomb structures give rise to the existence of Dirac points in 2-D systems. The existence of Dirac points in the periodic system was first reported by Wallace in the tight-binding model and demonstrated in the continuous systems by numerical and asymptotic approaches \cite{Ablowitz2013,Ablowitz2012,Avila2012,Wallace1947}. However, the rigorous justification on the existence of Dirac points for 2-D Schr\"odinger equation with a generic honeycomb potential was recently given by Fefferman and Weinstein \cite{fefferman2012honeycomb}. They used very simple conditions to characterize honeycomb potentials and developed a framework to rigorously justify the existence of Dirac points. Their framework paved the way for the mathematical analysis on such degenerate points, and their method has been successfully extended to other 2-D wave systems \cite{Drouot2020ubiquity,Drouot2020,Keller2018,Lee-Thorp2017}. There are also other rigorous approaches to demonstrate the existence of Dirac points.  Lee treated the case where the potential is a superposition of delta functions centered on sites of the honeycomb structure \cite{Lee2016}. Berkolaiko and Comech used the group representation theory to justify the existence and persistence of Dirac points \cite{Berkolaiko14}. The low-lying dispersion surfaces of honeycomb Schr\"odinger operators in the strong binding regime, and its relation to the tight-binding limit, was studied in \cite{Weinstein2016}. Ammari et al. applied the layer potential theory to honeycomb-structured Minnaert bubbles \cite{Ammari2018honeycomblattice}. Based on the rigorous justification of the existence of Dirac points, a lot of rigorous explanations on the related physical phenomena have been extensively investigated. For example, the effective dynamics of wave packets associated with Dirac points were studied in \cite{Ammari2020,Fefferman2014,hu2020linear,Watson18,Xie2019,xie2020wave}. The existence of edge states and associated dynamics are studied in \cite{Bal2019,Drouot2020,Fefferman2015}.

Despite successful applications on the aforementioned analysis of the Dirac points in 2-D systems, the advances in applications such as materials sciences, condensed matter physics, placed new theoretical demands that are not entirely met. Just as Kuchment pointed out in a recent overview article on periodic elliptic operators \cite{Kuchment2015}, "the story does not end here". One important missing piece is the analysis of 3-dimensional degenerate points which are referred to as Weyl points. Another piece is the conical points with higher order multiplicities.  In the literature, some special structures are proposed to admit Weyl points \cite{leng2020multiband,Tang17,Yang2017,Lu2020Double}. However, most constructions and demonstrations are based on either tight-binding models, numerical computations or formal asymptotic expansions. To the best of our knowledge, no similar construction and rigorous analysis as aforementioned literature have been given for Weyl points with higher-order multiplicities. Due to the importance and potential applications of Weyl points in quantum mechanics, photonics and mechanics, such generic analysis is highly desired. This is the goal of our current work.

This work is concerned with the $L^{2}(\R^3)$-spectrum of the following 3-dimensional Schr\"odinger equation
    \begin{equation} \lb{HV}
H=-\Delta + V(\bx), \q \bx\in \R^3,
    \end{equation}
where the potential $V(\bx)$ is real-valued and periodic. By Floquet-Bloch theory \cite{Kuchment1993,Kuchment2015,Kuchment2000}, the spectrum of $H$ in $L^{2}(\R^3)$ is the union of all energy bands $E_b(\bk), ~b\ge 1$ for all $\bk$ in the Brillouin zone. For some specific $V(\bx)$, two energy bands may intersect with each other conically at some $\bk_*$. This degenerate point $\bk_*$ in the three dimensional energy bands is called a Weyl point. There are different types of Weyl points depending on the multiplicity of degeneracy.

In this work, we shall give a simple construction of three-fold Weyl points, i.e., two energy bands intersect conically with an extra band between them. We shall also rigorously justify the existence of such degenerate points by using the strategies developed in \cite{fefferman2012honeycomb}. More specifically,  we first propose a very general class of admissible potentials which are characterized by several symmetries. Different from honeycomb potentials in which the inversion and $2\pi/3$-rotational symmetries are the indispensable ingredients, the potentials proposed in this paper have two rotational symmetries in addition to the inversion symmetry. These three symmetries together guarantee the three-fold degeneracy at certain high symmetry points and ensure conical structures in their vicinities.  Our analysis in this work involves many novel arguments on the eigenstructures at the high symmetry points in order to explain how the 3-fold degeneracy is protected by the underline symmetries and why the Fermi velocities corresponding to different branches are the same. These important arguments are relatively trivial in the honeycomb case \cite{fefferman2012honeycomb}. Our current work not only extends the theory developed in \cite{fefferman2012honeycomb} to 3-dimensional systems but also shines some light on the analysis of singular points with higher multiplicities. Our analysis also provides the starting point of future theoretical analysis on these higher order Weyl points, such as the existence of chiral surface states, Fermi arcs and so on\cite{Yang2017,Lu2020Double}.

This work is organized as follows. In Section 2, we first introduce the lattice $\Lambda$ and its dual lattice $\Lambda^{*}$ together with their 
fundamental cells $\Omega$ and $\Omega^{*}$, and then we precisely discuss the existence of high symmetry points $\bk_h$ in $\Omega^{*}$. Section 2 concludes with the Fourier analysis of $\Lambda$-periodic functions. Section 3 contains the definition of the admissible potentials characterized by several symmetries. We also review the relevant Floquet-Bloch theory for Schr\"odinger operators $H=-\Delta+V(\bx)$. In Section 4, we first propose required conditions of eigenstructures at high symmetry point $\bK$ for some eigenvalue $\mu_{*}$, i.e., $\textbf{H1-H2}$ and their consequences. We then prove the energy bands in the vicinity form a conical structure with an extra band in the middle. In Section 5, we justify that the required conditions $\textbf{H1-H2}$ do hold for nontrivial shallow admissible potentials. Specifically, we clearly show the significance of the $\mathcal{R}$ and $\mathcal{T}$ symmetries to preserve the multiplicity of eigenvalues of $H^{\e}=-\Delta+\e V(\bx)$ at $\bK$ while $\e$ is sufficiently small. Moreover, the justification is extended to generic admissible potentials. Section 6 discusses the instability of the Weyl points and perturbations of dispersion bands of when $V(\bx)$ is violated by an odd potential $W(\bx)$. Section 7 provides detailed numerical simulations of the energy bands and Weyl points in different cases for a special choice of admissible potential. In Appendix A, we present the proofs of certain Propositions and Lemmas in Section 4 and Section 5.

%
%

\subsection{Notations and conventions} \lb{nota}

Without specifications, we use the following notations and definitions.

\bu For $z\in \C$, $\ol{z}$ denotes the complex conjugate of $z$.

\bu For $\bx,\ \by\in\C^{n}$, $\langle \bx,\by \rangle:=\ol{\bx}\cdot\by=\ol{x}_{1} y_{1}+...+\ol{x}_{n}y_{n}$, and $|\bx|:= \sqrt{\langle \bx,\bx\rangle}$.

\bu For a matrix or a vector $A$, $A^{t}$ is its transpose and $A^{*}$ is its conjugate-transpose.

\bu $\Lambda\in\mathbb{R}^3$  denotes the lattice, and $\Lambda^{*}\subset(\mathbb{R}^3)^{*}=\mathbb{R}^{3}$ denotes the dual lattice of $\Lambda$. Moreover, $\bv_j,\ j=1,2,3$ are the basis vectors of $\Lambda^{*}$, while $\bq_{\ell},\ \ell=1,2,3$ are the dual basis vectors of $\Lambda^{*}$, which are chosen to satisfy $\bv_{j}\cdot \bq_{\ell}=2\pi \delta_{\ell j}$.

\bu $\langle f,g \rangle_{D}=\int_{D} \overline{f}g$ is the $L^{2}(D)$ inner product. In this work, the region $D$ of integration is assumed to be the fundamental cell $\Omega$ if it is not specified.

\bu $\nabla=(\partial_{x_{1}},\partial_{x_2},\partial_{x_3})^{T}$.

\bu $I$ denotes the $3\times3$ identity matrix.

\bu For $\ka=(\ka_x ,\ka_y ,\ka_z)\in \mathbb{R}^{3}$, $\ka^{\arg}$ represents $\frac{\ka_x \ka_y \ka_z}{|\ka|^{3}}$.

\section{Preliminaries} \lb{prel}




\subsection{The lattice $\Lambda$ and the rotation $R$} \lb{La-R}

Consider the following linearly independent vectors in $\R^3$ 
    \[
    \bv_1=\frac{a}{\sqrt{3}}\begin{pmatrix}1\\-1\\-1\end{pmatrix}, \q \bv_2=\frac{a}{\sqrt{3}}\begin{pmatrix}1\\1\\1\end{pmatrix},\q \bv_3=\frac{a}{\sqrt{3}}\begin{pmatrix}-1\\1\\-1\end{pmatrix}.
    \]
Here $a>0$ is the lattice constant. Define the lattice as follows
    $$
    \La=\mathbb{Z}\bv_{1} \oplus \mathbb{Z}\bv_{2} \oplus \mathbb{Z}\bv_{3}:=\{n_{1}\bv_{1} +n_{2}\bv_{2} +n_{3}\bv_{3}: n_{1}, n_{2}, n_{3}\in \mathbb{Z}\}.
    $$
The parameter $a$ then gives the distance between nearest neighboring sites.  The fundamental period cell of $\La$ is
    \be\lb{la}
\Om:= \{x_{1}\bv_{1}+ x_{2}\bv_{2}+ x_{3}\bv_{3}: 0\leq x_{i}\leq1,\ i=1,2,3\}.
    \ee

Let $\bq_{1}$, $\bq_{2}$, $\bq_{3}\in \R^3$ be the dual vectors of $\bv_{1}$, $\bv_{2}$, $\bv_{3}$, in the sense that
    \[
\bq_{\ell}\cdot\bv_{j}=2\pi \delta_{\ell j}, \q \ell,j=1,2,3.
    \]
Explicitly,
    \[
\bq_{1}=q\begin{pmatrix}1\\0\\-1\end{pmatrix},\q
\bq_{2}=q\begin{pmatrix}1\\1\\0\end{pmatrix},\q
\bq_{3}=q\begin{pmatrix}0\\1\\-1\end{pmatrix},
    \]
where $q=\frac{\sqrt{3}\pi}{a}$. Then the dual lattice of $\La$ is defined as
    \[
\Las=\mathbb{Z}\bq_{1}\oplus\mathbb{Z}\bq_{2}\oplus\mathbb{Z}\bq_{3}:=\{m_{1}\bq_{1}+m_{2}\bq_{2}+m_{3}\bq_{3}:m_{1},m_{2},m_{3}\in \mathbb{Z}\}.
    \]
The fundamental period cell of $\Las$ is chosen to be
    \begin{equation*}
    \Om^{*}:=\{c_1 \bq_1 + c_2 \bq_2 + c_3 \bq_3 :c_i \in \z(-1/2,1/2\y],i=1,2,3\}.
    \end{equation*}



In this work, we are interested in the following rotation transformation $R$ in $\mathbb{R}^{3}$
    \be\label{matrix R}
R= \begin{pmatrix} 0&-1&0\\1&0&0\\0&0&-1\\ \end{pmatrix}.
    \ee
Obviously, $R^t R = R R^t = I$. Moreover,
    \be\label{matrix}
R^{*}= R^t=R^{-1}=\begin{pmatrix} 0&1&0\\-1&0&0\\0&0&-1\\ \end{pmatrix},
\andq R^4=I.
    \ee
By direct calculations, we can conclude the following proposition.

\begin{prop}\lb{rem21}

\noindent$(1)$ The eigenvalues of $R$ is $i^{\ell},\ell=1,2,3$, with the corresponding eigenvectors
    \be\lb{R-ev}
    \om_1=\f{1}{\sqrt{2}}(1,-i,0)^t, \qq \om_2=(0,0,1)^t,\qq \om_3=\f{1}{\sqrt{2}}(1,+i,0)^t=\ol{\om_1}.
    \ee
$(2)$ $R^{*}$ and $R$ satisfy
    \be\label{Rq}
    \bb{split}
    & R^{*}\bv_{1}=\bv_{4},\q R^{*}\bv_{2}=\bv_{1},\q R^{*}\bv_{3}=\bv_{2},\q R^{*}\bv_{4}=\bv_{3}, \\
    & R\bq_{1}=\bq_{2}-\bq_{1},\q R\bq_{2}=\bq_{3}-\bq_{1},\q R\bq_{3}=-\bq_{1},\\
    & R^* \bq_{1}=-\bq_{3},\q R^{*}\bq_{2}=\bq_{1}-\bq_{3},\q R^{*}\bq_{3}=\bq_{2}-\bq_{3}.
    \end{split}
    \ee
Thus both $R$ and $R^*$ leave $\La$ and $\Las$ invariant.
\end{prop}

    \begin{defn}\lb{hsp}
A point $\bk_{h}\in\R^3$ is defined to be a high symmetry point with respect to $R$ if
    \[ 
    R\bk_{h}-\bk_{h}\in\Las.
    \]
    \end{defn}

    \bb{rem}\lb{k+q}
 By understanding $\Las_{\bk_h}:=\bk_h+\Las$ as shifted lattices, we know that $\bk_h$ is a high symmetry point if and only if $R$ leaves $\Las_{\bk_{h}}$ invariant, i.e.,
    \[
    R(\Las_{\bk_h}) = \Las_{\bk_{h}}.
    \]

\ifl{\bf
Define, for $\bk\in \R^3$,  the shifted lattice
    \[
    \Las_\bk:= \bk+\Las=\z\{ \bk+\bq: \bq \in \Las\y\}.
    \]
Then $\bK$ is a high symmetry point if and only if
    \be \lb{hsp11}
    R(\Las_\bK) = \Las_\bK,
    \ee
i.e. $R$ leaves $\Las_\bK$ invariant.}
\fi
    \end{rem}

The following lemma indicates that inside the fundamental period cell $\Om^*$, there exist precisely four high symmetry points.
    \bb{lem} \lb{fps}
A point $\bk_h=c_1\bq_{1}+c_2\bq_{2}+c_3\bq_{3}\in \Omega^{*} $ 
is a high symmetry point with respect to $R$ if and only if the coefficients $(c_1,c_2,c_3)$ take the following $4$ cases
    \be\lb{4k}
(c_1,c_2,c_3)=(0,0,0), \
\z(1/2,1/2,1/2\y), \
\z(1/4,1/4,1/4\y), \
\z(-1/4,-1/4,-1/4\y).
    \ee
    \end{lem}

\Proof  By (\ref{Rq}), we have
    \[
R\bk_h=(-c_1-c_2-c_3)\bq_{1}+c_1\bq_{2}+c_2\bq_{3}.
    \]
Then
    \be \lb{K1}
R\bk_h-\bk_h=(-2c_1-c_2-c_3)\bq_{1}+(c_1-c_2)\bq_{2}+(c_2-c_3)\bq_{3}\in\Las
    \ee
is the same as
    \[
    \z(-2c_1-c_2-c_3,\, c_1-c_2,\, c_2-c_3\y)\in\mathbb{Z}^3.
    \]
Due to the restrictions $c_i\in  (-1/2,1/2]$, \x{K1} has four solutions listed as in \x{4k}. \qed

In this work, we only focus on the following specific high symmetry point
    \[
\bK:= -\frac{1}{4}(\bq_{1}+\bq_{2}+\bq_{3})=q\z(-\frac12,-\frac12,\frac12\y).
    \]
It follows from \x{Rq} and \x{K1} that
    \begin{equation}\lb{RK}
R^4\bK=\bK, \andq R^\ell\bK=\bK+\bq_{\ell}\mbox{ for } \ell=0,1,2,3.
    \end{equation}
Here $\bq_0:=\bz$.

\ifl
The study for the following high symmetry point is analogous
   \[
\bK':=\frac{1}{4}(\bq_{1}+\bq_{2}+\bq_{3})=q(\frac12,\frac12,-\frac12)^t.
    \]

     \begin{equation}\lb{RK}
R\bK=q(1,-1,-1)^{t}=\bK+\bq_{1},\q
R^{2}\bK=q(1,1,1)^{t}=\bK+\bq_{2},\q
R^{3}\bK=q(-1,1,-1)^{t}=\bK+\bq_{3}.
    \end{equation}

Introduce the zone $\Om$ by denoting its vertices as:
    \begin{equation}
A\equiv \pi(1,1,-1)^t,\q B\equiv\pi(1,-1,-1)^{t},\q C\equiv\pi(1,1,1)^{t},\q D\equiv\pi(-1,1,-1)^{t}.
    \end{equation}
Then $R^{*}$ maps $\Om$ to itself.
\fi

\subsection{$\La$-periodic, $\La$-pseudo-periodic functions and Fourier expansions} \lb{Per-QPer}

We say that a function $f(\bx): \R^3\to \C$ is $\La$-periodic if
    \be \lb{per}
    f(\bx +\bv) = f(\bx)\q \forall \bx\in \R^3, \ \bv\in \La.
    \ee
More generally, given a quasi-momentum $\bk\in \R^3$, we say that a function $F(\bx): \R^3\to \C$ is $\La$-pseudo-periodic with respect to $\bk$ if
    \be \lb{kper}
    F(\bx +\bv) =e^{i \bk \d \bv} F(\bx)\q \forall \bx\in \R^3, \ \bv\in \La.
    \ee
Let us introduce the Hilbert space
    \[
    L^2_\kla:= \z\{F(\bx)\in L^2_{\rm loc}(\R^3,\C): F(\bx)\mbox{ satisfies \x{kper}} \y\},
    \]
where the inner product is
    \[
    \inn{F}{G}:= \int_\Om \ol{F(\bx)} G(\bx) d\bx \mbox{ for } F, \ G \in L^2_\kla.
    \]

    Similarly, we define
    \[
    H^{s}_{\bk,\Lambda}=\{F(\bx)\in H^{s}(\mathbb{R}^3,\mathbb{C}):\ F(\bx)\ \mbox{satisfies \x{kper}}.  \}
    \]
In particular, for $\bk={\bf 0}$,
    \[
    L^2_\per= L^2(\R^3/\La) :=  \z\{f(\bx)\in L^2_{\rm loc}(\R^3,\C): f(\bx)\mbox{ staisfies \x{per}} \y\}
    \]
is the space of square-integrable $\La$-periodic functions. Obviously, $F(\bx)\in L^2_\kla$ if and only if
    \[
    f(\bx):= e^{-i \bk \d \bx} F(\bx) \in L^2_\per.
    \]
That is, the mapping
    \be \lb{ff}
    f(\bx) \longmapsto F(\bx) :=e^{i \bk \d \bx} f(\bx)
    \ee
gives a one-to-one correspondence between $L^2_\per$ and $L^2_\kla$. Moreover, it is easy to see that
    \[
    \inn{F}{G} =\inn{f}{g} \qqf f, \ g\in L^2_\per.
    \]
That is, the mapping \x{ff} is an isometry from $L^2_\per$ to $L^2_\kla$.

Due to the $\La$-periodicity of functions $f(\bx)\in L^{2}_\per$, they can be expanded as Fourier series of the form
    \be\lb{fep}
    f(\bx) = \sum_{\bq\in \Las} \hat f_\bq e^{i \bq\d \bx},
    \ee
where $\z\{\hat f_\bq\y\}_{\bq \in \Las}\subset l^{2}(\Lambda)$ is the sequence of Fourier coefficients, indexed using the discrete indexes $\bq$ from $\Las$. Explicitly,
    \be \lb{fc0}
    \hat f_\bq := \f{1}{|\Om|} \int_\Om f(\bx) e^{-i \bq \d \bx} d\bx,
    \ee
where $|\Om|$ denotes the volume of the cell $\Om$. Such a form \x{fep} of Fourier expansions is consistent with
Example \ref{example-potential} and is more convenient for later uses. Note that
    \[ 
    \z\{\hat f_\bq\y\}_{\bq \in \Las}\in l^2_{\Las},
    \]
the Hilbert space of square-summable complex sequences over the dual lattice $\Las$.

    \bb{rem}\lb{k-exp}
Given $\bk\in \R^3$, pseudo-periodic functions $F(\bx)=e^{i \bk \d \bx} f(\bx) \in L^2_\kla$ can be expanded as
    \be \lb{F-e}
    F(\bx) = e^{i\bk\cdot\bx}\sum_{\bq\in \Las} \hat f_\bq e^{i \bq \d \bx}=\sum_{\bq\in \Las} \hat f_\bq e^{i (\bk+\bq) \d \bx},
    \ee
where $\{\hat f_\bq\}$ is as in \x{fc0}.
    \end{rem}

Rotations $R$ and $R^{*}$ in \x{matrix R} and (\ref{matrix}) can yield a transformation $\mcr$ for functions $F(\bx) \in L^2_\kla$ by
    \[
\mcr[F](\bx):=F(R^{*}\bx)\q\mbox{for } \bx\in \R^3.
    \]

    \bb{lem}\lb{iso}
Let $\bk_h$ be a high symmetry point w.r.t. $R$. Then

\bu $\mcr$ maps $L^2_{\bk_h,\Lambda}$ to itself as a unitary operator.

\bu Define an affine transformation $\Mfr: \Las\to \Las$ by
    \begin{equation}\label{eq:mfr}
    R_{\bk_h} (\bq) := R\bq+R\bk_h-\bk_h\q\mbox{for } \bq \in \Las.
    \end{equation}
Then, for any $\ell\in\Z$, one has
    \be \lb{mfr-l}
    R_{\bk_h}^\ell (\bq) =R^\ell \bq + R^\ell \bk_h-\bk_h = R^\ell(\bq+\bk_h)- \bk_h \q\mbox{for} \bq \in \Las.
    \ee
In particular,
    \be \lb{R4}
    R_{\bk_h}^4 (\bq) = \bq \q\mbox{for}\ \bq \in\Las.
    \ee

\bu The action $\mcr$ on $L^2_{\bk_h,\Lambda}$ is given by
    \be \lb{RR}
    \mcr\z[\sum_{\bq\in \Las} \hat f_\bq e^{i (\bk_h+\bq) \d \bx}\y]  =\sum_{\bq\in \Las} \hat f_\bq e^{i R(\bk_h+\bq) \d \bx}
    =\sum_{\bq\in \Las} \hat f_\bq e^{i (\bk_h+R_{\bk_h}(\bq)) \d \bx}\ .
    \ee
    \end{lem}

\Proof \bu For $F(\bx)=e^{i\bK \d \bx} f(\bx)\in L^2_{\bk_h,\Lambda}$, we can use expansion \x{F-e} to obtain
    \be \lb{RF1}
    \begin{split}
    \mcr&\z[e^{i\bk_h \d \bx} f(\bx)\y] = \sum_{\bq\in \Las} \hat f_\bq e^{i (\bk_h+\bq) \d R^*\bx}= \sum_{\bq\in \Las} \hat f_\bq e^{i R(\bk_h+\bq) \d \bx}
    \\&=e^{i\bk_h\cdot\bx} \sum_{\bq\in \Las} \hat f_\bq e^{i (R\bq +R\bk_h-\bk_h) \d \bx}\ .
    \end{split}
    \ee
As $R$ leaves $\Las$ invariant and $R\bk_h-\bk_h\in\Las$, we have $R_{\bk_h} (\bq)=R\bq +R\bk_h-\bk_h\in \Las$ for all $\bq \in \Las$. Thus
    \[
    \sum_{\bq\in \Las} \hat f_\bq e^{i (R\bq +R\bk_h-\bk_h) \d \bx}\in L^2_\per \andq \mcr[F]\in L^2_{\bk_h,\Lambda}.
    \]

Moreover, for $F(\bx)=e^{i\bk_h \d \bx} f(\bx),\ G(\bx)=e^{i\bk_h \d \bx} g(\bx)\in L^2_{\bk_h,\Lambda}$, one has
    \[
    \bb{split}
    & \inn{\mcr[F]}{\mcr[G]}=\int_\Om \ol{F(R^* \bx)} G(R^* \bx)d\bx \\
    & = \int_\Om \ol{f(R^*\bx)} g(R^*\bx)d\bx= \int_{R^*(\Om)} \ol{f(\by)} g(\by)d\by =\inn{f}{g}\\
    & =\inn{F}{G},
    \end{split}
    \]
because $R^*$ is an orthogonal transformation and both $f(\by)$ and $g(\by)$ are $\La$-periodic in $\by\in \R^3$. This shows that $\mcr$ is unitary.

\bu Let us check \x{mfr-l} only for $\ell\in \N$. By \x{eq:mfr}, we have for $\bq\in \Las$
    \[
    R_{\bk_h}^\ell(\bq)=R^\ell\bq + \sum_{j=0}^{\ell-1} R^j(R\bk_h - \bk_h)= R^\ell\bq + R^\ell \bk_h-\bk_h = R^\ell(\bq+\bk_h)-\bk_h,
    \]
the desired equalities in \x{mfr-l}.

By letting $\ell=4$ in \x{mfr-l}, we obtain \x{R4} because $R^4=I$.

\bu Using \x{F-e} and \x{eq:mfr}, equality \x{RF1} can be written as \x{RR}.\qed

    \bb{rem}\lb{k-K}
For $\bk_{h}=\bz$, one has $R_\bz=R$. For $\bk_h=\bK$, one has from \x{RK} that
    \[
    R_{\bK}(\bq)\equiv R \bq + \bq_1\q\mbox{for } \bq \in \Las.
    \]
    \end{rem}

\subsection{Decompositions of periodic and pseudo-periodic functions} \lb{Four}

In the following discussions we only consider the special high symmetry point $\bK$.
Notice from \x{RK} that $R_\bK^4= I$ on $\Las$, and
    \[
    R_\bK^\ell \ne I \q \mbox{on $\Las$\q for $\ell=1,2,3.$}
    \]

Each orbit of the action $R_\bK$ on $\Las$ consists of precisely four points. Let us introduce
    \[
    \ssk:= \Las/R_\bK = \Las/\z\{\bq \sim \bq': \bq, \ \bq'\in \Las, \ \bq'= R_\bK^{\ell} (\bq) \mbox{ for some } \ell\in \Z  \y\}\hookrightarrow \Las.
    \]
Then functions $F(\bx)=e^{i\bK \d \bx} f(\bx)\in L^2_\Kla$ can be decomposed into
    \be\label{S-l}
    \begin{split}
F(&\bx) =\sum_{\bq\in\Las}\hat f_{\bq}e^{i(\bK+\bq)\cdot\bx}= \sum_{\bq\in\ssk}  \sum^{3}_{\ell=0} \hat f_{R_\bK^\ell (\bq)} e^{i (\bK+R_\bK^\ell (\bq)) \cdot\bx}\\&= \sum_{\bq\in\ssk}  \sum^{3}_{\ell=0} \hat f_{R_\bK^\ell (\bq)} e^{i R^\ell(\bK+\bq) \cdot\bx}
= \sum_{\bq\in\ssk} (\hat f_{\bq}e^{i(\bK+\bq)\cdot\bx}+\hat f_{R_\bK(\bq)}e^{iR(\bK+\bq)\cdot\bx}
 \\&+ \hat f_{R_\bK^{2}(\bq)} e^{iR^{2}(\bK+\bq)\cdot\bx} +\hat f_{R_\bK^{3}(\bq)}e^{iR^{3}(\bK+\bq)\cdot\bx}).
 \end{split}
    \ee

Since $R^4=I$ and $R^{*4}=I$, one has $\mcr^{4}=I$ on $L^2_\per$. Hence eigenvalues $\si$ of the unitary operator $\mcr$ must satisfy $\si^4=1$. In fact, one has
    \begin{equation}\lb{si}
\si= i^\ell, \qq \ell=0,1,2,3.
    \end{equation}
Then we have an orthogonal decomposition for $L^{2}_\per$
    \begin{equation}\lb{dec0}
L^{2}_\per=L^{2}_{\bz,1}\oplus L^{2}_{\bz,i}\oplus L^{2}_{\bz,-1}\oplus L^{2}_{\bz,-i},
    \end{equation}
where the eigenspaces are
    \[
L^{2}_{\bz,i^\ell}:=\z\{f\in L^{2}_\per:\mcr[f]=i^\ell f \y\}, \qq \ell=0,1,2,3.
    \]
Note that \x{dec0} also yields an orthogonal decomposition for the space $L^2_\Kla$
    \[
L^{2}_\Kla=L^{2}_{\bK,1}\oplus L^{2}_{\bK,i}\oplus L^{2}_{\bK,-1}\oplus L^{2}_{\bK,-i},
    \]
where
    \[
L^2_{\bK,i^\ell} := 
\left\{ e^{i \bK \d \bx} f(\bx): f(\bx ) \in L^2_{\bz,i^\ell}\right\} \equiv \z\{F\in L^2_\Kla: \mcr[F] = i^\ell F \y\}, \ \ell=0,1,2,3.
    \]

Let $\si$ be as in \x{si} and
    \[
    F(\bx) = \sum_{\bq\in\Las}\hat f_{\bq}e^{i(\bK+\bq)\cdot\bx}\in L^2_{\bK, \si}\ .
    \]
Then
    \be \lb{Fl}
    \mcr^\ell[F]= \si^\ell F\qqf \ell\in \Z.
    \ee
By \x{RR}, we have
    \[
    \mcr^\ell[F](\bx)  =\sum_{\bq\in \Las} \hat f_\bq e^{i R^\ell(\bK+\bq) \d \bx}
    =\sum_{\bq\in \Las} \hat f_\bq e^{i (\bK+R_\bK^\ell\bq) \d \bx}\equiv \sum_{\bq\in \Las} \hat f_{R_\bK^{-\ell}(\bq)} e^{i (\bK+\bq) \d \bx}  .
    \]
Since
    \[
    \si^\ell F(x) = \sum_{\bq\in \Las} \si^\ell \hat f_\bq e^{i (\bK+\bq) \d \bx},
    \]
we deduce from \x{Fl} that the Fourier coefficients $\hat f_\bq$ satisfy
    \[
    \hat f_{R_\bK^{-\ell}(\bq)}=\si^\ell \hat f_\bq \qqf \bq\in \Las,
    \]
i.e.,
    \be \lb{FCr}
    \hat f_{R_\bK^{\ell}(\bq)}=\si^{-\ell} \hat f_\bq \qqf \bq\in \Las,\ \ell\in \Z.
    \ee
Combining with general decomposition \x{S-l}, we have the following results.

    \begin{lem}\label{prop:expressions-of-4Phi}
Let $\si$ be as in \x{si}.

\bu $F(\bx)\in L^{2}_{\bK,\si}$ if and only if there exists $\{ \hat f _{\bq} \} _{\bq\in \ssk} \in l^{2}_{\ssk}$ such that
    \begin{align}
F(\bx)&= \sum_{\bq\in\ssk} \z(\hat f_{\bq}\sum^{3}_{\ell=0}\si^{-\ell}e^{iR^{\ell}(\bK+\bq)\cdot\bx}\y) \lb{decomp1}\\
&=\sum_{\bq\in \ssk}\hat f _{\bq}\z(e^{i(\bK+\bq)\cdot\bx}+\ol{\si}e^{iR(\bK+\bq)\cdot\bx}
+ \si^{2} e^{iR^{2}(\bK+\bq)\cdot\bx} +\si e^{iR^{3}(\bK+\bq)\cdot\bx}\y) \lb{decomp2}.
    \end{align}

\bu If $F(\bx)\in L^{2}_{\bK,\si}$, then $\overline{F(-\bx)}\in L^{2}_{\bK,\bar \si}$\ .
    \end{lem}

\Proof
\bu Note that $R_\bK^4=I$ and $\si$ satisfies $\si^4=1$. Substituting relations \x{FCr}, $\ell=0,1,2,3$, into \x{S-l}, we obtain equality \x{decomp1}.

As for equality \x{decomp2}, we need only to notice in \x{decomp1} that
    \[
    \si^0=1, \qq \si^{-1}= \bar \si, \qq \si^{-2} = \si^2, \qq \si^{-3}= \si\ .
    \]

\bu We use  expansion \x{decomp1} for $F(\bx)$ to obtain
    \[
    \begin{split}
    \overline{F(-\bx)}&= \ol{\sum_{\bq\in\ssk} \z(\hat f_{\bq}\sum^{3}_{\ell=0}\si^{-\ell}e^{iR^{\ell}(\bK+\bq)\cdot(-\bx)}\y)}\\
    &=  \sum_{\bq\in\ssk} \z(\ol{\hat f_{\bq}}\sum^{3}_{\ell=0}\bar\si^{-\ell}e^{-iR^{\ell}(\bK+\bq)\cdot(-\bx)}\y)\\
    &= \sum_{\bq\in\ssk} \z(\ol{\hat f_{\bq}}\sum^{3}_{\ell=0}\bar\si^{-\ell}e^{iR^{\ell}(\bK+\bq)\cdot \bx}\y),
    \end{split}
    \]
which is in $L^{2}_{\bK,\bar \si}$, following from the characterization \x{decomp1} for the eigenvalue $\bar \si$. \qed

\section{Eigenvalues of periodic Schr\"odinger operators} \label{FBT}

\subsection{Admissible potentials} \lb{Adm}

In this work, we introduce the following admissible potentials.

    \begin{defn}\label{def:definition-of-lattice}
{\rm (Admissible Potentials)} Let $V(\bx)\in C^{\infty}(\R^{3})$ be real-valued. We say that $V(\bx)$ is an admissible {\rm potential} with respect to $\La$ if $V(\bx)$ satisfies

(1) $V(\bx)$ is $\La$-periodic, $V(\bx+\bv)=V(\bx)$ for all $\bx\in \R^{3}$ and $\bv\in\La$.

(2) $V(\bx)$ is real-valued and even, i.e., $\ol{V(\bx)}=V(\bx)$, $V(-\bx)\equiv V(\bx)$ for $\bx \in \R^3$.

(3) $V(\bx)$ is $\mcr$-invariant, i.e.,
    \[
\mcr[V](\bx)= V(R^{*}\bx)\equiv V(\bx)\mbox{ for } \bx\in \R^3.
    \]

(4) $V(\bx)$ is $\mathcal{T}$-invariant, i.e.,
     \[
    \mathcal{T}[V](\bx)\equiv V(T^{*}\bx)=V(\bx),
    \]
where $T$ is the following matrix
    \be\lb{mat}
    T=
    \begin{pmatrix}
    -1&0&0\\
    0&0&-1\\
    0&1&0\\
    \end{pmatrix}.
    \ee
     \end{defn}

We remark that the requirements (2) in Definition \ref{def:definition-of-lattice} are the so-called $\mathfrak{PT}$-symmetry. Moreover, requirement (4) is a novel symmetry for $3$-dimensional potentials which will play an important role in the later analysis for Weyl points. Admissible potentials have the following properties.

    \begin{cor}
Let $V(\bx)$ be an admissible potential.  Then its Fourier coefficients $\hat V_\bq$ satisfy
    \[
    \hat V_{-\bq}=\hat V_\bq\in \R\q \forall \bq\in \Las,
    \]
and
    \[
    \hat V_{R^\ell \bq} = \hat V_\bq, \q  \hat V_{T^\ell \bq} = \hat V_\bq \q \forall \bq\in \Las,\ \ell\in \Z.
    \]
    \end{cor}

\begin{rem}
Let us consider the orthogonal matrix $T$ in $\x{mat}$. It is easy to see that $T$ maps the lattice $\Lambda^{*}$ to itself and $T^{*}=T^{-1}$. Moreover, $T$ acts on $\Lambda^{*}$ as follows
    \[
 \begin{split}
 &T\bq_1=\bq_3-\bq_1, \qq T\bq_2=-\bq_1, \qq T\bq_3=\bq_2-\bq_1,\\
 &T\bK=\bK+\bq_1, \qq T^2 \bK=\bK+\bq_3, \qq T^3 \bK=\bK+\bq_2.
 \end{split}
    \]
\end{rem}

Typical admissible potentials can be constructed using Fourier expansions.

\begin{exa}\label{example-potential}
{\rm
Let us define real, even potentials
    \begin{equation*}
    \begin{split}
    V_1(\bx) &:=\cos(\bq_{1}\cdot\bx)+\cos((\bq_{2}-\bq_{1})\cdot\bx) +\cos((\bq_{3}-\bq_{2})\cdot\bx)+\cos(\bq_{3}\cdot\bx), \\
    V_2(\bx) &:=  \cos(\bq_{2}\cdot\bx)+\cos((\bq_{3}-\bq_{1})\cdot\bx).
    \end{split}
    \end{equation*}

It is easy to see that these $V_i(\bx)$ are $\mathcal{R}$-invariant potentials. Thus, for any real coefficients $c_i$, the potential
    \begin{equation*}
V(\bx)= \sum_{i=1}^2 c_i V_i(\bx)
    \end{equation*}
is also $\mathcal{R}$-invariant. However, $V(\bx)$ is, in general, not $\mathcal{T}$-invariant. In fact, by noting that $T\bq_1 =\bq_1 -\bq_3$, we know that $V(\bx)$ is $\mathcal{T}$-invariant if and only if $c_1 =c_2$. Therefore
    \[ 
    V(\bx):=c(V_1(\bx)+V_2(\bx))
    \]
is an admissible potential as in Definition \ref{def:definition-of-lattice} for any nonzero real number $c$.
\qed }
\end{exa}

The role of the $\mathcal{R}$- and $\mathcal{T}$-invariance of admissible potentials $V(\bx)$ can be stated as the following commutativity with the Schr\"odinger operator $H$ of \x{HV} we are going to study.


\begin{lem}\lb{prot} 

$(1)$ Transformations $\mathcal{R}$ and $\mathcal{T}$ are isometric, i.e.,
    \begin{equation*}
\langle\mathcal{R}f(\bx),\mathcal{R}g(\bx)\rangle=\langle f(\bx), g(\bx)\rangle,\andq
\langle\mathcal{T}f(\bx),\mathcal{T}g(\bx)\rangle=\langle f(\bx), g(\bx)\rangle
    \end{equation*}
for all $f(\bx),\ g(\bx)\in L^{2}_{\bK,\Lambda}$.

$(2)$ The commutators $[H,\mathcal{R}]:=H \mathcal{R}-\mathcal{R}H$ and $[H,\mathcal{T}]:=H \mathcal{T}-\mathcal{T}H$ vanish on $H^{2}_{\bK,\Lambda}$.

\end{lem}

The proofs are direct.

\subsection{Periodic Schr\"odinger operators and Floquet-Bloch theory} \lb{FBT1}

Let $\La$ be the lattice defined in \x{la} and $V: \R^3\to \R$ be an admissible potential in the sense of Definition $\ref{def:definition-of-lattice}$. 
For each quasi-momentum $\bk\in\R^{3}$, we consider the Floquet-Bloch eigenvalue problem
    \begin{equation}\lb{HV-k}
    \bb{split}
H\Phi(\bx,\bk) &=\mu(\bk)\Phi({\bx,\bk}),\q \bx\in\R^{3},\\
\Phi(\bx+\bv,\bk)& =e^{i\bk \cdot\bv}\Phi(\bx,\bk),\q \bx\in\R^{3},\ \bv\in\La,
    \end{split}
    \end{equation}
where $\mu(\bk)$ is the eigenvalue and the second condition is the pseudo-periodic condition for $\Phi(\bx,\bk)$. By setting
    \[
\Phi(\bx,\bk)=e^{i\bk\cdot \bx}\phi(\bx,\bk), \q\mbox{or} \q \phi(\bx,\bk)=e^{-i\bk\cdot \bx}\Phi(\bx,\bk),
    \]
we know that problem \x{HV-k} is converted into  the following periodic eigenvalue problem
    \begin{equation}\lb{HV-k2}
    \begin{split}
H(\bk)\phi(\bx,\bk) &=\mu(\bk)\phi(\bx,\bk),\q \bx\in\R^{3},\\
\phi(\bx+\bv,\bk) & =\phi(\bx,\bk),\q \bx\in\R^{3},\ \bv\in\La.
\end{split}
    \end{equation}
Here the shifted Schr\"odinger operator $H(\bk)$ is defined via
    \begin{equation*}\lb{HVk}
    \begin{split}
    \nabla_\bk \phi(\bx) := & e^{-i \bk\d \bx} \nabla\z(e^{i \bk\d \bx} \phi(\bx)\y)=\nabla \phi(\bx) +i \bk \phi(\bx) =\z(\nabla +i \bk \y) \phi(\bx),\\
    H(\bk)\phi(\bx) := & e^{-i \bk\d \bx} \Delta\z(e^{i \bk\d \bx} \phi(\bx)\y) +V(\bx) \phi(\bx)\\
    = & -(\nabla+i\bk)\cdot(\nabla+i\bk)\phi(\bx)+V(\bx)\phi(\bx) \\
    \equiv  & -\nabla_\bk\d \nabla_\bk \phi(\bx) +V(\bx)\phi(\bx)\ .
    \end{split}
    \end{equation*}

The general properties of the Schr\"odinger operator with a periodic potential is given by the Floquet-Bloch theory. We end this section by listing some most important conclusions of this theory without including their proofs. We refer readers to \cite{Eastham1973,fefferman2012honeycomb,Kuchment2015,Kuchment2000,method1972Reed} for details.

    \begin{prop} \rm{(Floquet-Block theory)}  \lb{FT}

$(1)$ For any $\bk\in \Omega^{*}$, the Floquet-Bloch eigenvalue problem \x{HV-k2} has an ordered discrete spectrum
    \begin{equation*}
\mu_{1}(\bk)\leq\mu_{2}(\bk)\leq\mu_{3}(\bk)\leq\ldots
    \end{equation*}
such that $\mu_{b}(\bk)\rightarrow+\infty$ as $b\rightarrow+\infty$. Furthermore, there exist eigenpairs $\{\phi_{b}(\bx,\bk),$ $\mu_{b}(\bk)\}_{b\in \N}$ for each $\bk\in\Omega^{*}$ such that $\z\{ \phi_{b}(\bx,\bk)\y\}_{b\geq1}$ can be taken to be a complete orthonormal basis of $L^{2}_\per.$

Accordingly, problem \x{HV-k} has eigenpairs $\z\{\Phi_{b}(\bx,\bk),\ \mu_{b}(\bk)\y\}_{b\in \N}$, where
    \[
    \z\{ \Phi_{b}(\bx,\bk):= e^{i \bk \d \bx}\phi_{b}(\bx,\bk)\y\}_{b\in \N}
    \]
is a complete orthonormal basis of $L^{2}_\Kla.$

$(2)$ The eigenvalues $\mu_{b}(\bk)$, referred as dispersion bands, are Lipschitz continuous functions of $\bk\in \Omega^{*}$.

$(3)$ For each $b\geq1$, $\mu_b(\bk)$ sweeps out a closed real interval $I_{b}$ over $\bk\in\Omega^{*}$, and the union of $I_b$ composes of the spectrum of $H$ in
$L^{2}_\per$:
    \[
{\rm spec}(H)=\bigcup_{b\geq1,\bk\in\Omega^{*}}I_{b}, \qq \mbox{where } I_b=\bigl[\min_{\bk\in\Omega^{*}}\mu_{b}(\bk), \max_{\bk\in\Omega^{*}}\mu_{b}(\bk)\bigr]\ .
    \]

$(4)$ Given $ \bk\in\Omega^{*}$, $\Phi_{b}(\bx,\bk)$ is smooth in $\bx\in\Omega$. Moreover, the set of eigenfunctions $\bigcup_{b\geq1, \bk\in\Omega^{*}} {\Phi_{b}(\bx,\bk)}$ is a complete orthonormal set of $L^{2}(\mathbb{R}^3)$. Consequently, any $f(\bx)\in L^{2}(\mathbb{R}^3)$ can be written in the summation form
    \be\lb{fx1}
f(\bx)=\frac{1}{|\Omega^{*}|}\sum_{b\geq1}\int_{\Omega^{*}}\wi{f_b}(\bk)\Phi(\bx,\bk)\mathrm{d}\bk,
    \ee
where
    \[
    \wi{f_b}(\bk)=\inn{\Phi_b(\bx,\bk)}{f(\bx)}=\int_{R^{3}}\ol{\Phi_{b}(\bx,\bk)}f(\bx)\mathrm{d}\bx.
    \]
Here the summation \x{fx1} is convergent in the $L^2$-norm.
    \end{prop}

\section{Weyl points and conical intersections}
\label{construction}

In this section, we are going to prove the existence of Weyl points on the energy bands of Schr\"odinger operators with admissible potentials that we propose in Definition \ref{def:definition-of-lattice}. The strategy used in this work is inspired by the framework that Fefferman and Weinstein developed for Dirac points in 2-D honeycomb structures \cite{fefferman2012honeycomb}. More specifically, (1) we first propose required conditions of eigen structure at $\bK$ for some eigenvalue $\mu_{*}$, i.e., the conditions \textbf{H1-H2} below; (2) we then prove the energy bands in the vicinity form a conical structure with an extra band in the middle under these conditions; (3) we justify that the required conditions \textbf{H1-H2} do hold for nontrivial shallow admissible potentials; (4) we extend the justification of required conditions to generic admissible potentials.

Compared to the study on Dirac points for the 2-D honeycomb case, the main difficulties of our current work arise from two perspectives: higher dimension and higher multiplicity. To the best of our knowledge, we have not found rigorous analysis on such degenerate points in the literature. Higher dimension makes the calculations more cumbersome. On the other hand, the higher multiplicity forces us to deal with a larger bifurcation matrix which has more freedoms which we need to reduce, for instance, the relations among the entries of the matrix. Some new symmetry arguments are introduced to conquer these difficulties.

\subsection{Spectrum structure at the high symmetry point $\bK$} \label{s41}

In this section, we are interested in the three-fold degeneracy of the high symmetry point $\bK$. So let us consider the $\bK$-quasi periodic eigenvalue problem
    \begin{equation}\lb{mapro}
 \begin{split}
 H\Phi(\bx,\bK)&\equiv [-\Delta+V(\bx)]\Phi(\bx,\bK)=\mu_{*}\Phi(\bx,\bK),\q \bx\in\R^{3},\\
 \Phi({\bf x+v,\bK})&=e^{i\bK\cdot \bv}\Phi({\bf x,\bK}),\q \bx\in\R^{3}, \bv\in \La.\\
 \end{split}
     \end{equation}

We first assume that there exists an eigenvalue $\mu_*$ such that the following assumption is fulfilled.
    \begin{asm1}\lb{h1}
$\mu_{*}$ is a three-fold eigenvalue of $H$ in problem \x{mapro} with the corresponding eigenspace $\mathcal{E}_{\mu_*}$ such that
    \[
    \mathcal{E}_{\mu_*} \perp L^{2}_{\bK,1},\q \mbox{and} \q \dim\{\mathcal{E}_{\mu_*}\cap L^{2}_{\bK,i}\}=1.
    \]
     \end{asm1}
Then the following proposition characterizes the fine structure of the eigenspace $\mathcal{E}_{\mu_*}$.

    \begin{prop}\lb{asphi}
 Assume that $\mbox{\rm{\textbf{H1}}}$ holds. Then there exist functions $\Phi_{\ell}(\bx) \in L^{2}_{\bK,i^{\ell}},$ $ ~j=1,2,3$ such that $\{\Phi_1(\bx),\ \Phi_2(\bx),\ \Phi_3(\bx)=\overline{\Phi_1(-\bx)}\}$ form an orthonormal basis of $\mathcal{E}_{\mu_*}$.
    \end{prop}

A direct consequence of above proposition is that $\mu_{*}$ is an $L^{2}_{\bK,i^{\ell}}$-eigenvalue of multiplicity $1$ for each $\ell=1,2,3$.

In order to keep the structure of the paper, the detailed proof of Proposition \ref{asphi} is placed in Appendix A.

%
%
%
%

\subsection{Bifurcation matrices} \lb{bif}

Under the assumption $\textbf{H1}$, we always can find an orthonormal basis $\{\Phi_1(\bx),\Phi_2(\bx),$ $\Phi_3(\bx)\}$ for $\mathcal{E}_{\mu_*}$ as in Proposition \ref{asphi}. However, the choice is not unique and a gauge freedom for each eigenfunction $\Phi_{\ell}(\bx)$ is allowed.

Giving such a basis, let us define a complex-valued matrix $M(\ka)$ for $\ka \in \mathbb{R}^3/\{0\}$ by
    \[
M(\ka):=
\begin{pmatrix}
\langle\Phi_1,2i\ka\cdot\nabla\Phi_1 \rangle&\langle\Phi_1,2i\ka\cdot\nabla\Phi_2 \rangle&\langle\Phi_1,2i\ka\cdot\nabla\Phi_3 \rangle\\
\langle\Phi_2,2i\ka\cdot\nabla\Phi_1 \rangle&\langle\Phi_2,2i\ka\cdot\nabla\Phi_2 \rangle&\langle\Phi_2,2i\ka\cdot\nabla\Phi_3 \rangle\\
\langle\Phi_3,2i\ka\cdot\nabla\Phi_1 \rangle&\langle\Phi_3,2i\ka\cdot\nabla\Phi_2 \rangle&\langle\Phi_3,2i\ka\cdot\nabla\Phi_3 \rangle\\
\end{pmatrix}.
    \]
It is called the bifurcation matrix which appears naturally in the eigenvalue problem. We shall see in the later section that the leading order structure of the eigenvalues of $H(\bk)$ for $\bk$ in the vicinity of $\bK$ is closely related to $M(\ka)$. In this subsection, the main properties of $M(\ka)$ and their justifications are provided.
We want to remark that $M(\ka)$ depends on the choice of the basis set $\{\Phi_1(\bx),\Phi_2(\bx),\Phi_3(\bx)\}$ due to the gauge freedom. It is evident that $M(\ka)$ is Hermitian since $2i\ka\cdot \nabla$ is self-adjoint.

We consider the admissible potential $V(\bx)$ in the sense of Definition \ref{def:definition-of-lattice}. Recall that $[H, \mathcal{T}]=0$ can imply $\mathcal{T}\mathcal{E}_{\mu_*}=\mathcal{E}_{\mu_*}$. In other words, there exists a $3\times 3$ matrix $Q_\mathcal{T}$ such that
    \[
\begin{pmatrix}
\mathcal{T}\Phi_1\\
\mathcal{T}\Phi_2\\
\mathcal{T}\Phi_3\\
\end{pmatrix}
=Q_{\mathcal{T}}\begin{pmatrix}
\Phi_1\\
\Phi_2\\
\Phi_3\\
\end{pmatrix}
=
\begin{pmatrix}
c_{11}&c_{12}&c_{13}\\
c_{21}&c_{22}&c_{23}\\
c_{31}&c_{32}&c_{33}\\
\end{pmatrix}
\begin{pmatrix}
\Phi_1\\
\Phi_2\\
\Phi_3\\
\end{pmatrix}.
    \]

Recall from Lemma \ref{prot} that $\mathcal{T}: L^2_{\bK,\Lambda}\to L^2_{\bK,\Lambda}$ preserves the inner product, i.e., $\langle\mathcal{T}F,\mathcal{T}G \rangle=\langle F,G \rangle$ for all  $f,g\in L^2_{\bK,\Lambda}$. It immediately follows that $Q_{\mathcal{T}}$ is unitary, i.e., $Q_{\mathcal{T}}^{*}Q_{\mathcal{T}}=I$. In other words, $\{\mathcal{T}\Phi_1, \mathcal{T}\Phi_2, \mathcal{T}\Phi_3\}$ is also an orthonormal basis of $\mathcal{E}_{\mu_*}$ which defines a new bifurcation matrix $M^{\mathcal{T}}(\ka)$. Namely,  
    \[
M^{\mathcal{T}}(\ka)\equiv \begin{pmatrix}
\langle\mathcal{T}\Phi_1,2i\ka\cdot\nabla\mathcal{T}\Phi_1 \rangle&\langle\mathcal{T}\Phi_1,2i\ka\cdot\nabla\mathcal{T}\Phi_2 \rangle&\langle\mathcal{T}\Phi_1,2i\ka\cdot\nabla\mathcal{T}\Phi_3 \rangle\\
\langle\mathcal{T}\Phi_2,2i\ka\cdot\nabla\mathcal{T}\Phi_1 \rangle&\langle\mathcal{T}\Phi_2,2i\ka\cdot\nabla\mathcal{T}\Phi_2 \rangle&\langle\mathcal{T}\Phi_2,2i\ka\cdot\nabla\mathcal{T}\Phi_3 \rangle\\
\langle\mathcal{T}\Phi_3,2i\ka\cdot\nabla\mathcal{T}\Phi_1 \rangle&\langle\mathcal{T}\Phi_3,2i\ka\cdot\nabla\mathcal{T}\Phi_2 \rangle&\langle\mathcal{T}\Phi_3,2i\ka\cdot\nabla\mathcal{T}\Phi_3 \rangle\\
\end{pmatrix}.
    \]

Similarly, by using the symmetry $\mathcal{R}$, we can define another bifurcation matrix $M^{\mathcal{R}}(\ka)$ and the corresponding unitary transformation $Q_{\mathcal{T}}$. In fact, it is easy to obtain
    \be \lb{QR}
    Q_{\mathcal{R}}=\begin{pmatrix}
i&0&0\\
0&-1&0\\
0&0&-i\\
    \end{pmatrix}.
    \ee
However, the explicit form for $Q_{\mathcal{T}}$ is unknown to us.


One has the following relations for these bifurcation matrices.

\begin{prop}\lb{indu}
For any $\ka\in \mathbb{R}^3/\{0\}$, there hold
    \begin{align}
M(\ka)=M^{\mathcal{R}}(R\ka)=Q_{\mathcal{R}}^{*}M(R\ka)Q_{\mathcal{R}}, \lb{tmar}\\
M(\ka)=M^{\mathcal{T}}(T\ka)=Q_{\mathcal{T}}^{*}M(T\ka)Q_{\mathcal{T}}, \lb{tmat}
\end{align}
where $R$ and $T$ are the orthogonal matrices in \x{matrix R} and \x{mat}.
    \end{prop}

\Proof
We only give the proof to \x{tmat}, while the proof of \x{tmar} is similar.

By Lemma 2 in $\cite{Lee-Thorp2017}$, one has for $\ell,m=1,2,3$,
    \[
    \langle\Phi_{\ell},\nabla\Phi_{m}\rangle=\langle\mathcal{T}\Phi_\ell,\mathcal{T}\nabla\Phi_m \rangle=\langle \mathcal{T}\Phi_\ell,T^{*}\nabla\mathcal{T}\Phi_m\rangle.
    \]
Therefore
    \begin{equation}\lb{mi}
    \begin{split}
(M(\ka))_{\ell m}&=\langle\Phi_{\ell},2i\ka\cdot \nabla\Phi_m\rangle= \langle\mathcal{T}\Phi_\ell,2i\ka\cdot T^{*}\nabla \mathcal{T}\Phi_m \rangle
=\langle\mathcal{T}\Phi_\ell,2iT\ka\cdot\nabla\mathcal{T}\Phi_m\rangle\\&= \sum\limits^{3}_{\ell=1}\sum\limits^{3}_{m=1}\ol{c_{i\ell}}c_{jm}\langle\Phi_{\ell},2i(T\ka)\cdot\nabla\Phi_{m}\rangle=(M^{\mathcal{T}}(T\ka))_{\ell m}.
    \end{split}
    \end{equation}
By recalling that $Q_{\mathcal{T}}=(c_{ij})$, we know that \x{mi} is equality \x{tmat}.
\qed

By substituting \x{QR} into \x{tmar}, we obtain
    \be\lb{eqi}
\begin{split}
M^{\mathcal{R}}&(R\ka)=Q^{*}_{\mathcal{R}}
\begin{pmatrix}
\langle\Phi_1,2iR\ka\cdot\nabla\Phi_1\rangle&\langle\Phi_1,2iR\ka\cdot\nabla\Phi_2\rangle&\langle\Phi_1,2iR\ka\cdot\nabla\Phi_3\rangle\\
\langle\Phi_2,2iR\ka\cdot\nabla\Phi_1\rangle&\langle\Phi_2,2iR\ka\cdot\nabla\Phi_2\rangle&\langle\Phi_2,2iR\ka\cdot\nabla\Phi_3\rangle\\
\langle\Phi_3,2iR\ka\cdot\nabla\Phi_1\rangle&\langle\Phi_3,2iR\ka\cdot\nabla\Phi_2\rangle&\langle\Phi_3,2iR\ka\cdot\nabla\Phi_3\rangle\\
\end{pmatrix}
Q_{\mathcal{R}}\\
&=\begin{pmatrix}
\langle\Phi_1,2i\ka\cdot R^{*}\nabla\Phi_1\rangle&i\langle\Phi_1,2i\ka\cdot R^{*}\nabla\Phi_2\rangle&-\langle\Phi_1,2i\ka\cdot R^{*}\nabla\Phi_3\rangle\\
-i\langle\Phi_2,2i\ka\cdot R^{*}\nabla\Phi_1\rangle&\langle\Phi_2,2i\ka\cdot R^{*}\nabla\Phi_2\rangle&i\langle\Phi_2,2i\ka\cdot R^{*}\nabla\Phi_3\rangle\\
-\langle\Phi_3,2i\ka\cdot R^{*}\nabla\Phi_1\rangle&-i\langle\Phi_3,2i\ka\cdot R^{*}\nabla\Phi_2\rangle&\langle\Phi_3,2i\ka\cdot R^{*}\nabla\Phi_3\rangle\\
\end{pmatrix}.\\
\end{split}
    \ee

Recall the transformation $R:\C^3 \to \C^3$  has eigenpairs listed in \x{R-ev}. We can then obtain the following structural result for the bifurcation matrix $M(\ka)$.

    \bb{thm} \lb{phaseK}
There exist $\ups_1, \ups_2, \ups_3 \in \mathbb{C}$ such that
    \be \lb{Mka}
M(\ka)=\begin{pmatrix}
0&\ka\cdot\upsilon_1\om_1&\ka\cdot\ol{\ups_{3}}\om_2\\
\ka\cdot\ups_{2}\om_1&0&\ka\cdot\ups_{2}\om_1\\
\ka\cdot\ups_{3}\om_2&\ka\cdot\ol{\ups_{2}}\om_3&0
\end{pmatrix},
    \ee
where $\om_{j},\ j=1,2,3$ are eigenvectors of $R$ listed in \x{R-ev}. Moreover, there have
    \be \lb{ups1}
    |\upsilon_1|=|\upsilon_2|=|\upsilon_3|,
    \ee
    \be \lb{ups2}
    \upsilon_1 \upsilon_2 \upsilon_3+\ol{\upsilon_1 \upsilon_2 \upsilon_3}=0.
    \ee
    \end{thm}

\Proof The proof is split into several steps.

1. Entries of $M(\ka)$. Note $(M(\ka))_{\ell j}=(M^{R}(R\ka))_{\ell j}$ $=(Q^{*}_{\mathcal{R}}M(R\ka)Q_{\mathcal{R}})_{\ell j}$ holds for $\ell,j=1,2,3$. By comparing the elements in $M^{\mathcal{R}}(\ka)$ displayed in \x{eqi} with $M(\ka)$, it is easily seen that for $\ka\in\mathbb{R}^3$, one has
    \[
    i^{j-\ell}\langle\Phi_{\ell},2i\ka\cdot R^{*} \nabla\Phi_{j}\rangle= \langle\Phi_{\ell},2i\ka\cdot\nabla\Phi_{j}\rangle \Longrightarrow
    \ka\cdot i^{j-\ell}\langle\Phi_{\ell},2i R^{*}\nabla\Phi_{j}\rangle=\ka\cdot \langle\Phi_{\ell},2i\ka\cdot\nabla\Phi_{j}\rangle.
    \]
Since $\ka\in \R^3$ is arbitrary, we claim that
    \be\lb{Phi14}
    R\langle\Phi_{\ell},2i \nabla\Phi_{j}\rangle=i^{j-\ell} \langle\Phi_{\ell},2i\nabla\Phi_{j}\rangle.
    \ee

Equalities in \x{Phi14} have shown that, for each pair $(\ell,j)$, $\langle\Phi_{\ell},2i \nabla\Phi_{j}\rangle$ is either the zero vector or an eigenvector of $R$ associated with the eigenvalue $i^{j-\ell}$.
If $\ell=j\in \{1,2,3\}$, we know that $i^{j-\ell}=1$ is not an eigenvalue of $R$ and therefore
    \[
    \langle\Phi_{\ell},2i \nabla\Phi_{\ell}\rangle=0 \q \mbox{for } \ell=1,2,3.
    \]
On the other hand, the other six equalities of \x{Phi14} imply that there exist constants $\ups_\ell,\ \tl \ups_\ell\in \C$ such that

    \be \lb{ups4}
    \z\{ \ba{l}
    \langle\Phi_{1}, 2i\nabla\Phi_{2}\rangle= \ups_1 \om_1, \qq \langle\Phi_{2}, 2i\nabla\Phi_{1}\rangle= \tl\ups_1 \om_3, \\
    \langle\Phi_{2}, 2i\nabla\Phi_{3}\rangle= \ups_2 \om_1, \qq \langle\Phi_{3}, 2i\nabla\Phi_{2}\rangle= \tl\ups_2 \om_3, \\
    \langle\Phi_{3}, 2i\nabla\Phi_{1}\rangle= \ups_3 \om_2, \qq \langle\Phi_{1}, 2i\nabla\Phi_{3}\rangle= \tl\ups_3 \om_2.
    \ea \y.
    \ee
Since $M(\ka)=(M(\ka))^{*}$ and $\ol{\om_3}=\om_1$, we have necessarily
    \(
    \tl \ups_\ell=\ol{\ups_\ell}
    \)
for $\ell=1,2,3.$

2. Proof of $|\upsilon_1|=|\upsilon_2|$. 
According to the definition of $\Phi_2(\bx) \in L^{2}_{\bK,-1}$, we have
    \[
\mathcal{R}[\Phi_2](\bx)=\Phi_2(R^{*}\bx)=-\Phi_2(\bx).
    \]
Thus
    \[
    \bb{split}
    &\Phi_2(R^{*}(-\bx))=-\Phi_2(-\bx),\\
&\ol{\Phi_2(R^{*}(-\bx))}=-\ol{\Phi_2(-\bx)},\\
& \mathcal{R}[\ol{\Phi_2(-\bx)}]=-\ol{\Phi_2(-\bx)}.
\end{split}
    \]
The last equality means that $\ol{\Phi_2(-\bx)}\in L^{2}_{\bK,-1}$. Since $\dim(\mathcal{E}_{\mu_{*}}\bigcap L^{2}_{\bK,-1})=1$ by $\textbf{H1}$ and $\ol{\Phi_2(-\bx)}$ is also $L^{2}$-normalized, therefore
    \[
    \Phi_2(\bx)\equiv e^{i\theta}\ol{\Phi_2(-\bx)}\q\mbox{for some }\th\in \R.
    \]
From this, we deduce that $\nabla\Phi_{2}(\bx) \equiv -e^{i\theta}\ol{\nabla \Phi_2(-\bx)}$ and
    \begin{align*}
    \langle\Phi_{1}, 2i\nabla\Phi_{2}\rangle&=\ol{\int {\Phi_1(\bx)} \d 2i e^{-i\theta}\nabla\Phi_2(-\bx) d\bx}\\
    &= \ol{\int {\Phi_1(-\bx)} \d 2i e^{-i\theta}\nabla\Phi_2(\bx) d\bx}\\
    &= e^{i\theta}\ol{\inn{\Phi_{3}}{2i\nabla\Phi_{2}}} \qq \mbox{(by changing $\bx$ to $-\bx$)}\\
    &= e^{i\theta}\inn{\Phi_{2}}{2i\nabla\Phi_{3}}.
    \end{align*}
From the definition of $\ups_\ell$ in \x{ups4}, we obtain $\upsilon_1 \om_1=e^{i\theta} \upsilon_2 \om_2$ and $|\upsilon_1|=|\upsilon_2|$.

3. Proof of \x{ups1} and \x{ups2}.
The proof of $|\upsilon_2|=|\upsilon_3|$ is different. For any $\ka\in \R^3$, we consider the characteristic polynomial of the bifurcation matrix $M(\ka)$
    \[
    p(a,\ka):=\det \z(aI+M(\ka)\y)=
\det \begin{pmatrix}
a&\langle\Phi_1,2i\ka\cdot\nabla\Phi_2 \rangle&\langle\Phi_1,2i\ka\cdot\nabla\Phi_3 \rangle\\
\langle\Phi_2,2i\ka\cdot\nabla\Phi_1 \rangle&a&\langle\Phi_2,2i\ka\cdot\nabla\Phi_3 \rangle\\
\langle\Phi_3,2i\ka\cdot\nabla\Phi_1 \rangle&\langle\Phi_3,2i\ka\cdot\nabla\Phi_2 \rangle&a\\
\end{pmatrix}.
    \]
It is a cubic polynomial of $a$ with coefficients depending on $\ka$. Since $Q_{\mathcal{T}}$ is unitary, it follows from \x{tmat} that
    \[
    \begin{split}
    p(a,\ka)&=\det \z(a I+ Q_{\mathcal{T}}^{*}M(T\ka)Q_{\mathcal{T}}\y) =\det\z( Q_{\mathcal{T}}^{*}\z(a I+M(T\ka)\y)Q_{\mathcal{T}}\y) \\ &=\det \z( a I+M(T\ka)\y).
    \end{split}
    \]
Thus $p(a,\ka)$ satisfies the following invariance
    \be\lb{con1}
p(a,\ka)\equiv p(a,T\ka).
    \ee

\ifl
By Lemma $\ref{indu}$, $\det M(a,\ka)=0$ imply $\det C^{*}(aI+M(T\ka))C=0$, since $C$ is invertible, we claim
\[
\det M(a,\ka)=0 \Longleftrightarrow \det (aI+M(T\ka))=0.
\]
Therefore, for fixed $\ka\in\mathbb{R}^3$, the three roots of $p(a,\ka)=\det M(a,\ka)$ solve $p(a,\ka)=\det (aI+M(T\ka))=0$, since $p(a,\ka)$ is also a cubic equation with highest order term $a^3$, we have
\fi

In particular, by taking $\ka=e_2:=(0,1,0)$ in \x{con1}, one has from \x{Mka} that
    \be \lb{p1}
    p(a,e_2)\equiv a^3-|\upsilon_2|^{2} a.
    \ee
Similarly, one has $T e_2= (0,0,1)=e_3$ and by using \x{Mka} again, we have
    \be \lb{p2}
    p(a,T e_2)\equiv a^3-|\upsilon_3|^{2} a+\frac{1}{2}(\upsilon_{1} \upsilon_{2} \upsilon_{3}+\ol{\upsilon_{1} \upsilon_{2} \upsilon_{3}}).
    \ee
By comparing the coefficients of \x{p1} and \x{p2}, we deduce from the invariance \x{con1} that there hold $|\upsilon_2|=|\upsilon_3|$ and equality \x{ups2}. Together with equality $|\upsilon_1|=|\upsilon_2|$ in the above step, we have obtained all equalities in \x{ups1} and \x{ups2}.   \qed

\ifl
Therefore, the coefficients of each term of polynomial $p(a,\ka)$ are identical with $p(a,T\ka)$. Set $\ka_1=(0,1,0)$, then
$T\ka_1=(0,0,1)$, by substituting $\ka_1$ and $T\ka_1$ into \x{eq:equality_threelambda}, we obtain
    \begin{align}
&p(a,\ka_1)=a^3-|\upsilon_2|^{2} a=0,\\
&p(a,T\ka_1)=a^3-|\upsilon_3|^{2} a=\frac{1}{2}(\upsilon_{1} \upsilon_{2} \upsilon_{3}+\ol{\upsilon_{1} \upsilon_{2} \upsilon_{3}}).
    \end{align}
By comparing the coefficients between $p(a,\ka)$ and $p(a,T\ka)$,  we proceed to derive
    \begin{align}
&|\upsilon_1|=|\upsilon_2|=|\upsilon_3|=a,\\
&\upsilon_{1} \upsilon_{2} \upsilon_{3}+\ol{\upsilon_{1} \upsilon_{2} \upsilon_{3}}=0. \lb{argc}
    \end{align}
\x{argc} indicates $\arg(\upsilon_{1} \upsilon_{2} \upsilon_{3})=\frac{\pi}{2}+k\pi\ k=0,1$.
\fi

We have also the following gauge invariance for $\upsilon_1 \upsilon_2 \upsilon_3$ and $|\upsilon_\ell|$.

\begin{cor}\lb{inv-g}

$(1)$ The quantity $\upsilon_1 \upsilon_2 \upsilon_3$ is gauge invariant in the sense that it does not depend on the choice of the orthonormal basis of $\mathcal{E}_{\mu_{*}}$.

$(2)$ The quantity $|\upsilon_1|=|\upsilon_2|=|\upsilon_3|$ is also gauge invariant.

\end{cor}

\Proof
Let $\z\{\Phi_\ell(\bx): \ell=1,2,3\y\}$ and $\z\{ \hat \Phi_\ell(\bx): \ell=1,2,3\y\}$ be two sets of orthonormal  eigenfunctions as in Proposition \ref{asphi}. Then there exist $\tau_\ell\in \R$ such that \(
    \tau_3 = -\tau_1,
    \)
and
    \[
    \hat \Phi_\ell(\bx)= e^{i\tau_{\ell}}\Phi_{\ell}(\bx), \q \ell=1,2,3.
    \]
By direct calculations, one has
    \[
\begin{split}
&\hat\ups_1\om_1 =\langle\hat \Phi_1(\bx),2i\nabla\hat \Phi_2(\bx)\rangle =e^{-i\tau_1+i\tau_2}\langle\Phi_{1}(\bx),2i\nabla\Phi_2(\bx)\rangle,\\
&\hat\ups_2\om_1=\langle\hat \Phi_2(\bx),2i\nabla\hat \Phi_3(\bx)\rangle =e^{-i\tau_2-i\tau_1}\langle\Phi_{2}(\bx),2i\nabla\Phi_3(\bx)\rangle,\\
&\hat\ups_3\om_2=\langle\hat \Phi_3(\bx),2i\nabla\hat \Phi_1(\bx)\rangle =e^{i\tau_1+i\tau_1}\langle\Phi_{3}(\bx),2i\nabla\Phi_1(\bx)\rangle.
\end{split}
    \]
Therefore
    $$\hat\ups_1=e^{-i\tau_1+i\tau_2}\upsilon_1, \q
\hat\ups_2=e^{-i\tau_2-i\tau_1}\upsilon_2, \q
\hat\ups_3=e^{2i\tau_1}\upsilon_3.
    $$
These yield the invariance
    \be \lb{vvs}
    \hat \ups_1 \hat \ups_2 \hat \ups_3 = \ups_1 \ups_2 \ups_3.
    \ee

For (2), by taking the norms in \x{vvs} and using equalities \x{ups1}, we obtain
    \[
    |\hat \upsilon_\ell|^3=|\upsilon_\ell|^3.
    \]
This leads to the desired invariance of $|\upsilon_\ell|$.
\qed

Due to the equalities in Theorem \ref{phaseK} and the invariance in Corollary \ref{inv-g}, let us define
    \be \lb{upsf}
    \upsilon_{_{\mathcal{F}}} :=|\upsilon_\ell|\in[0,+\oo),\ \ell=1,2,3.
    \ee
The quantity $\upsilon_{_{\mathcal{F}}}$ of \x{upsf} is referred to as {\it the Fermi velocity} in quantum mechanics.

Now we introduce another standing assumption in this paper, which can be simply stated as
    \begin{asm2}\lb{h2}
$\upsilon_{_{\mathcal{F}}}\neq 0$.
    \end{asm2}

\subsection{Conical structure of the spectrum near $\bK$}\lb{cs1}

With the eigenstructure at $\bK$, we are able to obtain the corresponding eigenstructure when quasi-momentum $\bk$ is near $\bK$. The results are stated as follows.

   \begin{thm}\label{thm:3+1}

Suppose that $V(\bx)$ is an admissible potential in the sense of Definition $\ref{def:definition-of-lattice}$ and consider the Schr\"odinger operator $H= -\Delta+V(\bx)$.  Assume that there exists $b>1$ such that $\mu_{b-1}=\mu_{b}=\mu_{b+1}=\mu_*$ is an $L^2_{\bK,\Lambda}$-eigenvalue of $H$ and the assumptions \rm{\textbf{H1-H2}} are fulfilled.

Then, for sufficiently small but nonzero $(\ka_x,\ka_y,\ka_z) \in \R^3$, eigenvalues of $H$ satisfy
    \be\lb{mu1}
\begin{split}
\mu_{b+1}(\bK+\ka)&=\mu_{*}+\xi_{+}\upsilon_{_{\mathcal{F}}}|\ka|+ o(|\ka|),\\
\mu_{b}(\bK+\ka)&=\mu_{*}+\xi_{0}\upsilon_{_{\mathcal{F}}}|\ka|+ o(|\ka|),\\
\mu_{b-1}(\bK+\ka)&=\mu_{*}+\xi_{-}\upsilon_{_{\mathcal{F}}}|\ka|+ o(|\ka|),
\end{split}
    \ee
where $\upsilon_{_{\mathcal{F}}}$ is the Fermi velocity defined before, and $\xi_{+}\geq\xi_0\geq\xi_{-}$ are the three (real) roots of the following cubic equation
    \be \lb{cubic}
    \xi^3-\xi+2\ka^{\arg}=0,\qq \ka^{\arg}:=\frac{\ka_x \ka_y \ka_z}{|\ka|^{3}}.
    \ee
    \end{thm}

\Proof The proof is based on the Lyapunov-Schmidt reduction. Thanks to the eigenstructure at $\bK$ and the explicit form of the bifurcation matrix which we established in last section, we now only need to do a perturbation expansion and a rigorous justification. Compared to the 2-D honeycomb case \cite{fefferman2012honeycomb}, we encounter more complicated computations on the bifurcation. We complete it in several steps.

1. Decomposition of spaces. For $\bk=\bK$, we have
    \[ 
    \phi_{\ell}(\bx)=e^{-i\bK\cdot\bx}\Phi_{\ell}(\bx,\bK) \in L^2_{\bz, i^\ell}\subset L^2_\per,\qq \ell=1,2,3,
    \]
such that
    \[
    H(\bK) \phi_\ell =\mu^{(0)} \phi_\ell,\qq \ell=1,2,3,
    \]
where $\mu^{(0)}:= \mu_{*}$. These define a space
    \[
    \mcx=\mcx_\bK := {\rm span} \{\phi_1,\phi_2,\phi_3\}.
    \]

Consider perturbation $\bk=\bK+\ka$, where $\ka\in \R^3$ is  small enough. From the defining equalities in \x{HVk}, one has
    \[
    H(\bK+\ka)  
    = H(\bK) -2i\ka\cdot(\nabla+i\bK)+\ka\cdot\ka
    =H(\bK) -2i\ka\cdot \nabla_\bK+\ka\cdot\ka.
    \]
To study eigenvalue problem \x{HV-k2}, let us decompose
    \[
    \psi(\bx,\bK+\ka)= \psi^{(0)}(\bx) + \psi^{(1)}(\bx), \qq \psi^{(0)} \in \mcx, \ \psi^{(1)}\in \mcx^\perp,
    \]
and write
    \[
    \mu(\bK+\ka) = \mu^{(0)}+\mu^{(1)}, \qq \mu^{(1)}\in \R.
    \]
Here the orthogonal complement $\mcx^\perp$ is taken from $L^2_\per$. Then
    \[
H(\bK+\ka) \psi(\bx,\bK+\ka)=\mu(\bK+\ka)\psi(\bx,\bK+\ka)
    \]
can be expanded as
    \begin{equation}\label{eq:3+1problem-system}
\begin{split}
&\z(H(\bK)-\mu^{(0)}I\y) \psi^{(1)}= F^{(1)}= F^{(1)}(\ka,\mu^{(1)}, \psi^{(0)}, \psi^{(1)}) \\
:=&  \z(2i\ka\cdot\nabla_\bK-\ka\cdot\ka+\mu^{(1)}\y) \psi^{(1)}
+\z(2i\ka\cdot\nabla_\bK-\ka\cdot\ka+\mu^{(1)}\y) \psi^{(0)}.
\end{split}
    \end{equation}

2. Splitting of the equation using the Lyapunov-Schmidt strategy. To solve Eq. \x{eq:3+1problem-system} using such a strategy, let us introduce the orthogonal projections
    \[
    \mathscr{Q}_{\parallel}: H^2(\R^3/\La) \to \mcx= {\rm span} \{\phi_1,\phi_2,\phi_3\}\ \mbox{and}\
    \mathscr{Q}_{\perp}:= I-\mathscr{Q}_{\parallel}: H^2(\R^3/\La) \to \mcx^\perp.
    \]
Applying $\mathscr{Q}_{\parallel}$ and $\mathscr{Q}_{\perp}$ to Eq. \x{eq:3+1problem-system}, we obtain an equivalent system
     \begin{align}
 (H(\bK)-\mu^{(0)}I)\psi^{(1)}&= Q_{\perp} F^{(1)}(\ka,\mu^{(1)}, \psi^{(0)}, \psi^{(1)}),\lb{S1}\\
 0 &= Q_{\parallel}F^{(1)} (\ka,\mu^{(1)}, \psi^{(0)}, \psi^{(1)}), \lb{S2}
     \end{align}

because
    \begin{equation}\label{relation:Q(perp-para)}
\mathscr{Q}_{\parallel}\psi^{(0)}=\psi^{(0)},\qq \mathscr{Q}_{\perp}\psi^{(1)}=\psi^{(1)}\andq \mathscr{Q}_{\parallel}\psi^{(1)}=\mathscr{Q}_{\perp}\psi^{(0)}=0.
    \end{equation}
By using \x{eq:3+1problem-system} for $F^{(1)}$, we have
    \begin{align}
    \mathscr{Q}_{\perp}F^{(1)}
    &=\mathscr{Q}_\perp\z(2i\ka \cdot \nabla_\bK-\ka\cdot\ka+\mu^{(1)}\y)\psi^{(1)}+ \mathscr{Q}_\perp\z(2i\ka\d \nabla_\bK\y)\psi^{(0)},\lb{F1}\\
    \mathscr{Q}_{\parallel} F^{(1)} &= \mathscr{Q}_{\parallel}\z(2i\ka\d\nabla_\bK\y)\psi^{(1)} + \mathscr{Q}_{\parallel}\z(2i\ka\d\nabla_\bK-\ka\cdot\ka+\mu^{(1)}\y)\psi^{(0)}.\lb{F2}
    \end{align}

\ifl
A solution to (\ref{eq:3+1problem-system}) could be obtained by solving the following system for $\psi^{(1)}$ and $\mu^{(1)}$
     \begin{equation}
 \begin{split}
 (H(\bK)-\mu^{(0)}I)\psi^{(1)}&=\mathscr{Q}_{\perp}F^{(1)}(\alpha,\beta,\gamma,\ka,\mu^{(1)},\psi^{(1)}),\\
 0&=\mathscr{Q}_{\parallel}F^{(1)}(\alpha,\beta,\gamma,\ka,\mu^{(1)},\psi^{(1)}).
 \end{split}
     \end{equation}
Specifically, that is
    \begin{equation}\label{eq:(H(K)-mu0)I-psi1}
\begin{split}
(H(\bK)-\mu^{(0)}I)\psi^{(1)}=&\mathscr{Q}_\perp(2i\ka \cdot \nabla_\bK-\ka\cdot\ka+\mu^{(1)})\psi^{(1)}+\\&\mathscr{Q}_\perp(2i\ka\nabla_\bK)\psi^{(0)},
\end{split}
    \end{equation}
    \begin{equation}
\mathscr{Q}_{\parallel}(2i\ka\nabla_\bK-\ka\cdot\ka+\mu^{(1)})\psi^{(0)}+\mathscr{Q}_{\parallel}(2i\ka\nabla_\bK)\psi^{(1)}=0.
    \end{equation}
    \fi

By the assumptions of the theorem on eigenfunctions of $H(\bK)$, one knows that, when restricted to $\mcx^\perp$, $H(\bK)-\mu^{(0)}I$ has a bounded inverse
    \begin{equation*}
\mathscr{E}=\mathscr{E}(\bK,\mu^{(0)}) =(H(\bK)-\mu^{(0)}I)^{-1}: \mcx^\perp \to \mathscr{Q}_\perp H^{2}(\R^{2}/\La).
    \end{equation*}
By (\ref{relation:Q(perp-para)}) and \x{F1}-\x{F2}, equation (\ref{S1}) is equivalent to
    \[
    \psi^{(1)} =\mathscr{E} \mathscr{Q}_\perp\z(2i\ka \cdot \nabla_\bK-\ka\cdot\ka+\mu^{(1)}\y)\psi^{(1)}= \mathscr{E} \mathscr{Q}_\perp\z(2i\ka\d \nabla_\bK\y)\psi^{(0)},
    \]
i.e.
    \be \lb{Eq11}
    \z(I - \mathscr{E} \mathscr{Q}_\perp\z(2i\ka \cdot \nabla_\bK-\ka\cdot\ka+\mu^{(1)}\y)\y)\psi^{(1)}= \mathscr{E} \mathscr{Q}_\perp\z(2i\ka\d \nabla_\bK\y)\psi^{(0)}.
    \ee
Due to the regularity, the mapping
    \[
f\mapsto \mathscr{T} \mathscr{Q}_\perp\z(2i\ka \cdot \nabla_\bK-\ka\cdot\ka+\mu^{(1)}\y) f
    \]
is a bounded operator defined on $H^{s}(\R^{2}/\La)$ for any $s$.

In the following we assume that $|\ka|+|\mu^{(1)}|$ is sufficiently small. Then the left-hand side of \x{Eq11} is invertible. Given any $\psi^{(0)}\in \mcx$, Eq. \x{Eq11} has then the unique solution in $\mathscr{Q}_{\perp}H^{2}(\R^{2}/\La):$
    \begin{equation}\lb{psi11}
\psi^{(1)}=\mcp_0 \psi^{(0)}:=\z(I - \mathscr{T} \mathscr{Q}_\perp\z(2i\ka \cdot \nabla_\bK-\ka\cdot\ka+\mu^{(1)}\y)\y)^{-1}\mathscr{T} \mathscr{Q}_\perp\z(2i\ka\d \nabla_\bK\y)\psi^{(0)}.
    \end{equation}
Here $\mcp_0=\mcp_0(\mu^{(1)},\ka): \mcx \to \mathscr{Q}_{\perp}H^{2}(\R^{2}/\La)$ is a bounded linear operator. Substituting \x{psi11} into equation (\ref{S2}) and making use of \x{F2}, we obtain an equation for the unknowns $(\mu^{(1)},\psi^{(0)})$
    \be\lb{A3}
    \mcm(\mu^{(1)},\ka)\psi^{(0)}+\wi{\mcm}(\mu^{(1)},\ka)\psi^{(0)}=0,
    \ee
where $\mcm(\mu^{(1)},\ka),\ \wi{\mcm}(\mu^{(1)},\ka): \mcx \to \mathscr{Q}_{\perp}H^{2}(\R^{2}/\La)$ are
    \begin{align*}
    \mcm(\mu^{(1)},\ka)&:= \mathscr{Q}_{\parallel}\z(2i\ka\d\nabla_\bK\y)\mcp_0(\mu^{(1)},\ka), \\
    \wi{\mcm}(\mu^{(1)},\ka)&:=  \mathscr{Q}_{\parallel}\z(2i\ka\d\nabla_\bK-\ka\cdot\ka+\mu^{(1)}\y).
    \end{align*}
Note that \x{A3} is a linear system mapping from $\psi^{(0)}\in \mcx$ to $\mcx$, with an unknown parameter $\mu^{(1)}\in \R$.

Since $\mcx$ is 3-dimensional, we can write $\psi^{(0)}$ in
    \be \lb{psi0}
\psi^{(0)}= \sum_{\ell=1}^3 \al_\ell \phi_\ell,\qq \al_\ell\in \C.
    \ee
In order that \x{A3} has a nonzero solution $\psi^{(0)}\in \mcx$, it is necessary and sufficient that the corrections $\mu^{(1)}=\mu(\bK+\ka) -\mu(\bK)$ for eigenvalues are determined by
    \begin{equation}\lb{Mmk}
\det E(\mu^{(1)},\ka)=0,
    \end{equation}
where $E(\mu^{(1)},\ka)$ is the $3\tm 3$ representation of left-hand side of \x{A3}, using the coordinates for $\psi^{(0)}$ as in \x{psi0}. Precisely,
    \be\lb{M01}
    E(\mu^{(1)},\ka) \equiv M(\mu^{(1)},\ka)+\wi{M}(\mu^{(1)},\ka),
    \ee

where
    \[
    \begin{split}
    &M(\mu^{(1)},\ka)=\z(M_{m,\ell j}(\mu^{(1)},\ka)\y)_{3\tm 3} := \z(\inn{\phi_\ell} {\mcm(\mu^{(1)},\ka)\phi_j}\y)_{3\tm 3},\\
    &\wi{M}(\mu^{(1)},\ka)=\z(M_{m,\ell j}(\mu^{(1)},\ka)\y)_{3\tm 3} := \z(\inn{\phi_\ell} {\wi{\mcm}(\mu^{(1)},\ka)\phi_j}\y)_{3\tm 3}.
    \end{split}
    \]

3. Explicit computation for nondegeneracy condition. We need to give a more explicit computation for equation \x{Mmk}.

To this end, by using \x{psi0} for $\psi^{(0)}$, we have from \x{psi11}
    \begin{equation}\label{eq:expression-psi}
\psi^{(1)}(\bx)=\psi^{(1)}(\bx,\ka,\mu^{(1)}) \equiv \sum_{\ell=1}^3\alpha_\ell c^{(\ell)}(\bx,\ka,\mu^{(1)}),
    \end{equation}
where $c^{(\ell)}(\bx,\ka,\mu^{(1)})$, $\ell=1,2,3$, are bounded by
    \begin{equation}\lb{cl}
\z\|c^{(\ell)}(\d,\ka,\mu^{(1)})\y\|_{H^{2}}\leq C(|\ka|+|\mu^{(1)}|)\q \mbox{for } |\ka|+|\mu^{(1)}|\ll 1.
    \end{equation}
Let us define
    \[
    C^{(\ell)}(\bx)=C^{(\ell)}(\bx,\ka,\mu^{(1)}):= e^{i\bK\cdot\bx}c^{(\ell)}(\bx,\ka,\mu^{(1)}).
    \]
Recall that
    \[ 
    \Phi_\ell(\bx)=e^{i \bK \d \bx} \phi_\ell(\bx),\q \nabla_{\bK}\phi_{j}(\bx)
    =e^{-i\bK\cdot\bx}\nabla\Phi_{j}(\bx),\q \inn{\Phi_{\ell}}{\Phi_{j}}= \inn{\phi_{\ell}}{\phi_{j}}= \delta_{\ell j}.
    \]
Moreover, as $\mathscr{Q}_{\parallel}\psi^{(1)}=0$, we have from \x{eq:expression-psi} that $\mathscr{Q}_{\parallel}c^{(\ell)}=0$, i.e., $c^{(\ell)}\in \mcx^\perp$. Thus
    \[
\inn{\Phi_{\ell}}{C^{(j)}}=\inn{\phi_\ell}{c^{(j)}}=0.
    \]

The matrix-valued functions $M(\mu^{(1)},\ka)$ and $\wi{M}(\mu^{(1)},\ka)$ in \x{M01} are
    \[
    \begin{split}
    &M(\mu^{(1)},\ka)=\\
    &\begin{pmatrix}
    \mu^{(1)}+\langle \Phi_{1},2i\ka\d\nabla\Phi_{1}\rangle
    &\langle \Phi_{1},2i\ka\d\nabla\Phi_{2}\rangle
    &\langle \Phi_{1},2i\ka\d\nabla\Phi_{3}\rangle\\
    \langle \Phi_{2},2i\ka\d\nabla\Phi_{1}\rangle
    &\mu^{(1)}+\langle\Phi_{2},2i\ka\d\nabla\Phi_{2}\rangle
    &\langle \Phi_{2},2i\ka\d\nabla\Phi_{3}\rangle\\
    \langle \Phi_{3},2i\ka\d\nabla\Phi_{1}\rangle
    &\langle \Phi_{3},2i\ka\d\nabla\Phi_{2}\rangle
    & \mu^{(1)}+\langle\Phi_{3},2i\ka\d\nabla\Phi_{3}\rangle\\
    \end{pmatrix}\\
    &=\mu^{(1)}I+M(\ka),
    \end{split}
    \]

    \[
    \begin{split}
    &\wi{M}(\mu^{(1)},\ka)=\\
    &\begin{pmatrix}
    \ka\cdot\ka+\langle \Phi_{1},2i\ka\cdot\nabla C^{(1)}\rangle
    &\langle \Phi_{1},2i\ka\cdot\nabla C^{(2)}\rangle
    &\langle \Phi_{1},2i\ka\cdot\nabla C^{(3)}\rangle\\
    \langle \Phi_{2},2i\ka\cdot\nabla C^{(1)}\rangle
    &\ka\cdot\ka+\langle \Phi_{2},2i\ka\cdot\nabla C^{(2)}\rangle
    &\langle \Phi_{2},2i\ka\cdot\nabla C^{(3)}\rangle\\
    \langle \Phi_{3},2i\ka\cdot\nabla C^{(1)}\rangle
    &\langle \Phi_{3},2i\ka\cdot\nabla C^{(2)}\rangle
    &\ka\cdot\ka+\langle \Phi_{3},2i\ka\cdot\nabla C^{(3)}\rangle
    \end{pmatrix}.
    \end{split}
    \]
By noticing \x{cl}, we know that
    \[
    \wi{M}(\mu^{(1)},\ka)_{_{\ell j}}=\mathcal{O}(|\ka|\cdot|\mu^{(1)}|+|\ka|^{2}).
    \]



4. Bifurcation of eigenvalues. By the results in Theorem \ref{phaseK}, $M(\mu^{(1)},\ka)$ simplifies to
    \begin{align*}
\mu^{(1)}I+M(\ka)&= \begin{pmatrix}
\mu^{(1)}&\langle \Phi_{1},2i\ka\d\nabla\Phi_{2}\rangle&\langle \Phi_{1},2i\ka\d\nabla\Phi_{3}\rangle\\
\langle \Phi_{2},2i\ka\d\nabla\Phi_{1}\rangle&\mu^{(1)}&\langle \Phi_{2},2i\ka\d\nabla\Phi_{3}\rangle\\
\langle \Phi_{3},2i\ka\d\nabla\Phi_{1}\rangle&\langle \Phi_{3},2i\ka\d\nabla\Phi_{2}\rangle&\mu^{(1)}\\
\end{pmatrix}\\
&=
\begin{pmatrix}
\mu^{(1)}&\upsilon_{1}(\ka_{x}-i\ka_{y})&\ol{\upsilon_{3}}(\ka_{z})\\
\ol{\upsilon_{1}}(\ka_{x}+i\ka_{y})&\mu^{(1)}&\upsilon_{2}(\ka_{x}-i\ka_{y})\\
\upsilon_{3}(\ka_{z})&\ol{\upsilon_{2}}(\ka_{x}+i\ka_{y})&\mu^{(1)}\\
\end{pmatrix}.
    \end{align*}
Thus the bifurcation equation \x{Mmk} is
    \begin{equation}\label{eq:equality_threelambda}
\z(\mu^{(1)}\y)^{3}-\mu^{(1)}\z[|\upsilon_{3}|^{2}\ka_{z}^{2}+\frac{|\upsilon_{1}|^{2}+ |\upsilon_{2}|^{2}}{2} (\ka_{x}^{2}+\ka_{y}^{2})\y] -h(\ka)-g(\mu^{(1)},\ka)=0,
    \end{equation}
where
    \[
    \begin{split}
    &h(\ka)= \frac{1}{2}\upsilon_{1} \upsilon_{2} \upsilon_{3}(\ka^{2}_{x}-2i\ka_{x} \ka_{y}-\ka^{2}_{y}) \ka_{z}+\ol{\upsilon_{1}\upsilon_{2} \upsilon_{3}}(\ka^{2}_{x}+2i\ka_{x}\ka_{y}-\ka^{2}_{y})\ka_{z},\\
    &g(\mu^{(1)},\ka)=\mathcal{O}(|\ka|^{\alpha}|\mu^{(1)}|^{\beta}),\q  \alpha+\beta\geq4.
    \end{split}
    \]

The definitions and properties of $\upsilon_i$ are displayed in Theorem $\ref{phaseK}$. Note 
that $\upsilon_1 \upsilon_2 \upsilon_3$ is purely imaginary, thus we may set $\arg(\upsilon_1 \upsilon_2 \upsilon_3)=\frac{3\pi}{2}$, because the case $\arg(\upsilon_1 \upsilon_2 \upsilon_3)=\frac{\pi}{2}$ is similar. Hence (\ref{eq:equality_threelambda}) simplifies to
    \begin{equation}\lb{pka}
(\mu^{(1)})^3-\mu^{(1)}\upsilon^{2}_{_{\mathcal{F}}}|\ka|^2=-2\upsilon^{3}_{_{\mathcal{F}}}\ka_x \ka_y \ka_z+g(\mu^{(1)},\ka).
    \end{equation}
We then follow the arguments as done for Proposition 4.2 in \cite{fefferman2012honeycomb}. Setting $\mu^{(1)}=\xi \ups_{\mathcal{F}}|\ka|+o(|\ka|)$ and substituting into \x{pka}, we observe that $\xi$ solves the cubic equation \x{cubic}.
\ifl    \begin{equation}\label{eq:eq-of-eig}
\xi^{3}+2\ka^{\arg}-\xi=0,
    \end{equation}
where $\ka^{\arg}=\frac{\ka_x \ka_y \ka_z}{|\ka|^3}$.
\fi

By the Cauchy-Schwarz inequality, one has $|\ka^{\arg}|\leq\frac{\sqrt{3}}{9}$. We therefore conclude that equation \x{cubic} has precisely three real solutions $\xi_{+}\geq\xi_{0}\geq\xi_{-}$ by using the discriminant of cubic equations. Moreover, $\xi_{+}+ \xi_{0}+ \xi_{-}=0$. Actually, the Floquet-Bloch eigenvalue problem has three dispersion hypersurfaces
    \[
\begin{split}
\mu_{b+1}&=\mu_{*}+\xi_{+}\upsilon_{_{\mathcal{F}}}|\ka|+o(|\ka|),\q\\
\mu_{b}&=\mu_{*}+\xi_{0}\upsilon_{_{\mathcal{F}}}|\ka|+o(|\ka|), \q\\
\mu_{b-1}&=\mu_{*}+\xi_{-}\upsilon_{_{\mathcal{F}}}|\ka|+o(|\ka|).\\
\end{split}
    \]
Consequently, we have the desired results \x{mu1} and the proof of the theorem is complete.
\qed

From the Theorem \ref{thm:3+1}, we see that the three bands intersect at the degenerate point $(\bK, \mu_*)$.

Note that the roots $\xi_*=\xi_*(\ka/|\ka|)$ of equation \x{cubic} depend only on the directions of $\ka$, not on the sizes $|\ka|$ of the quasi-momenta $\ka$.

We want to point out that there is a special direction along which two energy bands adhere to each other to leading order. Specifically, if $\ka \in \mathbf{n_*}\mathbb{R}_+$ with  $\mathbf{n_*}=\z(\frac{\sqrt{3}}{3}, \frac{\sqrt{3}}{3}, \frac{\sqrt{3}}{3}\y)$, the solutions of  (\ref{cubic}) take the form
    \[
\xi_{+}=\xi_{0}=\frac{\sqrt{3}}{3},\qq \xi_{-}=-\frac{2\sqrt{3}}{3}.
    \]
The result indicates that the three-fold degeneracy splits into a two-fold eigenvalue and a simple eigenvalue in the vicinity of the Weyl point $\bK$. We remark here that it is not clear whether the double degeneracy persists by including higher order terms of $|\ka|$. This is an interesting problem but is beyond the scope of the current work.

At the end of this section, we characterize the lower dimensional structure of the three energy bands near the Weyl point $\bK$. According to the expressions of dispersion bands $\mu(\ka)$ in \x{mu1}, we study a special case of dispersion equation \x{cubic} as follows. If $\ka^{\arg}=0$, or equivalently, either of $\ka_x,\ \ka_y,\ \ka_z$ vanishes, the bifurcation equation (\ref{cubic}) has solutions
    \[
    \xi_{+}=1,\qq\xi_{0}=0,\qq \xi_{-}=-1.\qq
    \]
In the transverse plane which is perpendicular to one axis direction, the three dispersion surfaces form a standard cone with a flat band in the middle, see Figure 2 in Section 7. This is exactly the band structure of the Lieb lattice in the tight binding limit \cite{Keller2018,Mukherjee14}. To the best of our knowledge, this structure has not been rigorously proved. We demonstrate its existence for our potentials in lower reduced planes.

Generally speaking, in the reduced plane, the three dispersion bands do not behave the same as the above case. Note that $(-\ka)^{\arg} =-\ka^{\arg}$. Let us fix a direction $\textbf{n}$. Then
    \[
\xi^{\textbf{n}}_{+}=-\xi^{-\textbf{n}}_{-},\qq \xi^{\textbf{n}}_{0}=-\xi^{-\textbf{n}}_{0},\qq
\xi^{\textbf{n}}_{-}=-\xi^{-\textbf{n}}_{+},
    \]
where the superscripts indicate the different choices of bifurcation equations depending on the directions $\textbf{n}$ or $-\textbf{n}$. We can actually construct three analytical branches of dispersion curves and each branch is a straight line to leading order. In fact, let us define
    \[
\begin{split}
E_{1}(\lambda)&=\mu_{b+1}(\bK+\lambda\textbf{n})=\mu_{*}+\xi^{\textbf{n}}_{+}\lambda\upsilon_{_{\mathcal{F}}}+o(|\lambda|), \\
E_{2}(\lambda)&=\mu_{b}(\bK+\lambda\textbf{n})=\mu_{*}+\xi^{\textbf{n}}_{0}\lambda\upsilon_{_{\mathcal{F}}}+o(|\lambda|),\\
E_{3}(\lambda)&=\mu_{b-1}(\bK+\lambda\textbf{n})=\mu_{*}+\xi^{\textbf{n}}_{-}\lambda\upsilon_{_{\mathcal{F}}}+o(|\lambda|).\\
\end{split}
    \]
Then for a fixed direction $\textbf{n}$, the three branches $E_j(\lambda),~ j=1,2,3$ are analytical in $\lambda$.

Next we allow $\textbf{n}$ to vary in a transverse plane. Namely,  let $\textbf{n}_1$ and $\textbf{n}_2$ be two orthonormal vectors and consider the dispersion surfaces in the plane spanned by $\textbf{n}_1$ and $\textbf{n}_2$. Then
    \[
\mu(\bK+\lambda_1 \textbf{n}_1 +\lambda_2\textbf{n}_2)=\mu_{*}+\lambda\ups_{_\mathcal{F}}\xi^{\hat{\lambda}}_{i}+o(|\lambda|),
    \]
where $|\lambda|$ denotes the length of $(\lambda_1,\lambda_2)$. Note, while $\lambda$ is fixed, $\ka^{\arg}$ is a continuous variable with respect to $\frac{\lambda_1}{\lambda}$, thus $\xi^{\hat{\lambda}}_{i}$ depends on $\frac{\lambda_1}{\lambda}$ continuously. Consequently, \x{mu1} exactly admits a cone (may not be standard and isotropic) adhered by an extra surface in the middle (see  Section 7 for related figures).

\section{Justification of Assumptions $\textbf{H1}$ and $\textbf{H2}$}
\label{justif}

Theorem \ref{thm:3+1} states that as long as H1-H2 hold, the Schr\"{o}dinger operator with an admissible potential always admits a 3-fold Weyl point at the high symmetry point $\bK$. In this section, we shall justify the two assumptions \textbf{H1-H2} can actually hold generally. We first examine shallow potentials in which case we can treat the small potential as a perturbation to the Laplacian operator. Then we can conduct the perturbation theory. The main difficulty is to prove the 3-fold degeneracy persists at any order of the asymptotic expansion. We remark that in the 2-D honeycomb case \cite{fefferman2012honeycomb}, the 2-fold degeneracy is naturally protected by the inversion symmetry. But that is not enough for higher multiplicity. What are the required arguments on the 3-fold degeneracy? We will answer this question in our analysis by imposing novel symmetry arguments.

\subsection{Weyl points in shallow potential case}

We first consider the Floquet-Bloch eigenvalue problem for the operator $H^{\e}=-\Delta+\e V(\bx)$, where $\e$ is possibly small and $V(\bx)$ is a nonzero admissible potential. Without loss of generality, we consider the case that $\e$ is positive. Then the $\bK$-pseudo-periodic eigenvalue problem on the four eigenspaces of $L^{2}_{\bK,i^{\ell}},\ \ell\in\{1,2,3,4\}$ takes the form
    \begin{equation}\label{eq:potential-system}
 \begin{split}
 H^{\e}\Phi_{\ell}(\bx,\bK)&\equiv [-\Delta+\e V(\bx)]\Phi^{\e}_{\ell}(\bx,\bK)=\mu^{\e}(\bK)\Phi^{\e}_{\ell}(\bx,\bK),\q\bx\in\R^{3} ,\\
 \Phi^{\e}_{\ell}({\bf x+v,K})&=e^{i\bK\cdot \bv}\Phi^{\e}_{\ell}({\bf x,K}),\q\bx\in\R^{3},\ \bv \in \La,\\
 \mcr[\Phi^{\e}_{\ell}(\bx,\bK)]&=i^{\ell}\Phi^{\e}_{\ell}(\bx,\bK), \q \ell\in\{1,2,3,4\}.\\
 \end{split}
    \end{equation}

We first study the special case that $\e=0$.
Note that $R$ is orthogonal and
    \[
    |\bK|=|R\bK|=|R^{2}\bK|=|R^{3}\bK|=\frac{3}{4} q^2.
    \]
By letting $\mu^{(0)}=|\bK|^2=\frac{3}{4} q^2$, we know that $e^{iR^\ell \bK\cdot\bx}$ are eigenfunctions associated with $\mu^{(0)}$. Thus $\mu^{(0)}$ is an eigenvalue of $H^{0}$ of multiplicity at least 4. To show that the multiplicity of $\mu^{(0)}$ is exactly 4, for $\bm=(m_{1},m_{2},m_{3})\in \Z^3$ and $\bq=m_{1}\bq_{1} +m_{2}\bq_{2}+ m_{3}\bq_{3}\in \Las$, the equation
    \[
    |\bK +\bq|^2 = |\bK|^2
    \]
will lead to
    \[
    [(2m_{1}+2m_{2}-1)^{2}+(2m_{1}+2m_{3}-1)^{2}+(2m_{2}+2m_{3}-1)^{2}]q^{2}=3q^{2}.
    \]
 Since $m_{1},\ m_{2},\ m_{3}$ are integers, it is
    \[
(2m_{1}+2m_{2}-1)^{2}=(2m_{1}+2m_{3}-1)^{2}=(2m_{2}+2m_{3}-1)^{2}=1,
    \]
with the precisely 4 solutions
    \[
    \bm=(0,0,0),\q (1,0,0),\q(0,1,0),\q(0,0,1).
    \]
For these $\bm$, $\bK+\bq$ correspond to $R^\ell\bK=\bK+\bq_\ell$, $\ell=1,2,3,4$, cf. \x{RK}.

Summarizing the above calculations, we have

    \bb{prop}\lb{multi}
The Laplacian $H^0\equiv-\Delta$ admits a real four-fold eigenvalue $\mu^{(0)}=|\bK|^2=\frac{3}{4} q^2$ at $\bK$, with the eigenspace spanned by $\z\{e^{iR^{\ell}\bK\cdot\bx}:\ell=1,2,3,4\y\}$.
    \end{prop}

Notice from \x{RK} that $R^\nu \bK = \bK+ \bq_\nu$. Let us take the following eigenfunctions associated with $\mu^{(0)}$
    \be \lb{Phi-l0}
    \begin{split}
    \Phi_{\ell}^{0}(\bx)&= \Phi_\ell^{0}(\bx,\bK):=\frac{1}{\sqrt{4|\Omega|}}\z(e^{i\bK\cdot\bx}+\ol{i^\ell}e^{iR\bK\cdot\bx}+ i^{2\ell} e^{iR^{2}\bK\cdot\bx} +i^\ell e^{iR^{3}\bK\cdot\bx}\y)\nn\\
    &= \frac{1}{\sqrt{4|\Omega|}}\sum_{\nu=0}^3 i^{-\ell \nu} e^{i R^\nu \bK\cdot\bx} \in L^{2}_{\bK,i^\ell}=
\frac{1}{\sqrt{4|\Omega|}}\sum_{\nu=0}^3 i^{-\ell \nu} e^{i(\bK+ \bq_\nu)\cdot\bx},
\end{split}
    \ee
where $\ell=1,2,3,4$, cf. \x{decomp2}. It is easily seen that
     \[
     \inn{\Phi_\ell^{0}}{\Phi_j^{0}} = \da_{\ell j}, \qq \ell,\, j=1,2,3,4.
     \]

Based on the results in Proposition \ref{multi}, we can justify Assumptions $\textbf{H1}$ and $\textbf{H2}$ when $\e>0$ is sufficiently small.

    \begin{thm}\label{thm:dispersion-3+1}
Let $V(\bx)$ be an admissible potential. Suppose that the Fourier coefficient $V_{1,0,0}>0$. Then there exists a constat $\varepsilon_0>0$ such that for any $\e\in (0,\e_0)$, $H^\e=-\Delta+\e V(\bx)$ fulfills the assumptions {\rm\textbf{H1}} and {\rm\textbf{H2}}. Moreover, one has
    \begin{align}
    \mu_{*}=\mu^{\e}_{\ell}&= |\bK|^{2}+\e(V_{0,0,0}-V_{1,0,0})+\mathcal{O}(\e^{2}),\q \ell=1,2,3, \lb{muss}\\
    \z|\ups^{\e}_{_\mathcal{F}}\y|&= q+\mathcal{O}(|\varepsilon|)>0.\lb{la-i1}
    \end{align}
Hence the lowest three energy bands intersect at the three-fold Weyl point $(\bK,\mu_{*})$.
    \end{thm}

\begin{rem}
The requirement $V_{1,0,0}>0$ in Theorem $\ref{thm:dispersion-3+1}$ can be replaced by $V_{1,0,0}<0$. In the latter case, one has the second, third and fourth bands intersect at the Weyl point $(\bK,\mu_{*})$.
\end{rem}

The proof of Theorem \ref{thm:dispersion-3+1} is inspired by the methods in \cite{Lee-Thorp2017}, where the 2-fold Dirac points in the $2$-D honeycomb structure is studied. The main difficulty in the present case is the justification of the three-fold degeneracy of the perturbed eigenvalue $\mu_{*}$ at $\bK$. Recall that the two-fold degeneracy is protected by the $\mathfrak{PT}$-symmetry of $V(\bx)$ in the $2$-D honeycomb case. The potential in our work also possesses the $\mathfrak{PT}$-symmetry so that a two-fold eigenvalue $\mu_{*}$ at $\bK$ is guaranteed. However, this is not adequate to admit the three-fold degeneracy of $\mu_{*}$. In fact, we need to combine $\mathcal{T}$-symmetry to ensure that another eigenvalue is the same as $\mu_{*}$ at $\bK$. This is the main difference compared to the analysis of the previous work. In the following proof, we only list the key calculations and point out the new ingredients.

\bigskip

We begin to prove Theorem \ref{thm:dispersion-3+1}.

1. Recall that $\mu^{(0)}=|\bK|^{2}$ is the eigenvalue of the Laplacian $-\Delta$ of multiplicity $4$. Moreover, $\mu^{(0)}$ is also a simple $L^{2}_{\bK,i^{\ell}}$-eigenvalue for $\ell\in\{ 1,2,3,4\}$, with the corresponding eigenstates $\Phi^{0}_{\ell}$. Let us decompose $\Phi^{\e}_{\ell}(\bx,\bK)\in L^{2}_{\bK,i^{\ell}}$ as
    \[
    \Phi^{\e}_{\ell}(\bx,\bK) =\Phi^{(0)}_{\ell}(\bx,\bK) +\e\Phi^{(1)}_{\ell}(\bx,\bK).
    \]
Similar to \cite{xie2020wave}, by applying Lyapunov-Schmidt reduction to (\ref{eq:potential-system}), we obtain the expression for $\mu^{\e}$ for sufficiently small $\e$
    \begin{equation}\label{eq:expansion-eigenvalue}
\mu^{\e}=\mu_\ell^\e \equiv \mu^{(0)}+\e\langle\Phi^{(0)}_{\ell},V(\bx)\Phi^{(0)}_{\ell}\rangle+\mathcal{O}(\e^{2}), \quad \ell\in\{1,2,3,4\} .
    \end{equation}

We now turn to the calculation of $\langle\Phi^{(0)}_{\ell},V(\bx)\Phi^{(0)}_{\ell}\rangle$. By using the $\mathcal{R}$-invariance of $V(\bx)$, it follows that
    \begin{equation}\label{cv}
\begin{split}
V_{0,0,0}&=\langle e^{i\bK\cdot\by},V(\by)e^{i\bK\cdot\by}\rangle=\langle e^{iR\bK\cdot\by},V(\by)e^{iR\bK\cdot\by}\rangle\\
&=\langle e^{iR^{2}\bK\cdot\by},V(\by)e^{iR^{2}\bK\cdot\by}\rangle
=\langle e^{iR^{3}\bK\cdot\by},V(\by)e^{iR^{3}\bK\cdot\by}\rangle,\\
V_{1,0,0}&=\langle e^{i\bK\cdot\by},V(\by)e^{iR\bK\cdot\by}\rangle=\langle e^{iR\bK\cdot\by}, V(\by)e^{iR^{2}\bK\cdot\by}\rangle\\
& =\langle e^{iR^{2}\bK\cdot\by},V(\by)e^{iR^{3}\bK\cdot\by}\rangle=\langle e^{iR^{3}\bK\cdot\by},V(\by)e^{i\bK\cdot\by}\rangle,\\
V_{0,1,0}& =\langle e^{i\bK\cdot\by},V(\by)e^{iR^{2}\bK\cdot\by}\rangle=\langle e^{iR\bK\cdot\by},V(\by)e^{iR^{3}\bK\cdot\by}\rangle\\
& =\langle e^{iR^{2}\bK\cdot\by},V(\by)e^{i\bK\cdot\by}\rangle=\langle e^{iR^{3}\bK\cdot\by},V(\by)e^{iR\bK\cdot\by}\rangle,\\
V_{0,0,1} &=\langle e^{i\bK\cdot\by},V(\by)e^{iR^{3}\bK\cdot\by}\rangle=\langle e^{iR\bK\cdot\by},V(\by)e^{i\bK\cdot\by}\rangle\\
& =\langle e^{iR^{2}\bK\cdot\by},V(\by)e^{iR\bK\cdot\by}\rangle=\langle e^{iR^{3}\bK\cdot\by},V(\by)e^{iR^{2}\bK\cdot\by}\rangle,
\end{split}
    \end{equation}
where
    $$
    V_{\alpha, \beta, \gamma}=\int_{\Om}e^{-i(\alpha\bq_{1}+\beta\bq_{2}+\gamma\bq_{3})}V(\by)d\by.
    $$
By inserting the expansion $(\ref{Phi-l0})$ of $\Phi^{(0)}_{\ell}(\bx,\bK)$ and the coefficients (\ref{cv}) into (\ref{eq:expansion-eigenvalue}), and noticing that $V(\bx)$ is even, it follows that
    \begin{equation}\lb{mue}
    \mu^{\e}_\ell=
\begin{cases}
&\mu^{(0)}+\e(V_{0,0,0}-V_{1,0,-1})+\mathcal{O}(\e^{2}),  \q \ell=1,3,\\
&\mu^{(0)}+\e(V_{0,0,0}+V_{1,0,-1}-2V_{1,0,0})+\mathcal{O}(\e^{2}), \q \ell=2,\\
&\mu^{(0)}+\e(V_{0,0,0}+V_{1,0,-1}+2V_{1,0,0})+\mathcal{O}(\e^{2}), \q \ell=4.
\end{cases}
    \end{equation}

2. 
Since $V(\bx)$ is $\mathcal{T}$-invariant, we have $V_{1,0,-1}=V_{1,0,0}>0$. In particular, \x{mue} is simplified to
    \begin{equation}\lb{mue1}
 \begin{split}
\mu^{\e}_1 =&\mu^{\e}_3=\mu^{(0)}+\e(V_{0,0,0}-V_{1,0,0})+\mathcal{O}(\e^{2}),  \q \ell=1,3,\\
\mu^{\e}_2=&\mu^{(0)}+\e(V_{0,0,0}-V_{1,0,0})+\mathcal{O}(\e^{2}), \q \ell=2,\\
\mu^{\e}_4=&\mu^{(0)}+\e(V_{0,0,0}+3V_{1,0,0})+\mathcal{O}(\e^{2}), \q \ell=4.
\end{split}
    \end{equation}

Here one shall notice that the $\mathcal{O}(\e^2)$ terms in $\mu^{\e}_{1,2}$ and $\mu^{\e}_{3}$ are undetermined. This means that we could not assert that $\mu^{\e}_{1,3}=\mu^{\e}_{2}$. However, it follows from \x{mue1} that these eigenvalues are  ordered so that
  \[
  \mu^{\e}_{1}=\mu^{\e}_{3}\approx \mu^{\e}_{2}<\mu^{\e}_{4}\ .
  \]

Let $\mathcal{E}_{\mu^{\e}_1}$ denote the eigenspace of $H^{\e}\Phi^{\e}=\mu^{\e}_{1}\Phi^{\e}$. Then the above analysis shows that
  \be\lb{prope}
  \mathcal{E}_{\mu^{\e}_{1}}\subset L^{2}_{\bK,i}\oplus L^{2}_{\bK,-i}\oplus L^{2}_{\bK,-1},\andq
  2\leq\dim \mathcal{E}_{\mu^{\e}_1}\leq3\ .
  \ee

The next step is to verify that $\mu^{\e}_{1}$ is really a three-fold eigenvalue, i.e., $\dim \mathcal{E}_{\mu^{\e}_{1}}=3$, with the help of the following lemma.

    \begin{lem}\lb{strut}
We assert that
$\mathcal{T}\Phi^{\e}_{1}\notin L^{2}_{\bK,i}\oplus L^{2}_{\bK,-i}$ for $\e$ is sufficiently small.
    \end{lem}

The detailed proof of Lemma \ref{strut} is displayed in Appendix B.

We continue the proof for Theorem \ref{thm:dispersion-3+1}. Recall that $[H,\mathcal{T}]=0$. Thus
     \[
 \mathcal{T}(-\Delta+\e V(\bx))\Phi^{\e}_{1}=(-\Delta+\e V(\bx))\mathcal{T}\Phi^{\e}_{1}=\mu^{\e}_{1}\mathcal{T}\Phi^{\e}_{1}\ .
     \]
Therefore 
$\mathcal{T}\Phi^{\e}_{1}\in \mathcal{E}_{\mu^{\e}_{1}}$. By Lemma $\ref{strut}$, we deduce that $\mathcal{T}\Phi^{\e}_{1}\notin L^{2}_{\bK,i}\oplus L^{2}_{\bK,-i}$. 
Hence
    $$
    \{\Phi^{\e}_{1}(\bx),\ \ol{\Phi^{\e}_{1}(-\bx)},\ \mathcal{T}\Phi^{\e}_{1}(\bx)\}
    $$
are linearly independent eigenfunctions in $\mathcal{E}_{\mu^{\e}_{1}}$. Thus $\dim \mathcal{E}_{\mu^{\e}_{1}}\geq3$. By \x{prope}, we conclude that $\dim \mathcal{E} _{\mu^{\e}_{1}}=3$ and $\mu^{\e}_{1}$ is a three-fold eigenvalue. Moreover, result \x{muss} follows from \x{mue1}.

3. We then embark on the proof of \x{la-i1}. In analogy with the construction of $\ups_{\ell}$ in Theorem \ref{phaseK}, we introduce $\ups^{(0)}_{\ell}$ by
    \[
 \ups^{(0)}_{1}\om_1=\langle\Phi^{(0)}_{1},2i\nabla\Phi^{(0)}_{2} \rangle \q \ups^{(0)}_{2}\om_1=\langle\Phi^{(0)}_{2},2i\nabla\Phi^{(0)}_{3}\rangle \q \ups^{(0)}_{3}\om_3=\langle\Phi^{(0)}_{3},2i\nabla\Phi^{(0)}_{1}\rangle\ .
    \]
Actually in the following we will present the full calculations for each $\ups^{(0)}_{\ell}$ under the above choice of $\Phi^{(0)}_{\ell}$. By discussions given in \cite{fefferman2012honeycomb}, it is standard to apply the Lyapunov-Schmidt reduction to approximate $\upsilon_{\ell}$ while $0<\e\ll 1$. The result is
    $$
    \langle \Phi_{\ell},2i\nabla\Phi_j\rangle= \langle \Phi_{\ell}^{(0)},2i\nabla\Phi_j^{(0)} \rangle +\mathcal{O}(|\e|)\ .
    $$
Here
    \[
\Phi^{(0)}_{\ell}(\bx)=\frac{1}{\sqrt{4|\Omega|}}\sum_{\nu=0}^3 i^{-\ell \nu} e^{i(\bK+ \bq_\nu)\cdot\bx}, \qq \ell=1,2,3,4.
    \]
Thus we can directly deduce that
    \begin{equation*}
\begin{split}
\ol{\Phi_{\ell}^{(0)}(\bx)} &=\frac{1}{\sqrt{4|\Omega|}} \sum_{\nu=0}^3 (-i)^{-\ell \nu} e^{-i(\bK+ \bq_\nu)\cdot\bx},\\
\nabla \Phi_{j}^{(0)}(\bx) & = \frac{-2}{\sqrt{4|\Omega|}} \sum_{\mu=0}^3 i^{-j \mu} e^{i(\bK+ \bq_\mu)\cdot\bx}(\bK+ \bq_\mu),\\
\ol{\Phi_{\ell}^{(0)}(\bx)} 2i \nabla \Phi_{j}^{(0)}(\bx) & =\frac{-2}{4|\Omega|}\sum_{\nu=0}^3 \sum_{\mu=0}^3(-i)^{-\ell \nu} i^{-j \mu} e^{i(\bq_\mu-\bq_\nu)\cdot\bx}(\bK+ \bq_\mu)\ .
\end{split}
    \end{equation*}
Therefore, by setting $d_{\ell j} :=(-i)^{-\ell} i^{-j}\equiv i^{\ell-j}$, one has
    \begin{align*}
    \inn{\Phi_{\ell}^{(0)}}{2i \nabla \Phi_{j}^{(0)}}&= -\f{1}{2} \sum_{\nu=0}^3 (d_{\ell j})^\nu(\bK+ \bq_\nu)\\
    &= -\f{1}{2} \z(\sum_{\nu=0}^3 (d_{\ell j})^\nu\y)\bK -\f{1}{2} \sum_{\nu=1}^3 (d_{\ell j})^\nu \bq_\nu\\
    &\equiv -\f{1}{2} \sum_{\nu=1}^3 (d_{\ell j})^\nu \bq_\nu,
    \end{align*}
where $(\ell,j)=(1,2), \ (2,3), \ (3,1)$. Since
    \(
    d_{12}=d_{23}=-i
    \)
    and
    \(
    d_{31}=-1,
    \)
one has
    \be\lb{vis}
    \begin{split}
    &\inn{\Phi_{1}^{(0)}}{2i \nabla \Phi_{2}^{(0)}}= \inn{\Phi_{2}^{(0)}}{2i \nabla \Phi_{3}^{(0)}}= -\f{1}{2}\sum_{\nu=1}^3 (-i)^\nu \bq_\nu\\ &=\f{1}{2} \z(i\bq_1+\bq_2-i \bq_3 \y)=\frac{\sqrt{2}}{2}(1+i)q\om_1,\\
    &\inn{\Phi_{3}^{(0)}}{2i \nabla \Phi_{1}^{(0)}}= -\f{1}{2} \sum_{\nu=1}^3 (-1)^\nu \bq_\nu=\f{1}{2} \z(\bq_1-\bq_2+\bq_3 \y)=-q\om_3\ .
    \end{split}
    \ee
From these we directly obtain $|\ups^{(0)}_{1}|=|\ups^{(0)}_{2}|=|\ups^{(0)}_{3}|=q$. This completes the proof of \x{la-i1}.
\qed

\begin{rem}
Note from \x{vis} that $\ups^{(0)}_{1}=\ups^{(0)}_2 =e^{i\frac{\pi}{4}}q$ and $\ups^{(0)}_{3}=-q$. Thus
    \begin{equation}\lb{argup}
\frac{\ups_1 \ups_2 \ups_3}{|\ups_1 \ups_2 \ups_3|}=\frac{\ups^{(0)}_{1} \ups^{(0)}_2 \ups^{(0)}_{3}+\mathcal{O}(|\e|)}{|\ups_1 \ups_2 \ups_3|}=
\frac{-q^3+\mathcal{O}(|\e|)}{q^3+\mathcal{O}(|\e|)}=-1+\mathcal{O}(|\e|)\ .
    \end{equation}
Recall that $\ups_1 \ups_2 \ups_3$ is gauge invariant and $\arg(\ups_1 \ups_2 \ups_3)$ is either $\frac{\pi}{2}$ or $\frac{3\pi}{2}$ by Theorem $\ref{phaseK}$. Thus we assert from \x{argup} that
    \[
    \arg(\ups_1 \ups_2 \ups_3)=\frac{3\pi}{2}
    \]
when $\e$ is sufficiently small.
\end{rem}

    \subsection{Remark on Weyl points in generic admissible potentials}\lb{generic}

Theorem \ref{thm:dispersion-3+1} studies the 3-fold Weyl points for the Schr\"odinger operator with shallow admissible potentials: $H^\e=-\Delta +\e V(\bx)$ for $\e\neq 0$ and small. In this subsection we make some remarks on the extension of these results to generic potentials, i.e., $\e=\mathcal{O}(1)$. Following the arguments established by Fefferman and Weinstein for the existence of Dirac points in 2-D honeycomb potentials, see \cite{fefferman2012honeycomb,Fefferman2014}, we claim that the assumptions {\bf H1} and {\bf H2} hold for some $(\bK, \mu_*)$ except for $\e$ in a discrete set $\mathcal{C}$ of $\R$.
Consequently, the conclusions of Theorem \ref{thm:dispersion-3+1} also hold, i.e., there always exists a 3-fold Weyl point, for the Schr\"odinger operator $H^\e=-\Delta +\e V(\bx)$ if $\e$ is not in the discrete set $\mathcal{C}$.

The main idea is based on an analytical characterization of $L^2_{\bK, \lambda}$-eigenvalue of $H^\e$. By a similar argument on the analytic operator theory and complex function theory strategy \cite{fefferman2012honeycomb,Fefferman2014}, it is possible to establish the analogous result.
Due to the length of this work, we omit the details and refer interesting readers to \cite{Fefferman2014,fefferman2012honeycomb}.

\section{Instability of the Weyl point under symmetry-breaking perturbations }\lb{sec6}

In the preceding sections, we have demonstrated that the admissible potentials generically admit a 3-fold Weyl point at $\bK$. The admissible potentials are characterized by the inversion symmetry, the $\mathcal{R}$-symmetry and the $\mathcal{T}$-symmetry. Actually we have seen the 3-fold degeneracy at $\bK$ and conical structure in its vicinity are consequence of combined actions of these symmetries. In this section, we shall discuss the instability of the 3-fold Weyl point $(\bK,\mu_*)$ if some symmetry is broken. More specifically, we only show the case where the inversion symmetry is broken which can be compared with the results to the 2-fold Dirac points in 2-D honeycomb case. The calculation of the case where $\mathcal{T}$-symmetry is broken is very cumbersome and we shall not give detailed discussion and only give numerical examples in Section 7.

Consider the perturbed eigenvalue problem
\begin{equation}\label{eq:stable}
(H+\delta V_{p}(\bx))\Psi^{\delta}(\bx,\bK)=\mu^{\delta}\Psi^{\delta}(\bx,\bK),
\end{equation}
where $V_{p}(\bx)$ is real and odd, 
and $\delta$ is the perturbation parameter which is assumed to be small.

We expand $\mu^{\delta}$ and $\Psi^{\delta}(\bx)$ near the 3-fold Weyl point $(\bK, \mu_*)$ as
    \[
\Psi^{\delta}(\bx)=\Psi^{(0)}(\bx)+\Psi^{(1)}(\bx), \andq  \mu^{\delta}=\mu_{*}+\mu^{(1)},
    \]
where $\Psi^{(0)}$ is the unperturbed eigenfunction corresponding to the the unperturbed eigenvalue $\mu_*$. We have stated in Theorem $\ref{thm:3+1}$ that
     \[
 \Psi^{(0)}(\bx)=\sum\limits^{3}_{i=1}\alpha_{\ell} \Phi_{\ell}(\bx)\ .
     \]
Calculations analogous to those in the proof of Theorem \ref{thm:3+1} can lead to a system of homogeneous linear equations for $\alpha_1, \alpha_2,\alpha_3$
    \[
(\mu^{(1)}I-\mathcal{M}_{1}-\mathcal{M}_{2})
\begin{pmatrix}
\alpha_1\\ \alpha_2\\ \alpha_3
\end{pmatrix}
=0,
    \]
where
    \[
\mathcal{M}_{1}=
\begin{pmatrix}
\langle \Phi_{1}(\bx) ,\delta V_{p}(\bx)\Phi_{1}(\bx) \rangle&\langle \Phi_{1}(\bx) ,\delta V_{p}(\bx)\Phi_{2}(\bx) \rangle&\langle \Phi_{1}(\bx) , \delta V_{p}(\bx)\Phi_{3}(\bx) \rangle\\
\langle \Phi_{2}(\bx) ,\delta V_{p}(\bx)\Phi_{1}(\bx) \rangle&\langle \Phi_{2}(\bx) ,\delta V_{p}(\bx)\Phi_{2}(\bx) \rangle&\langle \Phi_{2}(\bx) ,\delta V_{p}(\bx)\Phi_{3}(\bx) \rangle\\
\langle \Phi_{3}(\bx) ,\delta V_{p}(\bx)\Phi_{1}(\bx) \rangle&\langle \Phi_{3}(\bx) ,\delta V_{p}(\bx)\Phi_{2}(\bx) \rangle&\langle \Phi_{3}(\bx) ,\delta V_{p}(\bx)\Phi_{3}(\bx) \rangle\\
\end{pmatrix},
    \]
and $\mathcal{M}_2$ includes higher order terms.

Therefore $\mu^{\delta}$ is the solution for the perturbed eigenvalue problem \x{eq:stable} if and only if $\mu^{(1)}$ solves
\be\lb{truc}
\det(\mu^{(1)}I-\mathcal{M}_{1}-\mathcal{M}_{2})=0\ .
\ee

Following a standard perturbation theory for Floquet-Bloch eigenvalue problems, we obtain that the solutions of \x{truc} satisfy
     \[
 \mu^{(1)}=\wi{\mu}+o(\delta),
     \]
 where $\wi{\mu}$ is the leading order effect of the perturbation which solves the equation
    \begin{equation}\label{eq:stable-equation}
\det(\wi{\mu}I-\mathcal{M}_{1})
=0\ .
    \end{equation}
To understand the problem, it is key to compute the explicit form of $\mathcal{M}_1$. Note that
    \begin{equation}\lb{pvp1}
\begin{split}
&\langle \Phi_{1}(\bx) ,V_{p}(\bx)\Phi_{3}(\bx) \rangle=-\langle \Phi_{1}(-\by) ,V_{p}(\by)\Phi_{3}(-\by) \rangle\\
=& -\langle \ol{\Phi_{3}(\by)} ,V_{p}(\by)\ol{\Phi_{1}(\by)} \rangle=-\langle \Phi_{1}(\by)V_{p}(\by),\Phi_{3}(\by)\rangle,\\
&\langle \Phi_{2}(\bx) ,V_{p}(\bx)\Phi_{2}(\bx) \rangle=-\langle \Phi_{2}(-\by) ,V_{p}(\by)\Phi_{2}(-\by) \rangle\\
= &-\langle \ol{\Phi_{2}(\by)} ,V_{p}(\by)\ol{\Phi_{2}(\by)} \rangle=-\langle \Phi_{2}(\by)V_{p}(\by),\Phi_{2}(\by)\rangle\ .
\end{split}
    \end{equation}
Therefore
    $$
    \langle \Phi_{1}(\bx) ,V_{p}(\bx)\Phi_{3}(\bx) \rangle=\langle \Phi_{3}(\bx) ,V_{p}(\bx)\Phi_{1}(\bx) \rangle=\langle \Phi_{2}(\bx) ,V_{p}(\bx)\Phi_{2}(\bx) \rangle=0\ .
    $$
Similarly,
    \begin{equation}\lb{pvp2}
\begin{split}
&\langle \Phi_{1}(\bx) ,V_{p}(\bx)\Phi_{1}(\bx) \rangle=-\langle \Phi_{1}(-\by) ,V_{p}(\by)\Phi_{1}(-\by)\rangle\\
 =& -\langle \ol{\Phi_{3}(\by)} ,V_{p}(\by)\ol{\Phi_{3}(\by)}\rangle=-\langle \Phi_{3}(\by)V_{p}(\by),\Phi_{3}(\by)\rangle,\\
&\langle \Phi_{1}(\bx) ,V_{p}(\bx)\Phi_{2}(\bx) \rangle=-\langle \Phi_{1}(-\by) ,V_{p}(\by)\Phi_{2}(-\by) \rangle\\
=&-\langle \ol{\Phi_{3}(\by)} ,V_{p}(\by)\ol{\Phi_{2}(\by)}\rangle=-\langle \Phi_{2}(\by),V_{p}(\by)\Phi_{3}(\by)\rangle\ .\\
\end{split}
    \end{equation}
Combining \x{pvp1} and \x{pvp2}, we obtain
    \be\label{M1per}
\mathcal{M}_1=
\delta\begin{pmatrix}
-\ups^{\sharp}_1&\ups^{\sharp}_2&0\\
\ol{\ups^{\sharp}_2}&-\ups^{\sharp}_2\\
0&-\ol{\ups^{\sharp}_2}&+\ups^{\sharp}_1
\end{pmatrix}
,
    \ee
where $\ups^{\sharp}_1$ and $\ups^{\sharp}_2$ represent $\langle \Phi_{1}(\bx) ,V_{p}(\bx)\Phi_{1}(\bx) \rangle$ and $\langle \Phi_{1}(\bx) ,V_{p}(\bx)\Phi_{2}(\bx) \rangle$ respectively. Obversely, $\ups^{\sharp}_1$ is real.

Let us assume that both $\ups^{\sharp}_1$ and $\ups^{\sharp}_2$ are nonzero. Substituting \eqref{M1per} into \eqref{eq:stable-equation}, we obtain
    \begin{equation}\lb{eqss}
\wi{\mu}(\wi{\mu}^{2}-\delta^2(\ups^{\sharp}_1)^2)=2\delta^2 |\ups^{\sharp}_2|^2\wi{\mu}\ .
    \end{equation}
Then we can conclude from \x{eqss} that the 3-fold degenerate point $(\bK, \mu_*)$ splits into 3 simple eigenvalues under an inversion-symmetry-broken perturbation. More precisely,
    \[
\begin{split}
&\mu^{\delta}_1=\mu_{*}+\delta\sqrt{(\ups^{\sharp}_1)^2+2|\ups^{\sharp}_2|^2}+o(\delta), \\
&\mu^{\delta}_2=\mu_{*}+o(\delta), \\
&\mu^{\delta}_3=\mu_{*}-\delta\sqrt{(\ups^{\sharp}_1)^2+2|\ups^{\sharp}_2|^2}+o(\delta)\ .
\end{split}
    \]

The above analysis implies that the 3-fold Weyl point does not persist if the inversion symmetry of the system is broken. We also include the numerical simulations for a typical admissible potential with an inversion-symmetry-broken perturbation in Section 7, see Figure 2. It is seen that the 3 bands do not intersect at $\bK$ and there exist two local gaps.

We remark that if $\mathcal{T}$-symmetry is broken and inversion-symmetry persists, the 3-fold degenerate point split into a 2-fold eigenvalue and a simple eigenvalue, see Figure 3 in Section 7. The reason is that the inversion symmetry naturally protects the 2-fold degeneracy which is similar to the 2-D honeycomb case. Due to the length of this work, we shall not include the detailed calculations while some of main ingredients can be found in our analysis to the bifurcation matrix $M(\ka)$ in Section 4.

\section{Numerical results}\label{subsec:3+1}
\lb{sec7}

In this section, we use numerical simulations to demonstrate our analysis. The numerical method that we use is the Fourier Collocation Method \cite{Yang2010}. The potential that we choose is
     \begin{equation}\lb{cVx}
\begin{split}
V(\bx)= 5(&\cos(\bq_{1}\cdot\bx)+\cos((\bq_{2}-\bq_{1})\cdot\bx)
+\cos((\bq_{3}-\bq_{2})\cdot\bx)+\cos(\bq_{3}\cdot\bx)\\
&+\cos(\bq_{2}\cdot\bx)+\cos((\bq_{3}-\bq_{1})\cdot\bx)).
\end{split}
    \end{equation}
It is evident that $V(\bx)$ is an admissible potential in the sense of Definition $\ref{def:definition-of-lattice}$.

%
%

According to our analysis--Theorem \ref{thm:3+1} and Theorem \ref{thm:dispersion-3+1}, the first three energy bands intersect conically at $\bK$. In the following illustrations, we plot the figures of first three energy bands in vicinity of $\bK$. Since the full energy bands are defined in $\mathbb{R}^3$, it is not easy to visualize such high dimensional structure. We just show the figures in the reduced parameter space, i.e., energy curves with the quasi-momentum being along certain specific directions and energy surfaces with the quasi-momentum being in a plane.

\begin{figure}[htbp]
\centering
\subfigure[]{
\begin{minipage}{0.3\textwidth}
  \includegraphics[width=0.95\textwidth,height=3cm]{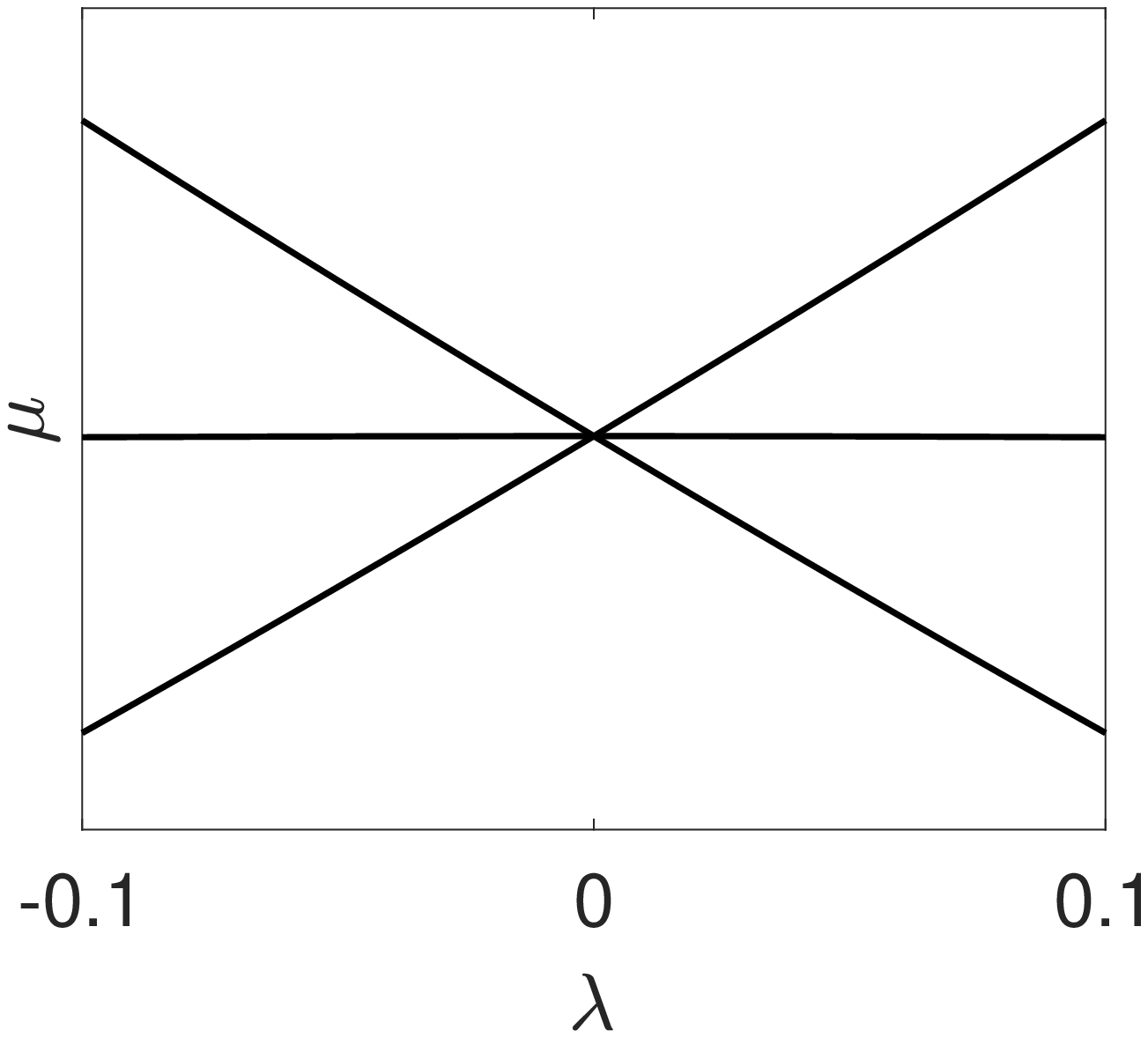}
  \end{minipage}
}
\subfigure[]{
\begin{minipage}{0.3\textwidth}
  \includegraphics[width=0.95\textwidth,height=3cm]{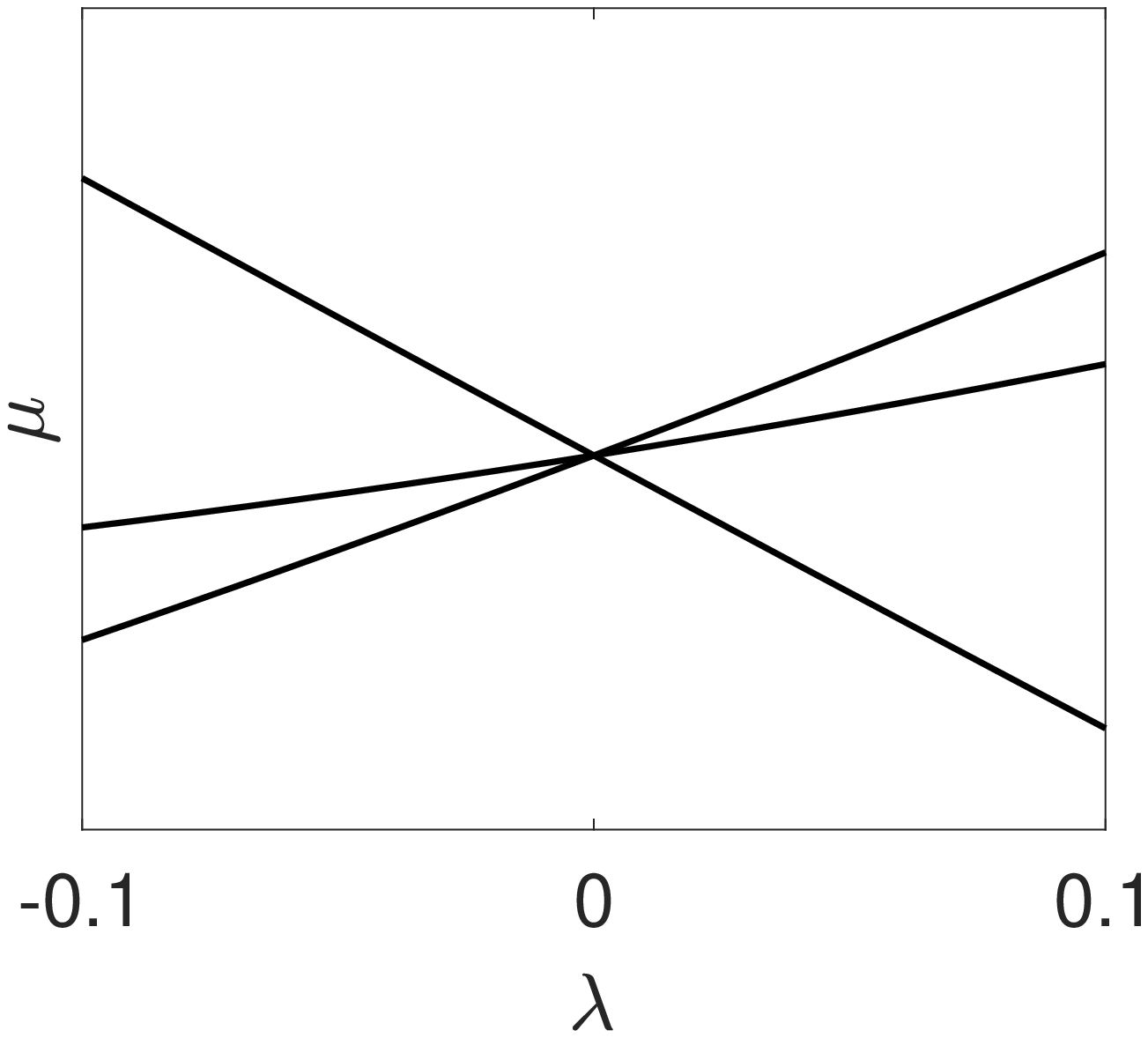}
  \end{minipage}
}
\subfigure[]{
\begin{minipage}{0.3\textwidth}
  \includegraphics[width=0.95\textwidth,height=3cm]{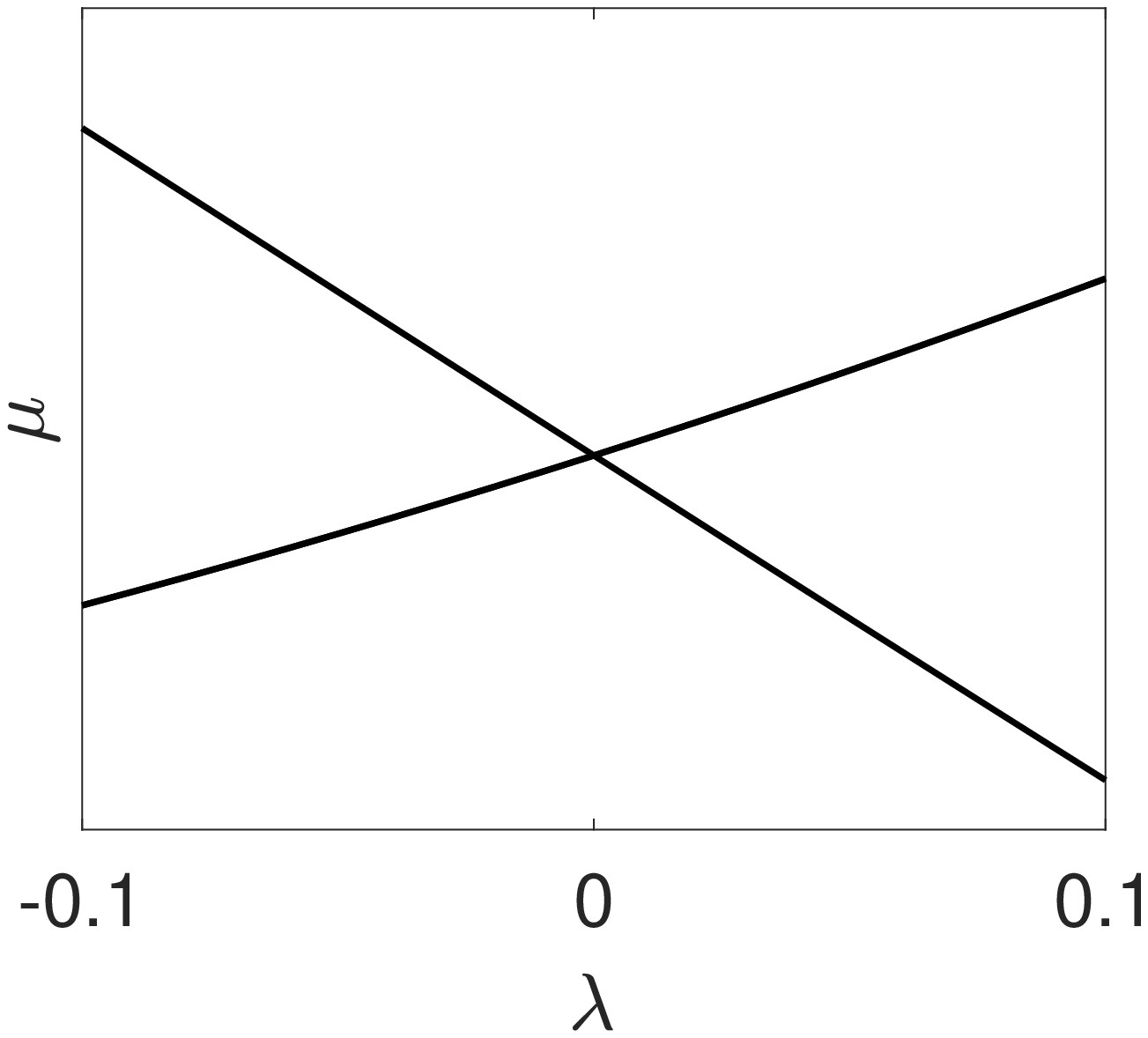}
  \end{minipage}
}

  \quad
  \subfigure[]{
\begin{minipage}{0.3\textwidth}
  \includegraphics[width=0.95\textwidth,height=3.5cm]{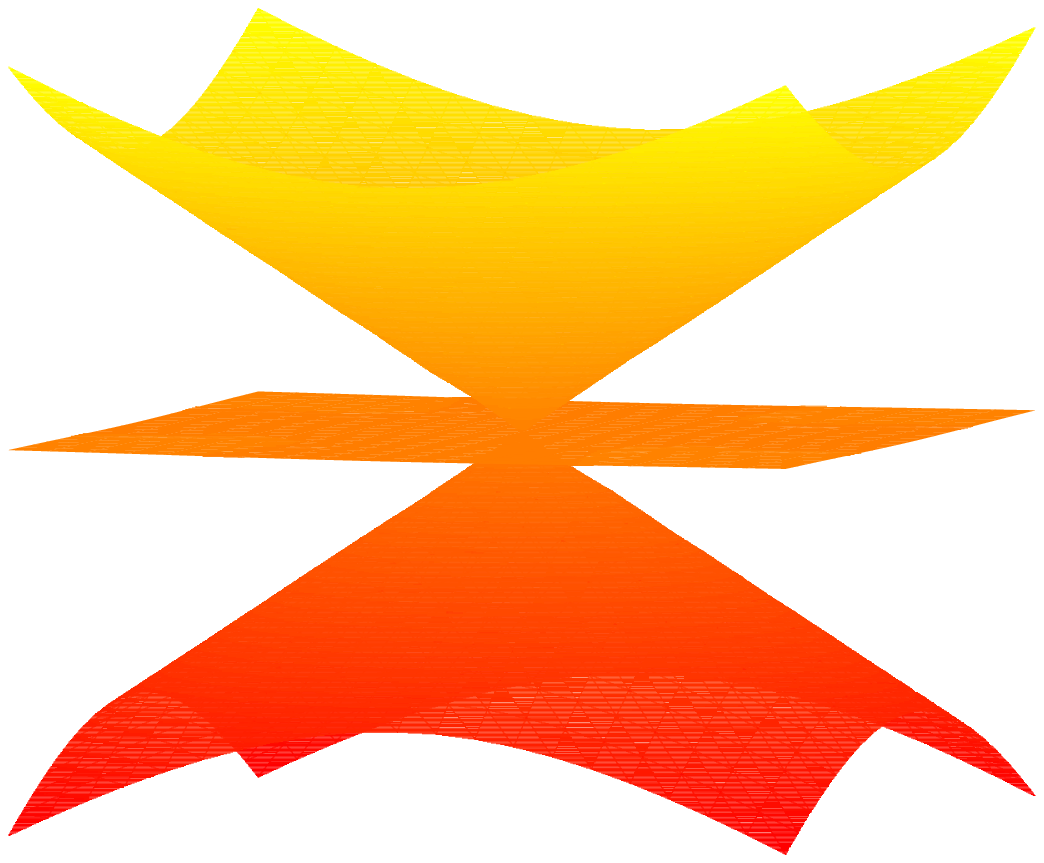}
  \end{minipage}
}
\subfigure[]{
\begin{minipage}{0.3\textwidth}
  \includegraphics[width=0.95\textwidth,height=3.5cm]{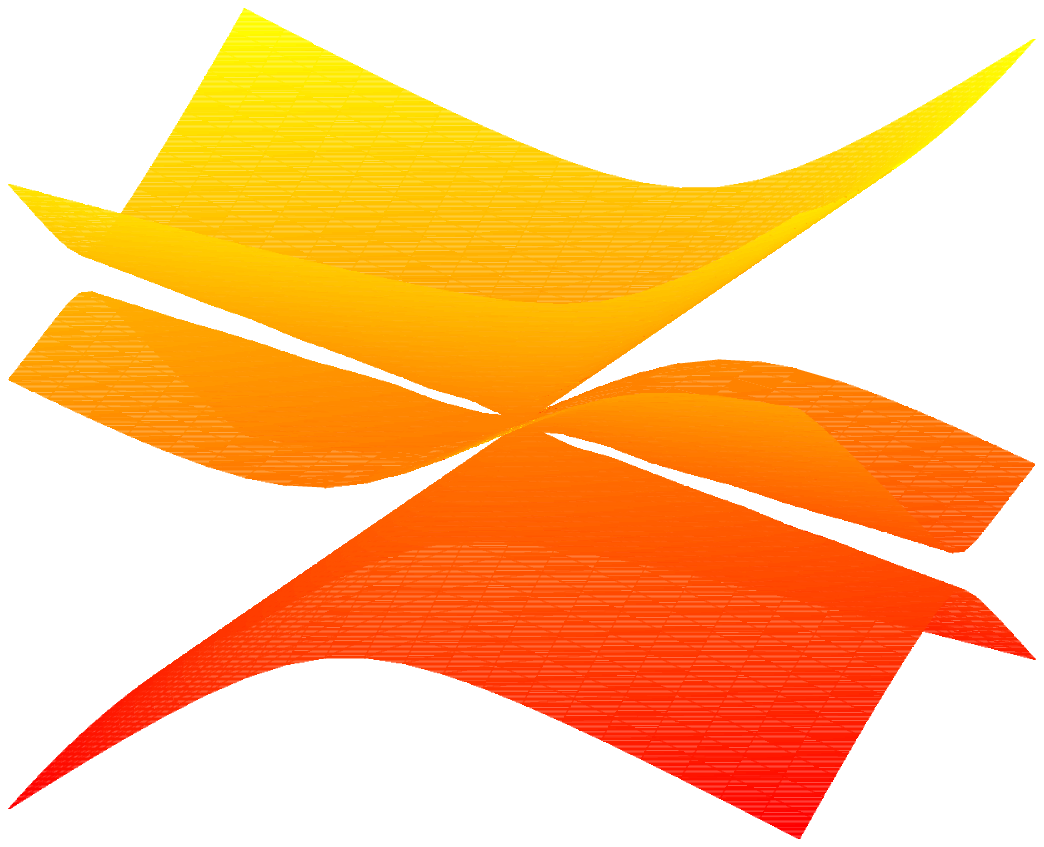}
  \end{minipage}
}
\subfigure[]{
\begin{minipage}{0.3\textwidth}
  \includegraphics[width=0.95\textwidth,height=3.5cm]{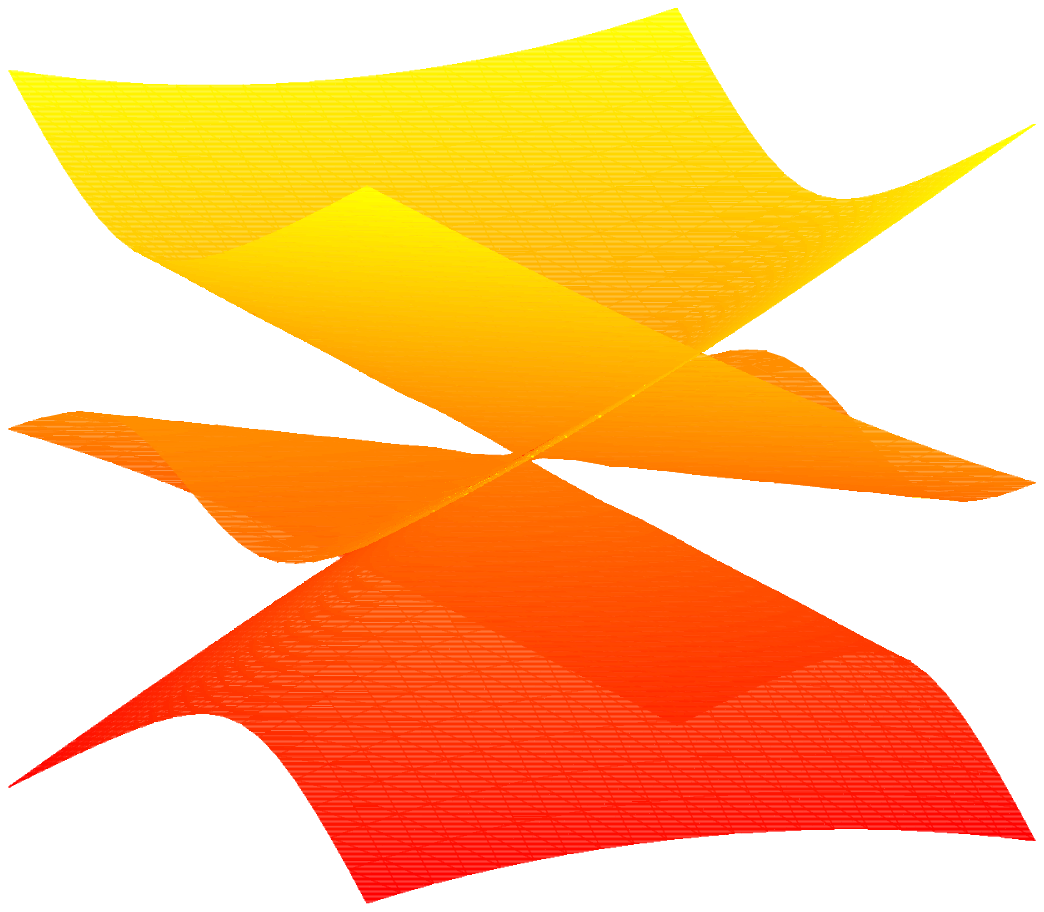}
  \end{minipage}
}
  \caption{The first three energy bands of $H=-\Delta +V(\bx)$ with $V(\bx)$ in \eqref{cVx} . {\bf Top Panel:} Energy curves $\mu(\bK+\lambda \textbf{n})$ in \x{bandc} with the quasi-momentum along a fixed direction $\textbf{n}$ being (a)$(1,0,0)$;(b)$\z(\frac{2}{3},\frac23,\frac13\y)$;(c)$ \z(\frac{\sqrt{3}}{3},\frac{\sqrt{3}}{3},\frac{\sqrt{3}}{3}\y)$ respectively.
  {\bf Bottom Panel:} Energy surfaces $\mu(\lambda_1,\lambda_2)$ with the quasi-momentum along two directions $\textbf{n}_1,\ \textbf{n}_2$, where $\textbf{n}_1$ is chosen to be $(1,0,0)$ and $\textbf{n}_2$ equals (d)$(0,0,1)$; (e)$\z(0,\frac{\sqrt{2}}{2},\frac{\sqrt{2}}{2}\y)$; (f)$\z(0,\frac35,\frac45\y)$. The three energy bands intersect conically at the origin, i.e., at the Weyl point.}\label{fig:3+1any direction}
\end{figure}

We plot dispersion bands $\mu(\bk)$ near $\bK$ along a certain direction $\textbf{n}$, i.e.,
    \be\lb{bandc}
\mu(\lambda)=\mu(\bK+\lambda\textbf{n}).
    \ee
The dispersion curves $\mu(\lambda)$ along three different directions are displayed on the top panel of Figure \ref{fig:3+1any direction} where we choose three different directions
    $$
\textbf{n}=(1,0,0),\q \z(\frac{2}{3},\frac23,\frac13\y),\q \z(\frac{\sqrt{3}}{3},\frac{\sqrt{3}}{3},\frac{\sqrt{3}}{3}\y).
    $$
In the first two cases, we see that the three straight lines intersect at $\lambda=0$, i.e., at the Weyl point. In the last example, we only see two straight lines intersect since one straight line is two-fold degenerate to leading order, see discussions in Section 5. The numerical simulations agree with our analysis given in Theorem \ref{thm:3+1}.

%

We also plot the energy surfaces with the quasi-momentum varying in along two directions, i.e.,
    \[
\mu(\lambda_1,\lambda_2)=\mu(\bK+\lambda_1\textbf{n}_1+\lambda_2\textbf{n}_2).
    \]

The dispersion surfaces $\mu(\lambda_1,\lambda_2)$ are displayed on the bottom panel of Figure \ref{fig:3+1any direction} where in all cases $\textbf{n}_1=(1,0,0)$ and
    $$
    \textbf{n}_2=(0,0,1),\q \z(0,\frac{\sqrt{2}}{2},\frac{\sqrt{2}}{2}\y),\q\z(0,\frac35,\frac45\y)
    $$
respectively. From the figure, we see that the three dispersion surfaces intersect at the Weyl point. The first and third bands conically intersect each other with the second band in the middle. This result also agrees well with our analysis.

%

%

We next verify the instability of conical singularity under certain symmetry-breaking perturbations.   A perturbation is added to the above admissible potential \eqref{cVx}. In other words, we consider the Schr\"{o}dinger
operator $H^{\delta}_{i}=-\Delta+V(\bx)+\delta V_{p_{i}}(\bx)$, where $V_{p_i}(\bx),\ i=1,2$ denote the perturbation potential and $\delta$ a small parameter. In our simulations, we choose $\delta=0.01$.

\begin{figure}[htbp]
\centering
\subfigure[]{
\begin{minipage}{0.3\textwidth}
  \includegraphics[width=0.95\textwidth,height=3cm]{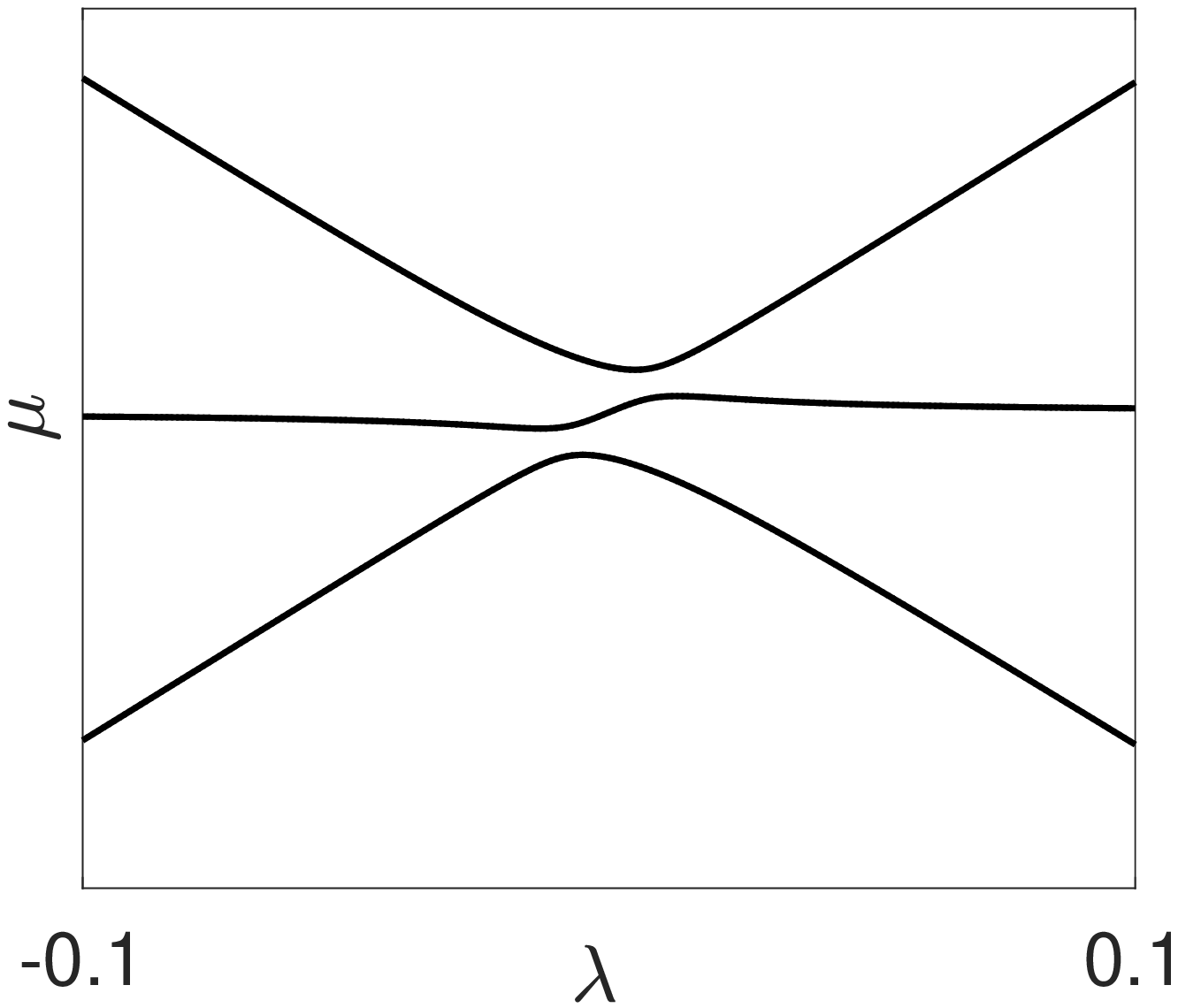}
  \end{minipage}
}
\subfigure[]{
\begin{minipage}{0.3\textwidth}
  \includegraphics[width=0.95\textwidth,height=3cm]{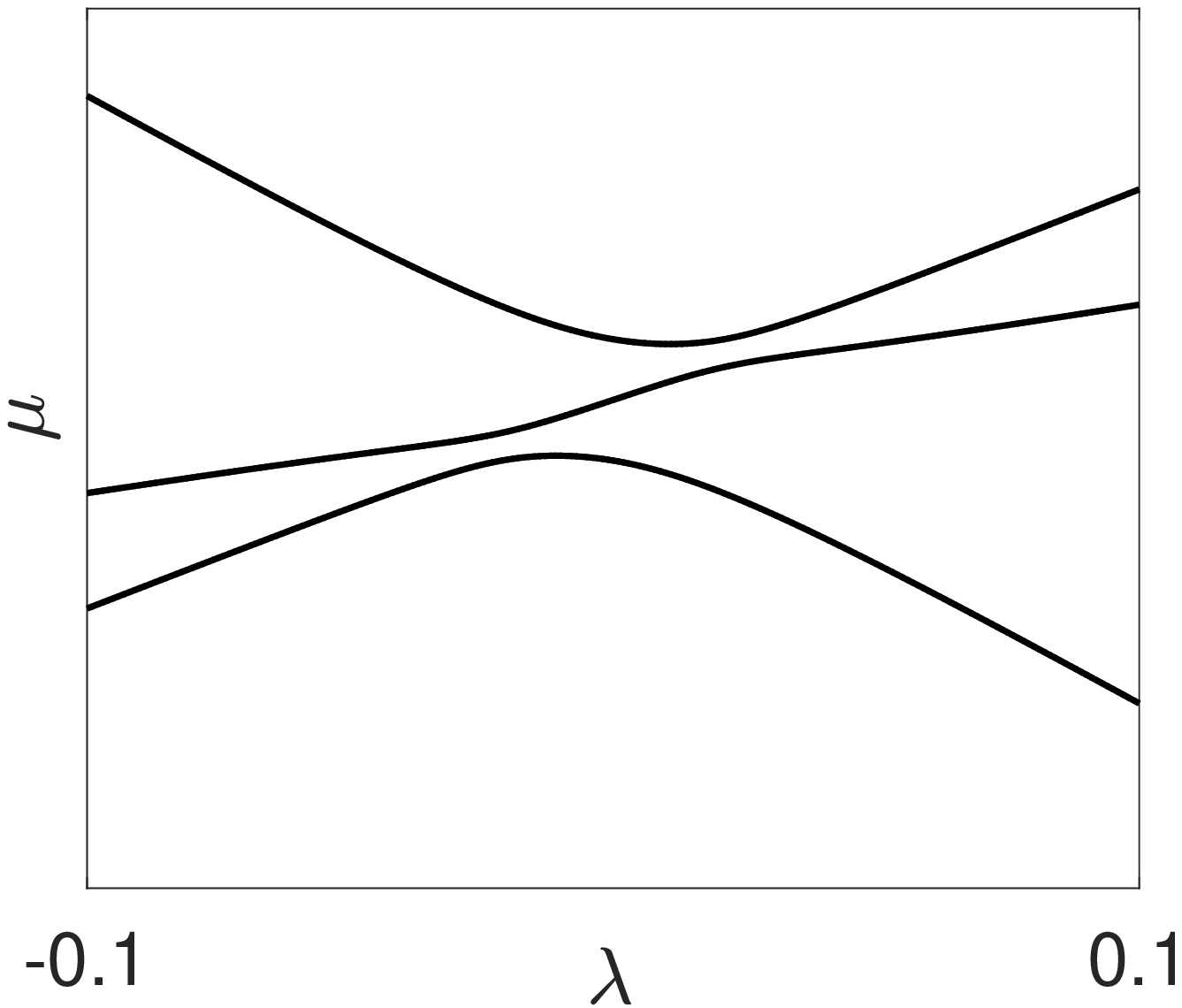}
  \end{minipage}
}
\subfigure[]{
\begin{minipage}{0.3\textwidth}
  \includegraphics[width=0.95\textwidth,height=3cm]{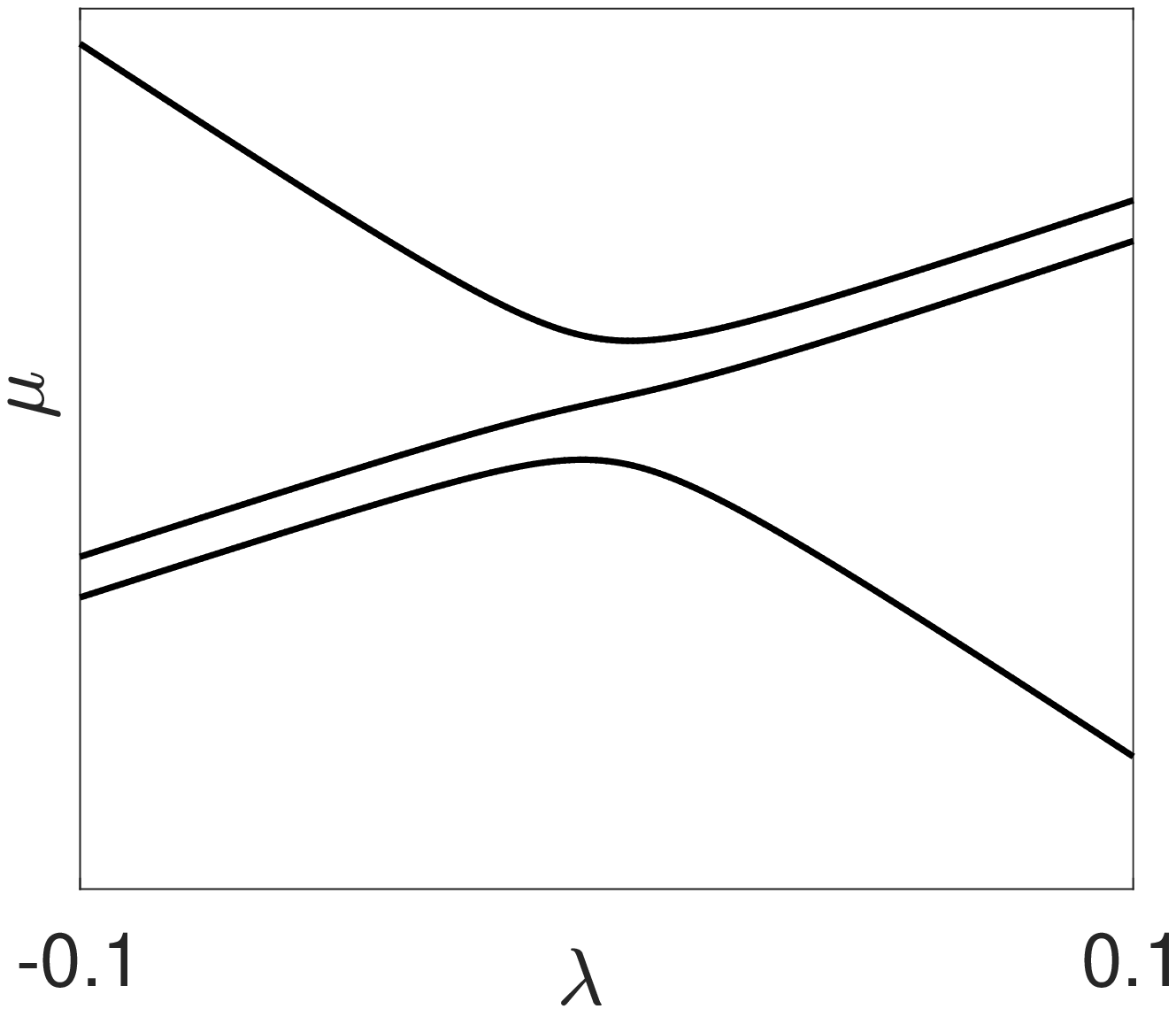}
  \end{minipage}
}

  \quad
  \subfigure[]{
\begin{minipage}{0.3\textwidth}
  \includegraphics[width=0.95\textwidth,height=3.5cm]{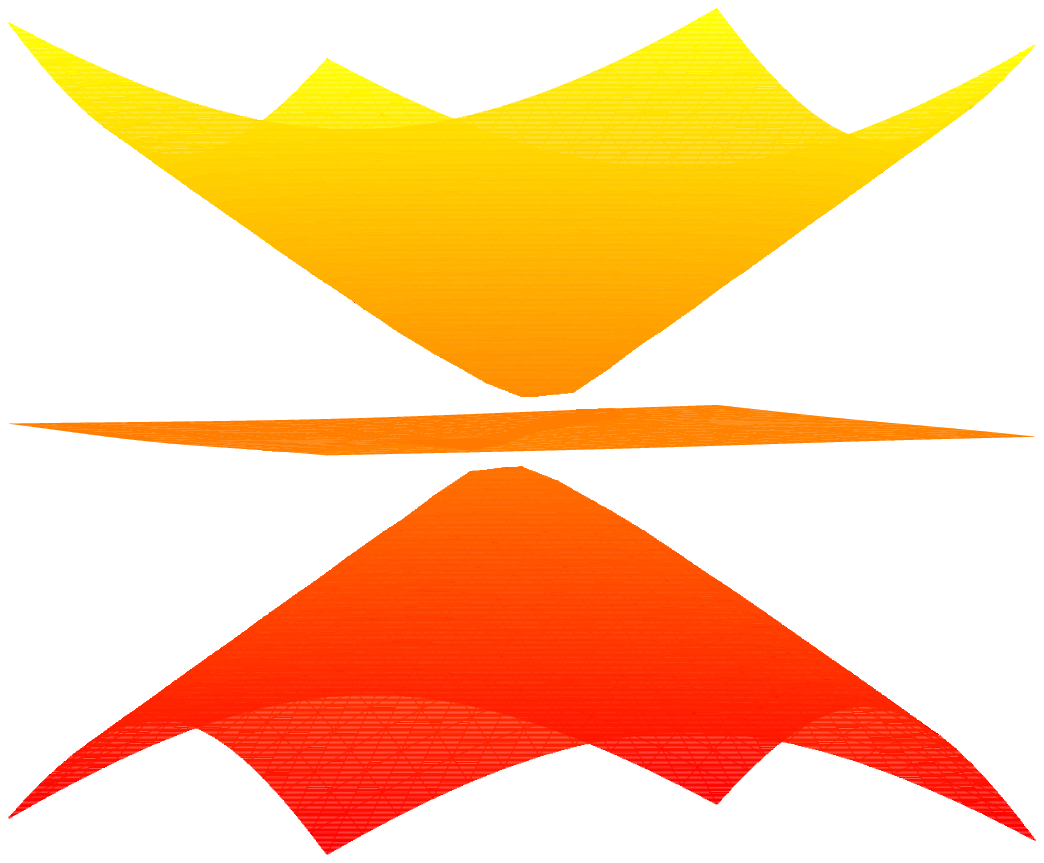}
  \end{minipage}
}
\subfigure[]{
\begin{minipage}{0.3\textwidth}
  \includegraphics[width=0.95\textwidth,height=3.5cm]{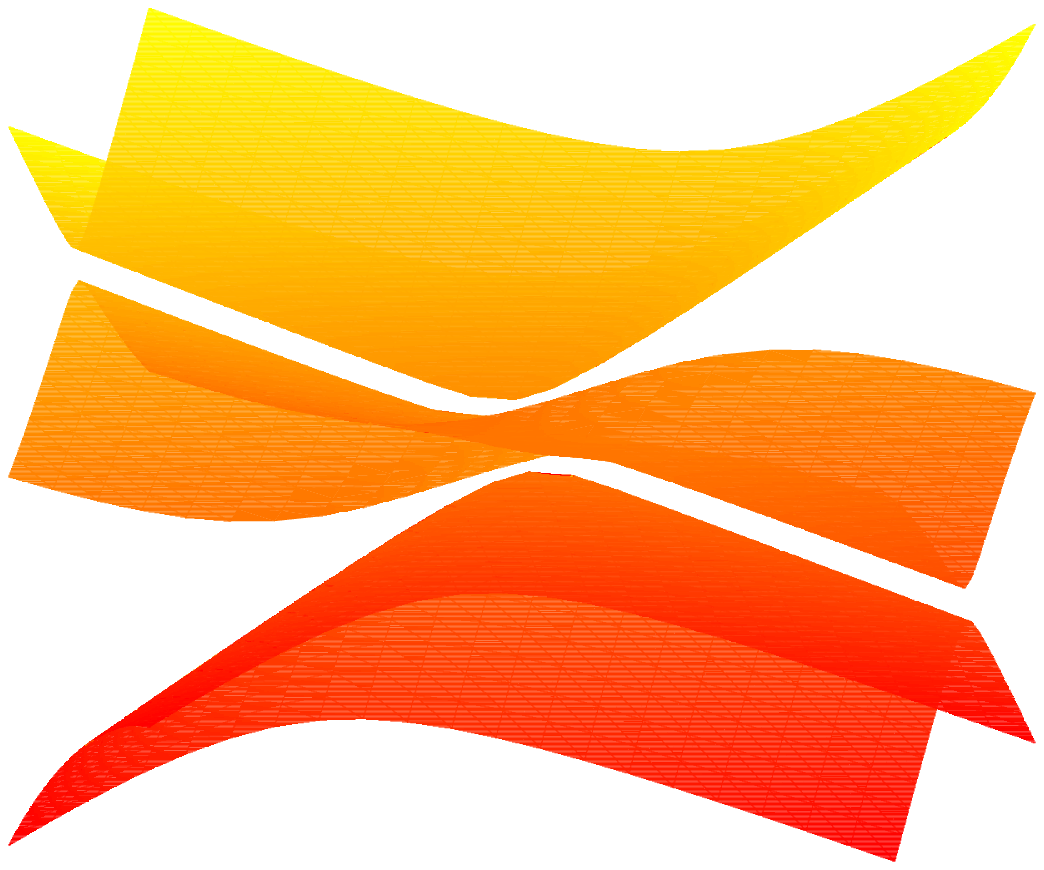}
  \end{minipage}
}
\subfigure[]{
\begin{minipage}{0.3\textwidth}
  \includegraphics[width=0.95\textwidth,height=3.5cm]{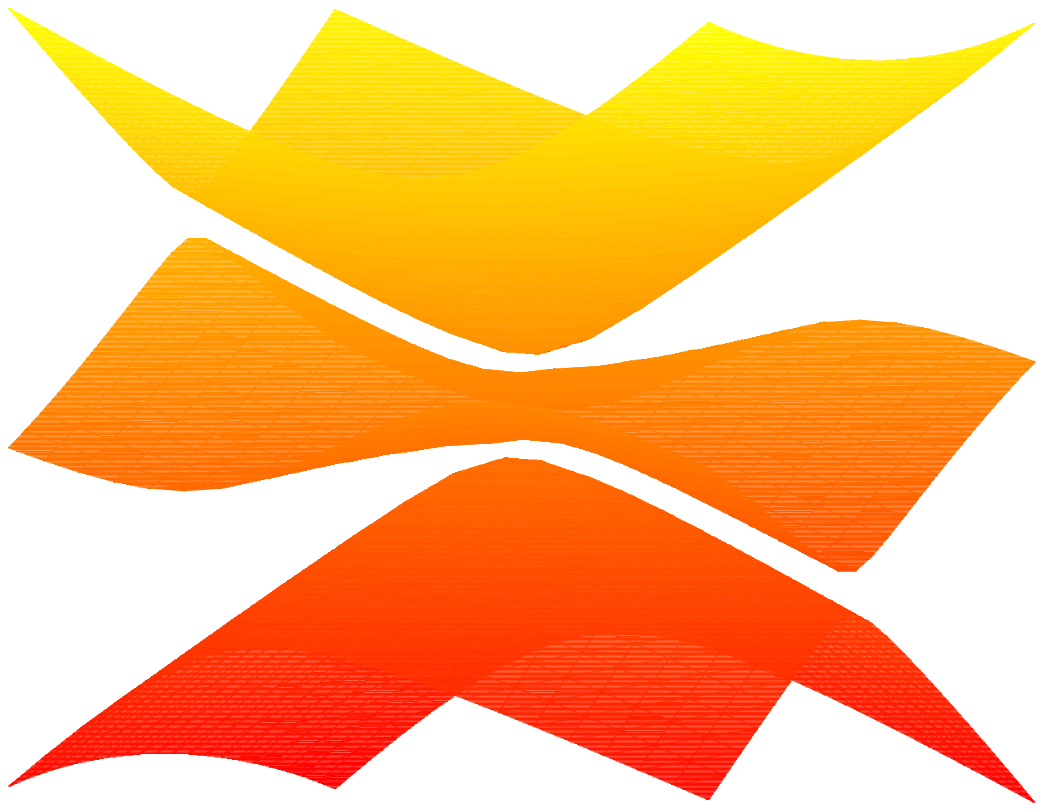}
  \end{minipage}
}
  \caption{The first three energy bands of $H^\delta=-\Delta +V(\bx)+\delta V_p(\bx)$ with the inversion-symmetry-breaking potential $V_p(\bx)$ in \eqref{Vpx}. The setup is the same as that in Fig. \ref{fig:3+1any direction}. The 3-fold degenerate point disappears and two local gaps open. }\label{fig:instability}
\end{figure}

\bu We first examine the role of $\mathfrak{PT}$-symmetry. The perturbation that we choose is
\begin{equation}\lb{Vpx}
\begin{split}
 V_{p_1}(\bx)= &\sin(\bq_{1}\cdot\bx)+\sin((\bq_{2}-\bq_{1})\cdot\bx)
+\sin((\bq_{3}-\bq_{2})\cdot\bx)+\sin(\bq_{3}\cdot\bx)\\
&+\sin(\bq_{2}\cdot\bx)+\sin((\bq_{3}-\bq_{1})\cdot\bx).
\end{split}
\end{equation}
Obviously, $V_{p_1}$ is odd and thus  breaks $\mathfrak{PT}$-invariance of $V(\bx)$. We plot the same energy band functions of $H^{\delta}_{1}$ as shown in Figure \ref{fig:instability}. We see that the three energy band functions separate with each other and two gaps open.

\bu To see the significance of $\mathcal{T}$-invariance in the formation of three-fold conical structures, we consider the perturbation $V_{p}(\bx)$ which breaks $\mathcal{T}$-invariance. In our simulation, we choose
 \begin{equation}\label{Vp2}
\begin{split}
V_{p_2}(\bx)= &\cos(\bq_{1}\cdot\bx)+\cos((\bq_{2}-\bq_{1})\cdot\bx)
+\cos((\bq_{3}-\bq_{2})\cdot\bx)+\cos(\bq_{3}\cdot\bx).
\end{split}
\end{equation}
Obviously, the perturbed potential \eqref{Vp2} possess $\mathcal{R}$-invariance and $\mathfrak{PT}$-invariance, but does not have the $\mathcal{T}$-symmetry since $T\bq_1=\bq_3-\bq_1$.

\begin{figure}[htbp]
\centering
\subfigure[]{
\begin{minipage}{0.3\textwidth}
  \includegraphics[width=0.95\textwidth,height=3cm]{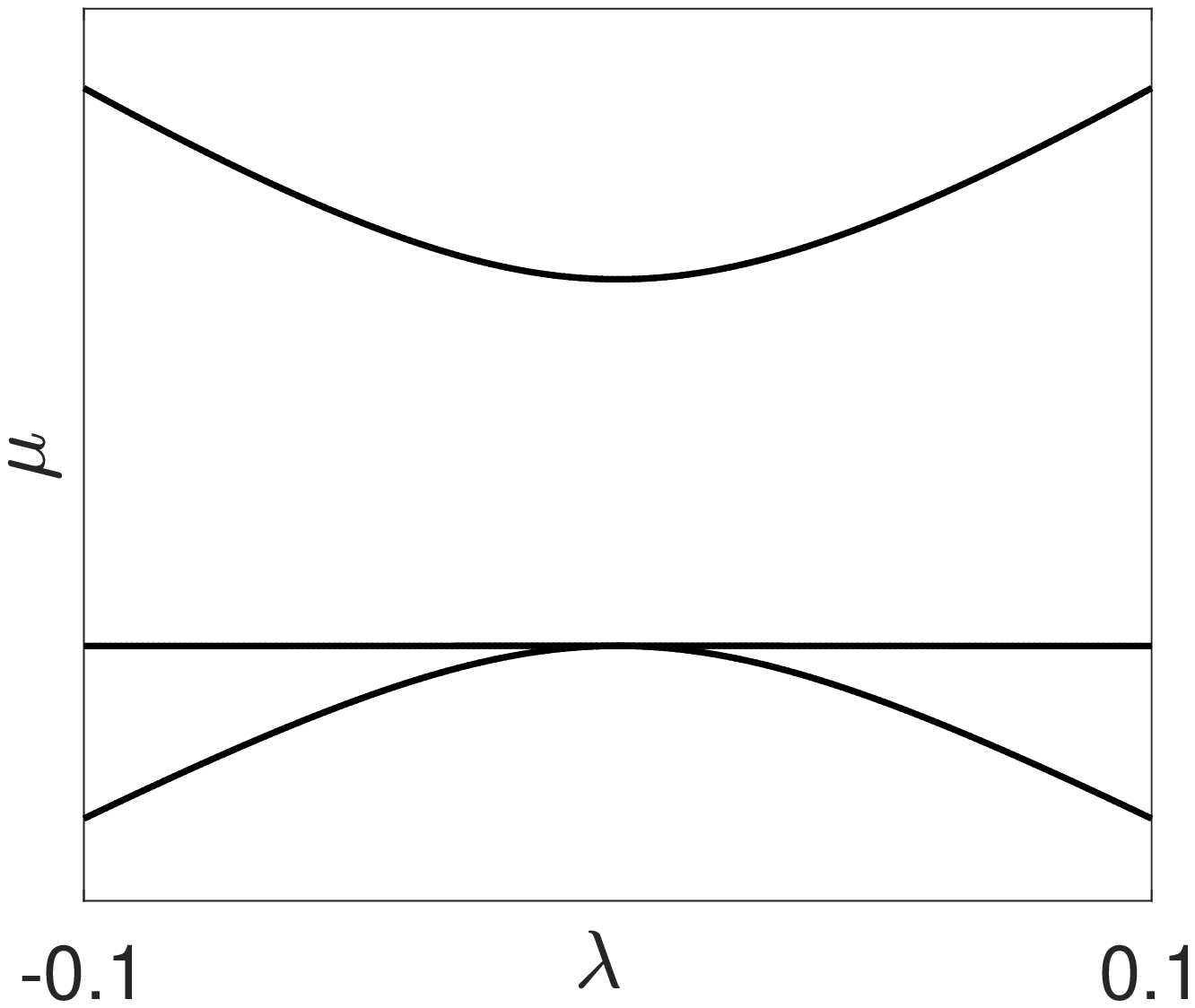}
  \end{minipage}
}
\subfigure[]{
\begin{minipage}{0.3\textwidth}
  \includegraphics[width=0.95\textwidth,height=3cm]{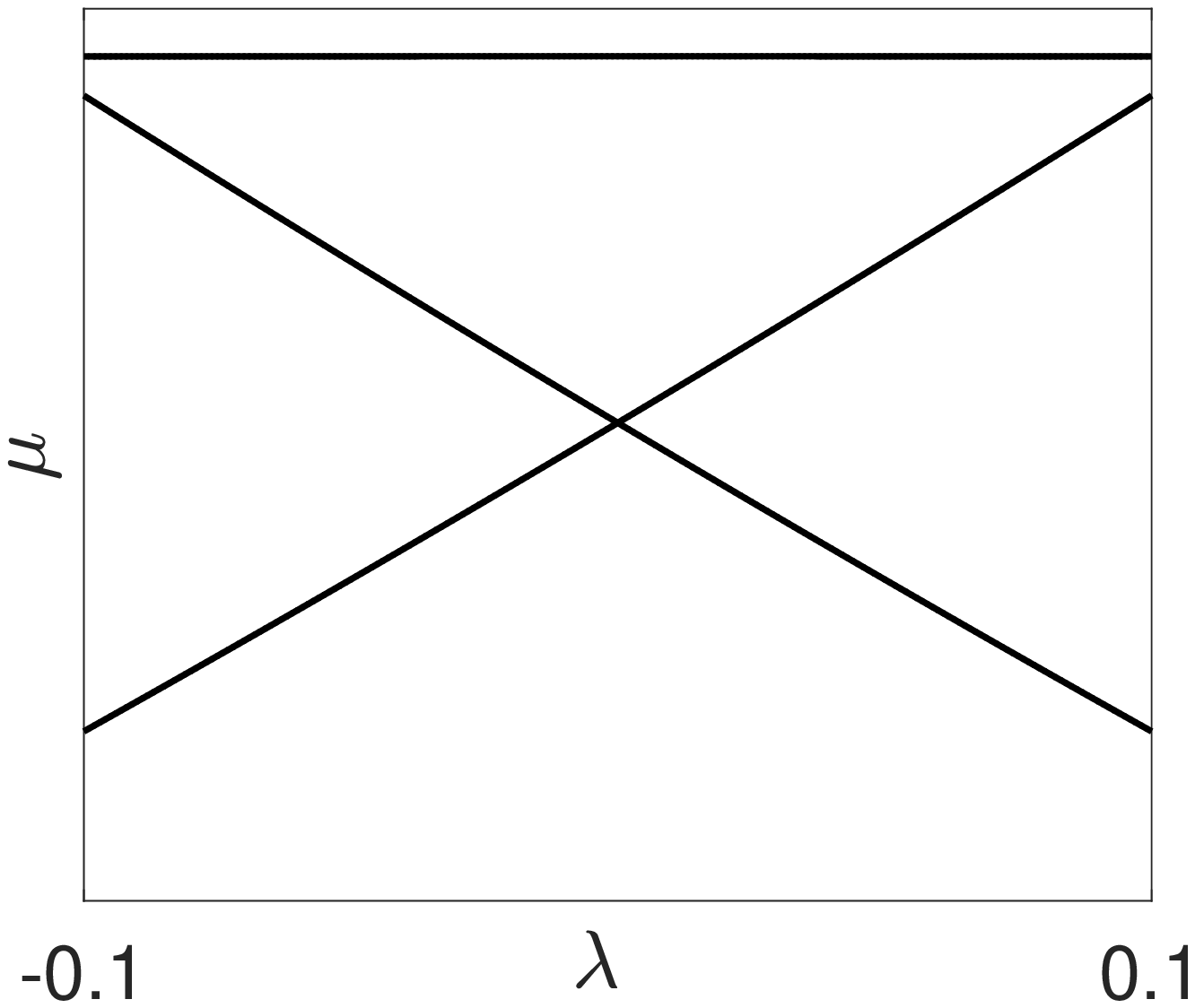}
  \end{minipage}
}
\subfigure[]{
\begin{minipage}{0.3\textwidth}
  \includegraphics[width=0.95\textwidth,height=3cm]{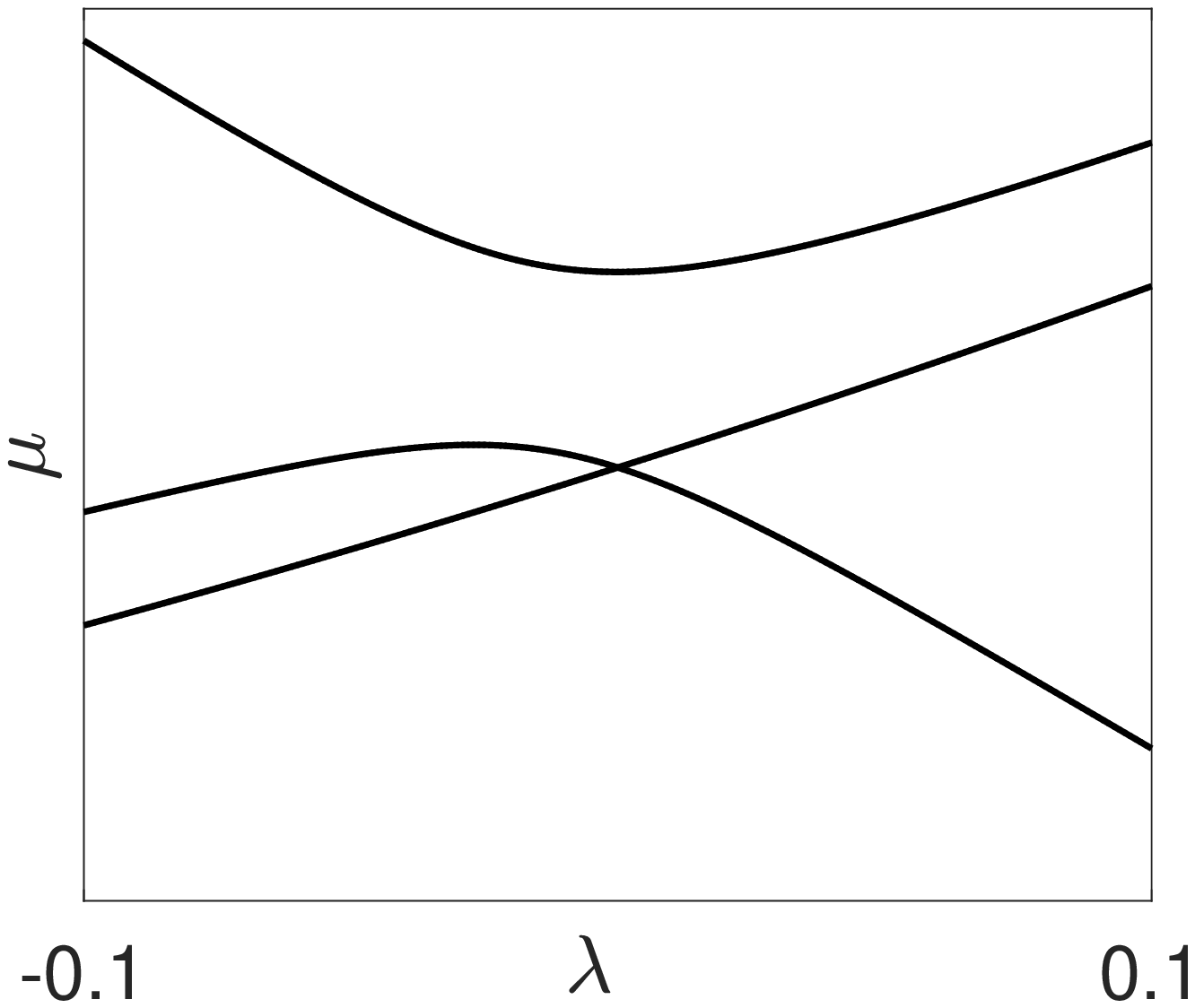}
  \end{minipage}
}

  \quad
  \subfigure[]{
\begin{minipage}{0.3\textwidth}
  \includegraphics[width=0.95\textwidth,height=3.5cm]{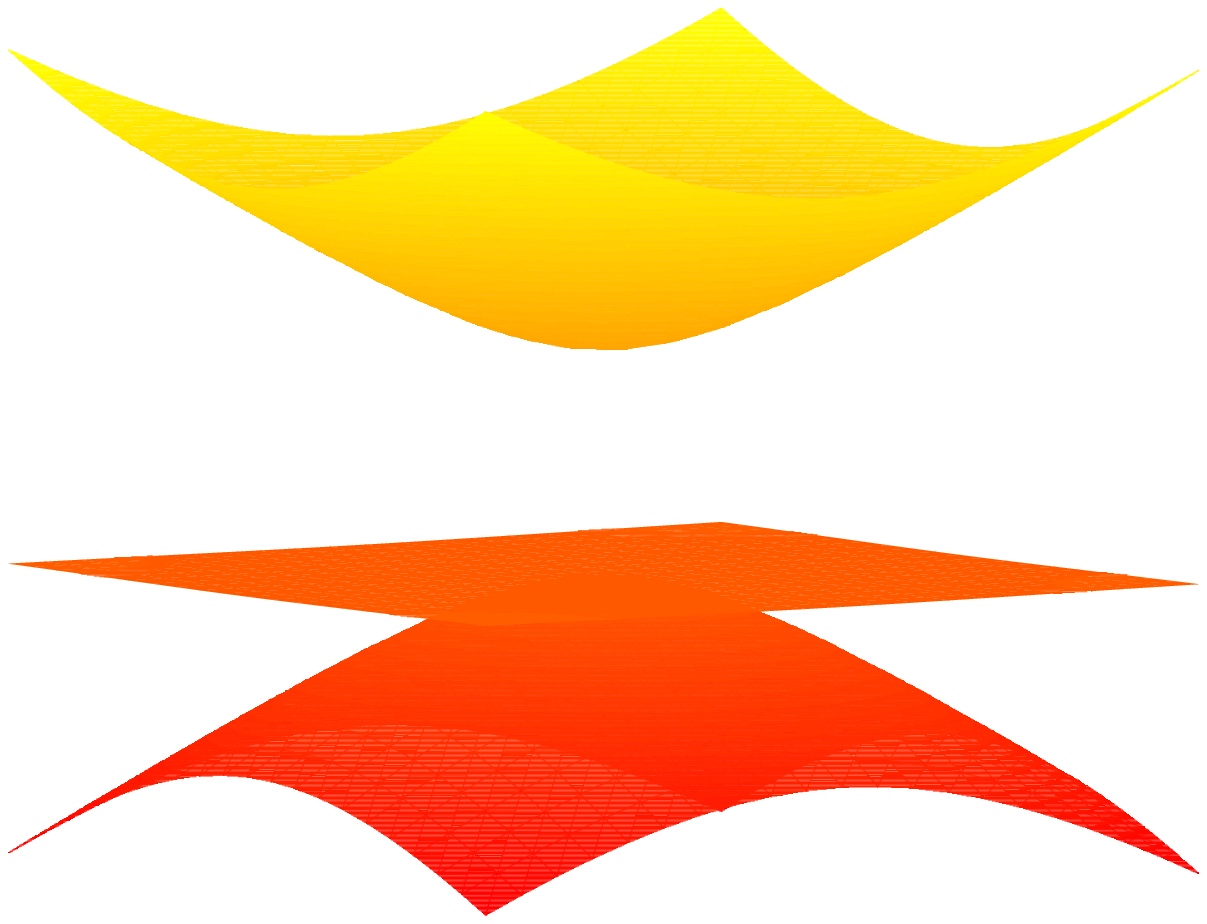}
  \end{minipage}
}
\subfigure[]{
\begin{minipage}{0.3\textwidth}
  \includegraphics[width=0.95\textwidth,height=3.5cm]{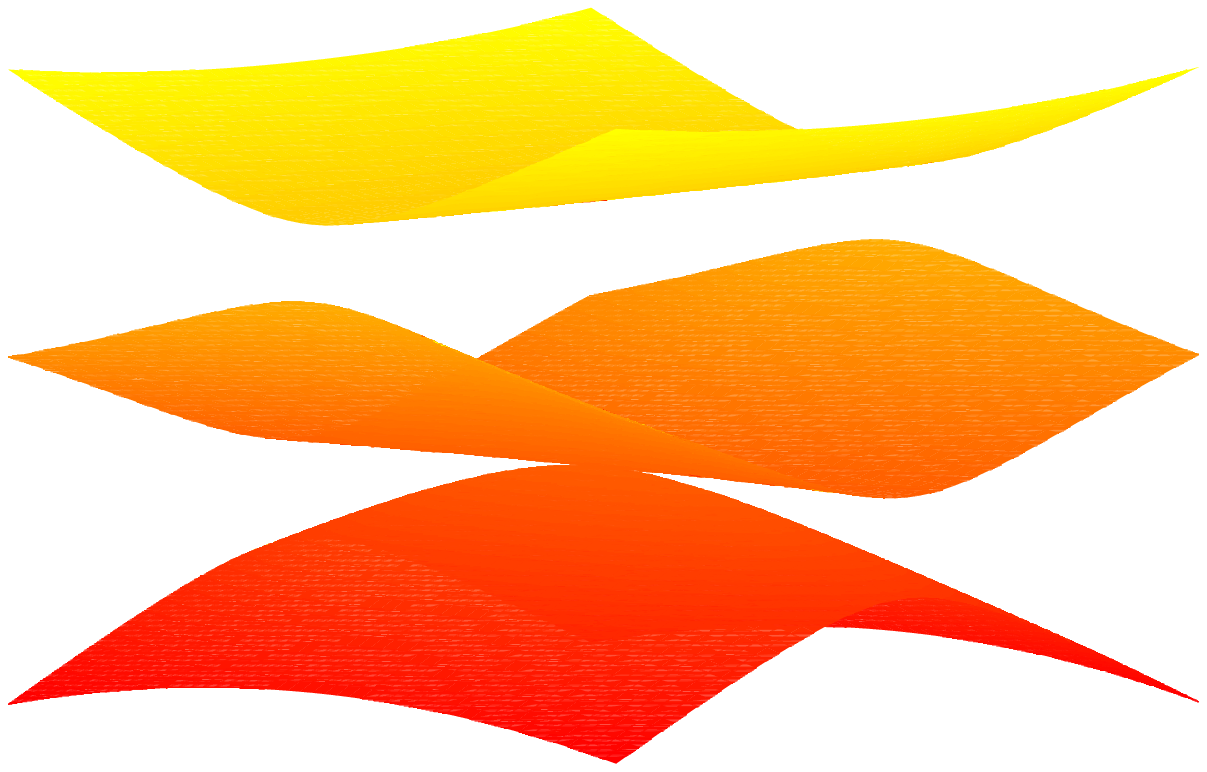}
  \end{minipage}
}
\subfigure[]{
\begin{minipage}{0.3\textwidth}
  \includegraphics[width=0.95\textwidth,height=3.5cm]{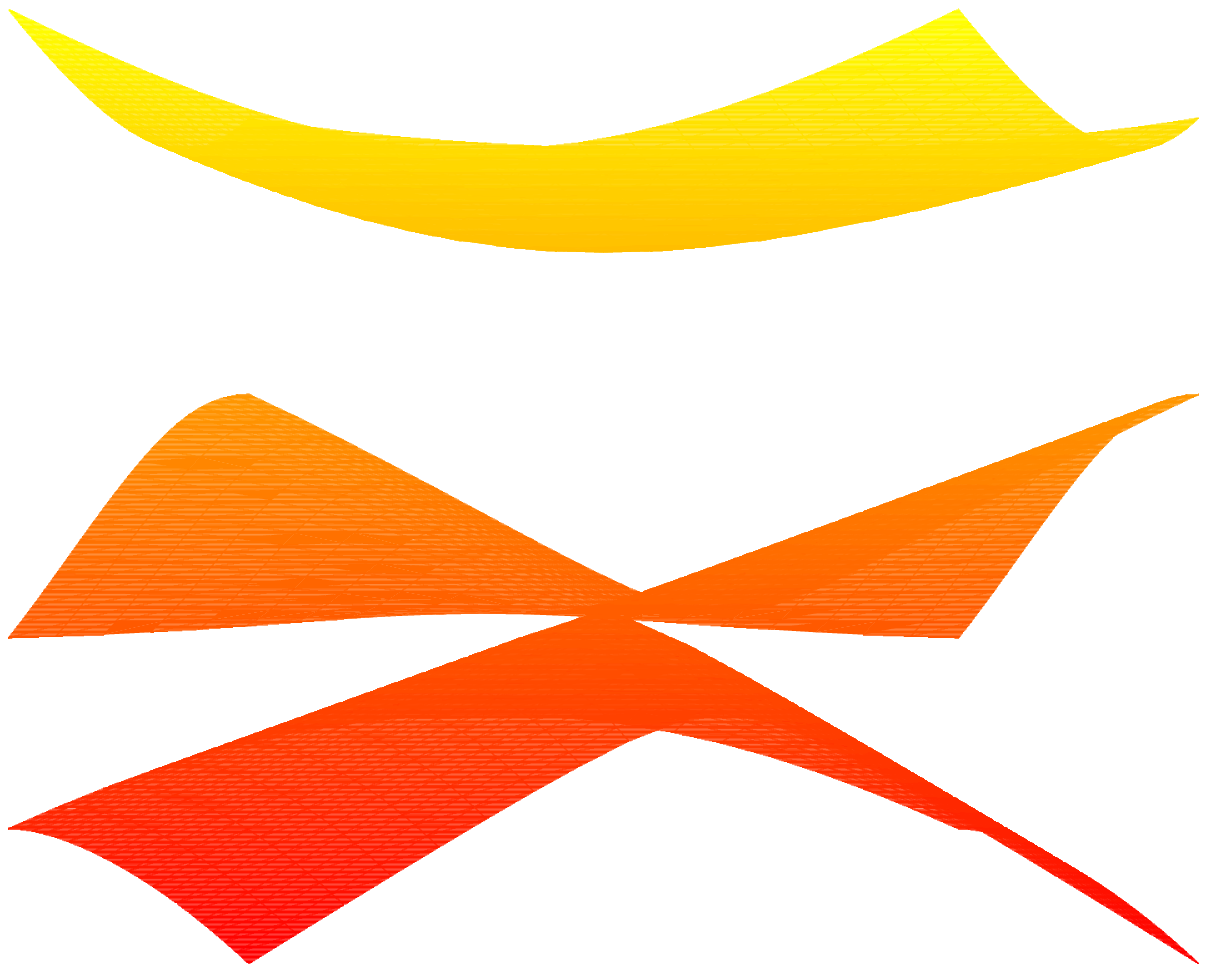}
  \end{minipage}
}
\caption{The first three energy bands of $H^\delta=-\Delta +V(\bx)+\delta V_p(\bx)$ with the $\mathcal{T}$-symmetry-breaking potential $V_p(\bx)$ in \eqref{Vp2}. The setup is the same as that in Fig. \ref{fig:3+1any direction}. The 3-fold degenerate point splits into a two-fold and a simple eigenvalues. The two-fold degeneracy comes from the inversion-symmetry of the system which we keep. There is no general conical structure near the perturbed two-fold degenerate point. }\label{tbrk}
\end{figure}

As before, we display the energy curves and surfaces near $\bK$ in Figure $\ref{tbrk}$. It is shown that the original three-fold degenerate cone structure disappears and breaks into one simple and one double eigenvalue. The nearby structure near the double eigenvalue is not naturally conical. It may correspond to other interesting phenomena but is beyond the scope of this paper.

\begin{appendix}
    \renewcommand{\thesection}{\Alph{section}}
    \section{Proof of Proposition \ref{asphi}}
    \lb{app-a}

    The purpose of this appendix is to give a detailed proof to Proposition \ref{asphi}. We first prove the following lemma.

    \begin{lem}\lb{even}
Let $\mu_{*}$ be an eigenvalue of $H(\bK)$ of eigenvalue problem $\x{mapro}$ with the corresponding eigenspace $\mathcal{E}_{\mu_{*}}$. If $\mathcal{E}_{\mu_{*}}\subset L^{2}_{\bK,i}\oplus L^{2}_{\bK,-i}$, then $\dim\mathcal{E}_{\mu_{*}}$ is even.
\end{lem}

\Proof Let $\Phi\in\mathcal{E}_{\mu}\subset L^{2}_{\bK,i}\oplus L^{2}_{\bK,-i}$. Then $\Phi(\bx)=c_1\Phi_1(\bx)+c_3\Phi_3(\bx)$ for some constants $c_1, \ c_3$, where $\Phi_1 \in L^{2}_{\bK,i}$  and $\Phi_3 \in L^{2}_{\bK,-i}$. We distinguish the following two cases.

\bu $c_1\cdot c_3=0$, say $c_3=0$ for instance. Then $\Phi(\bx)=c_1\Phi_1(\bx)$. Note that $\ol{\Phi(-\bx)}=c_1 \ol{\Phi_1(-\bx)}\in L^{2}_{\bK,-i}$ is linearly independent of $\Phi(\bx)$. Recall that $\{\ol{\Phi(-\bx)},\mu(\bK)\}$ is also an eigenpair of eigenvalue problem \x{HV-k}. We directly obtain $\ol{\Phi_1(-\bx)}\in \mathcal{E}_{\mu_{*}}$.

\bu $c_1\cdot c_3\neq0$. Applying $\mathcal{R}$ to $\Phi(\bx)$, one has $\mathcal{R}[\Phi](\bx)=ic_1\Phi_1(\bx)-ic_3\Phi_3(\bx)\in \mathcal{E}_{\mu_{*}}$. In the present case, it is easy to see that $\mathcal{R}[\Phi](\bx)$ is linearly dependent of $\Phi(\bx)$.

By the above analysis, we conclude that $\dim\mathcal{E}_{\mu}$ is even.\qed

Now we are ready to prove Proposition \ref{asphi}. Since $\Phi_{1}(\bx)\in L^{2}_{\bK,i}$ solves the Floquet-Bloch problem \x{mapro}, then $\Phi_{3}(\bx):=\ol{\Phi_{1}(-\bx)}\in L^{2}_{\bK,-i}$ is also an eigenfunction.

As $\dim\mathcal{E}_{\mu_{*}}=3$, there exists $\Phi'(\bx)\in\mathcal{E}_{\mu_{*}}$ such that $\Phi'(\bx)\notin L^{2}_{\bK,i}\oplus L^{2}_{\bK,-i}$. Hence
    \be\lb{phi} \Phi'(\bx)=c_1 \Phi''_1(\bx)+c_2 \Phi''_2(\bx)+c_3 \Phi''_3(\bx)\in \mathcal{E}_\mu,
    \ee
where $c_2\neq 0$ and $\Phi''_{\ell}\in L^{2}_{\bK,i^{\ell}}$ satisfy $\mathcal{R}[\Phi''_{\ell}]=
i^{\ell}\Phi''_{\ell}$ for $\ell=1,2,3$.

We aim at proving the assertion that
$\Phi''_2(\bx)\in \mathcal{E}_{\mu_{*}}$.

 \bu  Case 1: $c_1=c_3=0$. In this case the assertion is trivial from \x{phi}.

 \bu Case 2: One of $c_1$ and $c_3$ is nonzero and another is zero, say $c_1\ne 0$ and $c_3=0$. Then we have equalities
 \[
 \begin{split}
 & \Phi'(\bx)=c_1\Phi''_1(\bx)+c_2\Phi''_2(\bx)\in \mathcal{E}_{\mu}, \\
 & \mathcal{R}[\Phi'](\bx)=ic_1\Phi''_1(\bx)-c_2\Phi''_2(\bx)\in \mathcal{E}_{\mu},\\
 & \mathcal{R}[\Phi'](\bx)+\Phi'(\bx)=(i+1)c_1\Phi''_1(\bx)\in L^{2}_{\bK,i}\cap \mathcal{E}_{\mu}\ .
  \end{split}
  \] Since the multiplicity in $L^{2}_{\bK,i}$ is one, we conclude from the last equality that
$\Phi''_1(\bx)=\al \Phi_1(\bx)$ for some $\al$. Consequently, we conclude from the first equality that $\Phi''_2(\bx)\in\mathcal{E}_{\mu}$.

 \bu Case 3: Both $c_1$ and $c_3$ are nonzero. Basing on the decomposition \x{phi}, one has
  \[
  \mathcal{R}[\Phi'](\bx)=ic_1\Phi''_1(\bx)-c_2\Phi''_2(\bx)-ic_3\Phi''_3(\bx).
  \]
 Then
   \be\lb{Phi''}
   \Phi''(\bx):=\mathcal{R}[\Phi'](\bx)+\Phi'(\bx)=k_1\Phi''_1(\bx)+k_2\Phi''_3(\bx)\in L^{2}_{\bK,i}\oplus L^{2}_{\bK,-i}
   \ee
   where $k_1=c_1(i+1)$ and $k_2=c_3(1-i)$.

Assume that $\Phi''\not\in {\rm span}\{\Phi_1,\Phi_3\}$. Without loss of generality, we assume that $\Phi''_1(\bx)$ is linearly independent of $\Phi_1(\bx)$. Then $\mathcal{R}[\Phi''](\bx)+i\Phi''(\bx)=2ik_1\Phi''(\bx)\in L^{2}_{\bK,i}$, which would imply that $\mu(\bK)$ is not a three-fold eigenvalue. Thus we have shown that $\Phi''\in {\rm span}\{\Phi_1,\Phi_3\}$ is a linear combination of $\Phi_1$ and $\Phi_3$. It then follows from \x{Phi''} that $\Phi''_1=\al \Phi_1$ and $\Phi''_3=\bt \Phi_3$ for some constants $\al, \ \bt$. Going back to \x{phi}, we obtain  
   $\Phi'(\bx)=c'_1\Phi_1(\bx)+c_2\Phi''_2(\bx)+c'_3\Phi_3(\bx)$. This leads to the assertion $\Phi''_2(\bx)\in\mathcal{E}_{\mu_{*}}$.

The proof is complete. \qed
%
%
%



\section{Proof of Lemma $\ref{strut}$ }

In this appendix, we actually give a proof of a stronger conclusion. Assume that $\Phi^{c}(\bx)\in L^{2}_{\bK,i}\oplus L^{2}_{\bK,-i}$ has the form
\[\Phi^{c}(\bx)=\Phi_1(\bx)+\Phi_3(\bx),\]
where $\Phi_1(\bx)\in L^{2}_{\bK,i}$ and $\Phi_3(\bx)\in L^{2}_{\bK,-i}$ are of the form \eqref{decomp2}. That is,
\[
\begin{split}
\Phi_1(\bx)&=c_1 \Phi^{(0)}_1 (\bx)+\Phi^{h}_1(\bx)= c_1 \Phi^{(0)}_1 (\bx)+\sum_{\bq\in\ssk\setminus \{(0,0,0)\}}\Phi_{\bq}(e^{i(\bK+\bq)\cdot\bx}-ie^{iR(\bK+\bq)\cdot\bx}
\\ &- e^{iR^{2}(\bK+\bq)\cdot\bx} +ie^{iR^{3}(\bK+\bq)\cdot\bx}),\\
\Phi_3(\bx)&=c_3 \Phi^{(0)}_1 (\bx)+\Phi^{h}_3(\bx)=c_3 \Phi^{(0)}_3 (\bx)+\sum_{\bq\in\ssk\setminus \{(0,0,0)\}}\Phi_{\bq}(e^{i(\bK+\bq)\cdot\bx}+ie^{iR(\bK+\bq)\cdot\bx}\\
&-e^{iR^{2}(\bK+\bq)\cdot\bx} +i e^{iR^{3}(\bK+\bq)\cdot\bx})\ .
\end{split}
\]
By the symmetry, we have the following conclusion.

    \begin{lem}\lb{tsrt}
If $|c_1|+|c_3|>0$, then $\mathcal{T}\Phi^{c}(\bx)\notin L^{2}_{\bK,i}\oplus L^{2}_{\bK,-i}$.
\end{lem}

Note that Lemma $\ref{strut}$ is just a direct consequence of above conclusion. Indeed, let us recall that $\Phi^{\e}_1(\bx)=(1+\mathcal{O}(\e))\Phi^{(0)}_1(\bx)+\Phi^{h}_{1}(\bx)$. Thus for sufficiently small $\e$, $\Phi^{\e}_1(\bx)$ satisfies the conditions of Lemma $\ref{tsrt}$, i.e., $c_1=1+O(\e)\neq 0$.  So $\mathcal{T}\Phi^{\e}_1(\bx)\notin L^{2}_{\bK,i}\oplus L^{2}_{\bK,-i}$.
\qed

It remains to prove Lemma $\ref{tsrt}$. We begin the proof by considering the action
$\mathcal{T}$ on $\Phi_\ell^{(0)}(\bx)$. By employing $\mathcal{T}$ on $e^{iR^{\ell}\bK\cdot\bx}$ accordingly, we obtain
 \[
 \begin{split}
 &\mathcal{T}\Phi^{(0)}_1=\frac{1}{\sqrt{4|\Omega|}}(-e^{i\bK\cdot\bx}+e^{iR\bK\cdot\bx}+ie^{iR^2\bK\cdot\bx}-ie^{iR^3\bK\cdot\bx}),\\
 &\mathcal{T}\Phi^{(0)}_2=\frac{1}{\sqrt{4|\Omega|}}(e^{i\bK\cdot\bx}+e^{iR\bK\cdot\bx}-e^{iR^2\bK\cdot\bx}-e^{iR^3\bK\cdot\bx}),\\
 &\mathcal{T}\Phi^{(0)}_3=\frac{1}{\sqrt{4|\Omega|}}(-e^{i\bK\cdot\bx}+e^{iR\bK\cdot\bx}-ie^{iR^2\bK\cdot\bx}+ie^{iR^3\bK\cdot\bx}),\\
 &\mathcal{T}\Phi^{(0)}_4=\frac{1}{\sqrt{4|\Omega|}}(e^{i\bK\cdot\bx}+e^{iR\bK\cdot\bx}+e^{iR^2\bK\cdot\bx}+e^{iR^3\bK\cdot\bx}).
 \end{split}
 \]

Obviously $\mathcal{T}\Phi^{(0)}_4(\bx)=\Phi^{(0)}_4(\bx)$.  By direct calculations, we have the following relations between $\{\mathcal{T}\Phi^{(0)}_{\ell}(\bx)\}^{3}_{\ell=1}$ and $\{\Phi^{(0)}_{\ell}(\bx)\}^{3}_{\ell=1}$
\be\lb{Tact}
\begin{pmatrix}
\mathcal{T}\Phi^{(0)}_{1}\\
\mathcal{T}\Phi^{(0)}_2\\
\mathcal{T}\Phi^{(0)}_3
\end{pmatrix}
=
Q^{0}_{\mathcal{T}}
\begin{pmatrix}
\Phi^{(0)}_{1}\\
\Phi^{(0)}_2\\
\Phi^{(0)}_3
\end{pmatrix},
\ee
where
\be\lb{Cm}
Q^{0}_{\mathcal{T}}=
\begin{pmatrix}
-\frac{1}{2}&-\frac{1}{2}+\frac{i}{2}&-\frac{i}{2}\\
\frac{1}{2}+\frac{i}{2}&0&\frac{1}{2}-\frac{i}{2}\\
\frac{i}{2}&-\frac{1}{2}-\frac{i}{2}&-\frac{1}{2}
\end{pmatrix}.
\ee

Assume that $\mathcal{T}\Phi^{c}\in L^{2}_{\bK,i}\oplus L^2_{\bK,-i}$. Then
\[
\mathcal{T}\Phi^{c}=\mathcal{T}\Phi_1+\mathcal{T}\Phi_3 =c_1\mathcal{T}\Phi^{(0)}_{1}+c_3\mathcal{T}\Phi^{(0)}_3+\mathcal{T}\Phi^{h}_1+\mathcal{T}\Phi^{h}_3\ .
\]
By the relations in \x{Tact} and \x{Cm}, we have
\[
c_1(-\frac12 +\frac{i}{2})+c_3(-\frac{1}{2}-\frac{i}{2})=-(c_1+c_3)\frac{1}{2}+\frac{i}{2}(c_1-c_3)=0\ .
\]
This implies $c_1 =c_3 =0$, which contradicts with $|c_1|+|c_3|>0$. Therefore, $\mathcal{T}\Phi^{c}(\bx)\notin L^{2}_{\bK,i}\oplus L^{2}_{\bK,-i}$.
\qed

\end{appendix}
\bibliography{citations}

\begin{thebibliography}{10}

\bibitem{Ablowitz2013}
M.~Ablowitz, C.~Curtis, and Y.~Zhu.
\newblock Localized nonlinear edge states in honeycomb lattices.
\newblock {\em Phys. Rev. A}, 88:013850, Jul 2013.

\bibitem{Ablowitz09}
M.~Ablowitz, S.~Nixon, and Y.~Zhu.
\newblock Conical diffraction in honeycomb lattices.
\newblock {\em Phys. Rev. A}, 79:053830, May 2009.

\bibitem{Ablowitz2012}
M.~Ablowitz and Y.~Zhu.
\newblock Nonlinear waves in shallow honeycomb lattices.
\newblock {\em SIAM J. Appl. Math.}, 72(1):240--260, 2012.

\bibitem{Allaire08}
G.~Allaire, M.~Palombaro, and J.~Rauch.
\newblock Diffractive geometric optics for {B}loch wave packets.
\newblock {\em Arch. Ration. Mech. Anal.}, 202(2):373--426, 2011.

\bibitem{Ammari2018honeycomblattice}
H.~Ammari, H.~Fitzpatrick, E.O. Hiltunen, H.~Lee, and S.~Yu.
\newblock Honeycomb-lattice {M}innaert bubbles.
\newblock {\em SIAM J. Math. Anal.}, 52(6):5441--5466, 2020.

\bibitem{Ammari2020}
H.~Ammari, E.~Hiltunen, and S.~Yu.
\newblock A high-frequency homogenization approach near the {D}irac points in
  bubbly honeycomb crystals.
\newblock {\em Arch. Ration. Mech. Anal.}, 238(3):1559--1583, 2020.

\bibitem{Avila2012}
C.~Avila, H.~Schulz-Baldes, and C.~Villegas-Blas.
\newblock Topological invariants of edge states for periodic two-dimensional
  models.
\newblock {\em Math. Phys. Anal. Geom.}, 16(2):137--170, 2013.

\bibitem{Bal2019}
G.~Bal.
\newblock Continuous bulk and interface description of topological insulators.
\newblock {\em J. Math. Phys.}, 60(8):081506, 20, 2019.

\bibitem{Berkolaiko14}
G.~Berkolaiko and A.~Comech.
\newblock Symmetry and {D}irac points in graphene spectrum.
\newblock {\em J. Spectr. Theory}, 8(3):1099--1147, 2018.

\bibitem{Castro07}
A.~H. Castro~Neto, F.~Guinea, N.~M.~R. Peres, K.~S. Novoselov, and A.~K. Geim.
\newblock The electronic properties of graphene.
\newblock {\em Rev. Mod. Phys.}, 81:109--162, Jan 2009.

\bibitem{Drouot2020ubiquity}
A.~Drouot.
\newblock Ubiquity of conical points in topological insulators.
\newblock {\em arXiv:2004.07068}, 2020.

\bibitem{Drouot2020}
A.~Drouot and M.I. Weinstein.
\newblock Edge states and the valley {H}all effect.
\newblock {\em Adv. Math.}, 368:107142, 51, 2020.

\bibitem{Eastham1973}
M.~S.~P. Eastham.
\newblock {\em The spectral theory of periodic differential equations}.
\newblock Texts in Mathematics (Edinburgh). Scottish Academic Press, Edinburgh;
  Hafner Press, New York, 1973.

\bibitem{Fefferman2014}
C.~L. Fefferman, J.~P. Lee-Thorp, and M.~I. Weinstein.
\newblock Topologically protected states in one-dimensional continuous systems
  and {D}irac points.
\newblock {\em Proc. Natl. Acad. Sci. USA}, 111(24):8759--8763, 2014.

\bibitem{Fefferman2015}
C.~L. Fefferman, J.~P. Lee-Thorp, and M.~I. Weinstein.
\newblock Edge states in honeycomb structures.
\newblock {\em Ann. PDE}, 2(2):Art. 12, 80, 2016.

\bibitem{Weinstein2016}
C.L. Fefferman, J.~Lee-Thorp, and M.I. Weinstein.
\newblock Honeycomb {S}chr\"{o}dinger operators in the strong binding regime.
\newblock {\em Comm. Pure Appl. Math.}, 71(6):1178--1270, 2018.

\bibitem{fefferman2012honeycomb}
C.L. Fefferman and M.I. Weinstein.
\newblock Honeycomb lattice potentials and {D}irac points.
\newblock {\em J. Amer. Math. Soc.}, 25(4):1169--1220, 2012.

\bibitem{Goldman2012}
N.~Goldman, J.~Beugnon, and F.~Gerbier.
\newblock Detecting chiral edge states in the hofstadter optical lattice.
\newblock {\em Phys. Rev. Lett.}, 108:255303, Jun 2012.

\bibitem{Graf2012}
G.~Graf and M.~Porta.
\newblock Bulk-edge correspondence for two-dimensional topological insulators.
\newblock {\em Comm. Math. Phys.}, 324(3):851--895, 2013.

\bibitem{hu2020linear}
P.~Hu, L.~Hong, and Y.~Zhu.
\newblock Linear and nonlinear electromagnetic waves in modulated honeycomb
  media.
\newblock {\em Stud. Appl. Math.}, 144(1):18--45, 2020.

\bibitem{Keller2018}
R.~T. Keller, J.~L. Marzuola, B.~Osting, and M.~I. Weinstein.
\newblock Spectral band degeneracies of {$\frac{\pi}{2}$}-rotationally
  invariant periodic {S}chr\"{o}dinger operators.
\newblock {\em Multiscale Model. Simul.}, 16(4):1684--1731, 2018.

\bibitem{Kuchment1993}
P.~Kuchment.
\newblock {\em Floquet theory for partial differential equations}, volume~60 of
  {\em Operator Theory: Advances and Applications}.
\newblock Birkh\"{a}user Verlag, Basel, 1993.

\bibitem{Kuchment2015}
P.~Kuchment.
\newblock An overview of periodic elliptic operators.
\newblock {\em Bull. Amer. Math. Soc. (N.S.)}, 53(3):343--414, 2016.

\bibitem{Kuchment2000}
P.~Kuchment and S.~Levendorski\v{i}.
\newblock On the structure of spectra of periodic elliptic operators.
\newblock {\em Trans. Amer. Math. Soc.}, 354(2):537--569, 2002.

\bibitem{Lee2016}
M.~Lee.
\newblock Dirac cones for point scatterers on a honeycomb lattice.
\newblock {\em SIAM J. Math. Anal.}, 48(2):1459--1488, 2016.

\bibitem{Lee-Thorp2017}
J.~P. Lee-Thorp, M.~I. Weinstein, and Y.~Zhu.
\newblock Elliptic operators with honeycomb symmetry: {D}irac points, edge
  states and applications to photonic graphene.
\newblock {\em Arch. Ration. Mech. Anal.}, 232(1):1--63, 2019.

\bibitem{Mukherjee14}
S.~Mukherjee, A.~Spracklen, D.~Choudhury, N.~Goldman, P.~\"{O}hberg,
  E.~Andersson, and R.~Thomson.
\newblock Observation of a localized flat-band state in a photonic lieb
  lattice.
\newblock {\em Phys. Rev. Lett.}, 114:245504, Jun 2015.

\bibitem{Novoselov11}
K.~S. Novoselov.
\newblock Nobel lecture: Graphene: Materials in the flatland.
\newblock {\em Rev. Mod. Phys.}, 83:837--849, Aug 2011.

\bibitem{leng2020multiband}
M.~L. Patrick, X.~Liu, S.S. Tsirkin, T.~Neupert, and T.~Bzdu$\check{s}$ek.
\newblock Multi-band nodal links in triple-point materials.
\newblock {\em arXiv:2008.02807}, 2020.

\bibitem{Perez2014}
P.~Perez-Piskunow, G.~Usaj, C.~A. Balseiro, and L.~F.~Torres.
\newblock Floquet chiral edge states in graphene.
\newblock {\em Phys. Rev. B}, 89:121401, Mar 2014.

\bibitem{method1972Reed}
M.~Reed and B.~Simon.
\newblock {\em Methods of modern mathematical physics. {I}. {F}unctional
  analysis}.
\newblock Academic Press, New York-London, 1972.

\bibitem{Tang17}
P.~Tang, Q.~Zhou, and S.~Zhang.
\newblock Multiple types of topological fermions in transition metal silicides.
\newblock {\em Phys. Rev. Lett.}, 119:206402, Nov 2017.

\bibitem{Wallace1947}
P.~R. Wallace.
\newblock The band theory of graphite.
\newblock {\em Phys. Rev.}, 71:622--634, May 1947.

\bibitem{Watson18}
A.~Watson and M.I. Weinstein.
\newblock Wavepackets in inhomogeneous periodic media: propagation through a
  one-dimensional band crossing.
\newblock {\em Comm. Math. Phys.}, 363(2):655--698, 2018.

\bibitem{Xie2019}
P.~Xie and Y.~Zhu.
\newblock Wave packet dynamics in slowly modulated photonic graphene.
\newblock {\em J. Differential Equations}, 267(10):5775--5808, 2019.

\bibitem{xie2020wave}
P.~Xie and Y.~Zhu.
\newblock Wave packets in the fractional nonlinear schr\"{o}dinger equation
  with a honeycomb potential.
\newblock {\em arXiv:2006.05928}, 2020.

\bibitem{Yang2010}
J.~Yang.
\newblock {\em Nonlinear Waves in Integrable and Non-Integrable Systems}.
\newblock Society for Industrial and Applied Mathematics, USA, 2010.

\bibitem{Yang2017}
Z.~Yang, M.~Xiao, F.~Gao, L.~Lu, Y.~Chong, and B.~Zhang.
\newblock Weyl points in a magnetic tetrahedral photonic crystal: erratum.
\newblock {\em Optics Express}, 25:23725, 10 2017.

\bibitem{Lu2020Double}
T.~Zhang, Z.~Song, A.~Alexandradinata, H.~Weng, C.~Fang, L.~Lu, and Z.~Fang.
\newblock Double-weyl phonons in transition-metal monosilicides.
\newblock {\em Phys. Rev. Lett.}, 120:016401, Jan 2018.

\end{thebibliography}
\end{document}